\documentclass[9pt,twocolumn,twoside]{opticajnl}
\journal{opticajournal} 

\setboolean{shortarticle}{false}

\usepackage{graphicx}
\usepackage{dcolumn}
\usepackage{bm}

\usepackage{extarrows}
\usepackage{physics}

\usepackage{commath} 

\usepackage{siunitx}
\usepackage{comment}
\newcommand{\SIadj}[2]{\SI[number-unit-product={\text{-}}]{#1}{#2}}
\DeclareSIUnit\bar{bar}

\newcommand{\etal}[1]{\mbox{{#1} \textit{et al.}}}
\newcommand{\ie}{\textit{i}.\textit{e}.,\ }
\newcommand{\eg}{\textit{e}.\textit{g}.,\ }
\newcommand*{\Exp}{\mathrm{e}}
\newcommand*\diff{\mathop{}\!\mathrm{d}}

\title{Highly-efficient perturbative Raman shifting by engineering the nonlinear temporal response}

\newcommand{\RN}[1]{%
\textup{\uppercase\expandafter{\romannumeral#1}}%
} 
\newcommand{\rn}[1]{%
\textup{\lowercase\expandafter{\romannumeral#1}}%
} 

\usepackage[version=4]{mhchem} 

\author[1,*]{Yi-Hao Chen}
\author[1]{Wenchao Wang}
\author[1]{Shaodong Hou}
\author[2]{Jose Enrique Antonio-Lopez}
\author[2]{Rodrigo Amezcua-Correa}
\author[1]{Chris Xu}
\author[1]{Frank Wise}

\affil[1]{School of Applied and Engineering Physics, Cornell University, Ithaca, New York 14853, USA}
\affil[2]{CREOL, The College of Optics and Photonics, University of Central Florida, Orlando FL 32816, USA}

\affil[*]{yc2368@cornell.edu}

\begin{abstract}
Raman scattering underlies a broad range of spectroscopic and light-generation techniques, yet its conventional description, based on the Raman gain spectrum, accurately describes only long-pulse, steady-state dynamics. We employ a time-domain theoretical approach to account for spectrotemporal aspects of continuous frequency-shifting, which are not addressed by prior theories. In particular, molecules with strong Raman responses do not produce an efficient soliton self-frequency shift in gas-filled hollow-core fibers. The time-domain analysis exposes temporal and spectral distortions from the Raman response that impact frequency-shifting detrimentally, and identifies how these distortions can be suppressed by reducing the Raman interaction to a perturbation on the electronic response. Experiments that employ gas mixtures with tunable Raman fractions of the nonlinear response demonstrate up to a four-fold increase in quantum efficiency (from \num{20} to \SI{80}{\percent}) compared to the pure molecular gas, and unity-efficiency Raman shifting will be possible. The time-domain framework uncovers phenomena that are inaccessible through the frequency-domain treatment of Raman scattering, and it applies to Raman interactions in solids, liquids, and gases across Raman temporal regimes.
\end{abstract}

\setboolean{displaycopyright}{false} 

\begin{document}
\maketitle

\section{Introduction}

Raman scattering is a third-order nonlinear optical process in which photons inelastically interact with excitations of a material. It is a central process in spectroscopy \cite{Maker1965,Ploetz2007,Min2025} and optical frequency shifters \cite{Hill1976,Boyraz2004,Sirleto2020}. Raman scattering has been traditionally and widely described in the frequency domain, where generation and amplification of fields are characterized by the Raman gain spectrum \cite{Lin2006}. This description, however, applies only in the steady-state regime, where the driving field varies slowly compared with the Raman response time. Calculations show that, in the short-pulse transient regime, the Raman gain is inherently time-dependent; it arises from the evolving phonon waves rather than a fixed spectral response \cite{Carman1970,Chen2024}. In this regime, the effective gain is governed jointly by the instantaneous phase-matching condition and the time-integrated pulse energy with the result that the gain spectrum varies across the pulse (this is illustrated in Fig.~3 of Ref.~\cite{Chen2024}). Focusing on Raman amplification also obscures the accompanying Raman-induced phase modulation, which strengthens self-phase modulation (SPM) in both the transient and steady-state regimes, and plays a key role in the temporal compression of the generated Stokes signal \cite{Konyashchenko2019,Chen2024} and supercontinuum generation \cite{Hosseini2018,Fan2020,Beetar2020}. For ultrashort pulses, the Raman polarization persists as a long-lived oscillatory index wave, often referred to as molecular modulation \cite{Sokolov2002,Bustard2008}, whose index modulation can nonlinearly shape subsequent pulses \cite{Bartels2001,Zhavoronkov2002,Saleh2015,Belli2018}. As the ultrashort driving pulse interacts only with a brief portion of the oscillatory wave, the frequency-domain description with the Raman gain spectrum or the index wave, which relies on the Fourier transform of the entire long-lasting response, does not provide an adequate representation of this interaction. The recent demonstration of Raman-enhanced spectral compression leverages the inertial delayed nonlinear index to smooth the phase modulation, yielding higher-fidelity compression than is achievable with a purely electronic nonlinearity \cite{Wang2026a}. These processes extend beyond the frequency-domain framework and require a time-domain description to capture their dynamics and ultimate effects.

In this work, we analyze continuous frequency shifting through the Raman-induced index modulation $\triangle\epsilon_R(t)$ in the time domain. Most existing treatments of continuous shifting rely on the frequency-domain picture of intrapulse Raman scattering, in which high-frequency components of a pulse transfer energy to lower-frequency components \cite{Dianov1985,Mitschke1986,Gordon1986}. The most-common setting has been systems with anomalous dispersion, where it manifests as the soliton self-frequency shift (SSFS) \cite{Mitschke1986,Gordon1986}. Because these analyses assume steady-state Raman processes (as is appropriate for glasses) with soliton pulses, Raman-driven redshifting is conflated with electronically-driven soliton dynamics, yielding expressions that, as shown below, fail in other Raman temporal regimes or in media with independently-tunable electronic and Raman nonlinearities (\eg mixtures of Raman-active and -inactive gases). A description that extends beyond the steady-state limit and the assumption of soliton pulses is required. Earlier studies have examined time-domain Raman responses in molecular systems 
\cite{Bartels2001,Zhavoronkov2002,Bartels2003,Saleh2015,Belli2018,Yan1987,Yan1987a,Ruhman1987,Ruhman1988,Dhar1994,Korn1998,Nazarkin1998,Nazarkin1999,Bartels2021}, but, with few exceptions \cite{Bartels2003,Korn1998,Nazarkin1998}, how Raman-induced temporal dynamics play out in propagation has not been addressed. As a result, the interplay between distinct Raman temporal regimes, and its consequences for continuous frequency shifting, have remained unexplored.

With the time-domain viewpoint, we identify previously-unrecognized dynamics that emerge during Raman frequency conversion. The Raman process can distort the temporal and spectral profiles of a pulse, which leads to reduced efficiency in frequency-conversion applications. Raman scattering is investigated in a gas-filled hollow-core fiber, where mixtures of molecular and noble gases are used to tailor the nonlinear optical response. As predicted theoretically, the SSFS is significantly enhanced by reducing the relative Raman contribution to the total nonlinearity (electronic plus Raman components), until the Raman contribution becomes a perturbation to the electronic response. This counterintuitive behavior follows from the reduction of Raman-induced temporal distortions of the pulse. With input pulses at \SI{1030}{\nm}, we observe wavelength shifts out to \SI{1650}{\nm} with quantum efficiencies up to \SI{80}{\percent} and pulse durations as short as \SI{60}{\fs}, both of which are several-fold improvements over the performance with the pure molecular gas. Lastly, a brief discussion of application of the time-domain analysis to discrete frequency-shifting concludes the paper.

The frequency-domain picture is based on the Raman gain spectrum. The Stokes field evolves according to ${\diff I^S/\diff z}={g_RI^PI^S}$, where ${I^{P/S}(z,t)}$ denotes the pump (P) or Stokes (S) intensity. Although this expression is time-dependent, all dynamics are effectively collapsed into one scalar gain coefficient $g_R$, which is fixed by the value of the Raman gain at the pump-Stokes frequency difference. More-complicated Maxwell-Bloch coupled equations, which involve the Raman coherence, follow the same concept, with the Raman-scattering parameters fundamentally obtained through the Raman gain spectrum \cite{Raymer1990,Bauerschmidt2015a,Bloembergen1964,Bloembergen1967,Raymer1981,Mostowski1981}. The foundation of these approaches is the steady-state slow-field assumption, taken relative to the Raman characteristic times (\ie with strong dephasing). As we will show at the end of Sec.~\ref{sec:Raman_temporal_regimes} below, only for the steady-state interaction with narrowband fields can the spectrotemporal Raman response be reduced to a Raman gain spectrum. On the other hand, time-domain formulations, which model from first principles the molecular vibration or rotation, have long been employed to study impulsive Raman interactions. However, these treatments have been applied primarily in pump-probe configurations with short propagation distances and limited frequency-shifting \cite{Yan1985,Yan1987,Yan1987a,Weiner1991,Dhar1994,Belli2018}. The transitional behavior that emerges beyond the impulsive limit---into the transient and steady-state regimes---has not been addressed. For example, we will see that Raman scattering can create an upper bound on the pulse duration during nonlinear propagation away from the impulsive limit. Moreover, prior analyses typically isolate the Raman contribution while neglecting the potentially-dominant electronic nonlinearity that continuously shapes the overall pulse evolution \cite{Carman1970}. Most importantly, these first-principle approaches rely on coupling to harmonic-oscillator models \cite{Korn1998,Nazarkin1998,Nazarkin1999,Boyd2008} or quantum rotational wave packets \cite{Martin1988,Nibbering1997,Chen2007,Hoque2011,Langevin2019}, and don't illuminate effects on the pulse propagation. Here, we employ a recently-developed approach \cite{Chen2024} that unifies rotational and vibrational Raman responses through the Raman-induced index modulation $\triangle\epsilon_R(t)$, which directly modulates the optical field. Ref.~\cite{Chen2024} primarily addressed the behavior of the Raman gain, \ie discrete-frequency interpulse Raman scattering, in the long-pulse transient and steady-state regimes. In the present work, we use the time-domain treatment to overcome the limitations outlined above and elucidate important Raman dynamics governed by the simultaneous evolution of the spectral and temporal components.

Despite over two decades of experimental work \cite{Ouzounov2003,Luan2004,Gerome2008,Chen2022,Tani2022,Eisenberg2025a,Lippl2026}, no theory of the SSFS in gases exists; the steady-state framework developed for glasses was simply assumed to apply. Yet ultrashort pulses in molecular gases operate far outside that regime, and frequency-domain steady-state arguments have led to explanations that are either mutually incompatible or not general. A spectral-cutoff interpretation \cite{Gorbach2008} helps explain the generation of non-solitonic radiation, but cannot account for the high-quality Raman solitons observed in \ce{H2},
where the ratio of full-width at half-maximum (FWHM) pulse duration to Raman period $\tau_{\text{FWHM}}/\tau_R$ is in the range of \numrange[range-phrase=--]{0.55}{0.69}
\cite{Chen2022,Lippl2026} and should, under its criterion, produce strong non-solitonic radiation. Likewise, the \SIadj{4.5}{\fs} Raman time inferred in \ce{H2} \cite{Lippl2026} is a result of fitting a steady-state formula and does not represent the true dynamics: we will show that the Raman time is pulse-duration dependent and vanishes for long pulses in weakly-dephasing media like \ce{H2}. The time-domain approach developed here resolves these inconsistencies, explains the physical origin of non-solitonic radiation, identifies the Raman fraction as the governing parameter of gas-phase Raman dynamics, and enables efficient SSFS. We expect that it will find significant application given the growing interest in gas-based frequency conversion in hollow-core fibers.

The time-domain framework delivers a general treatment of Raman scattering that is valid for solids, liquids, and gases, and which demonstrates the limits of the frequency-domain picture. It provides physical insight that has been inaccessible to prior literature on Raman scattering since its discovery \cite{Smekal1923,Raman1928,Raman1928a,Raman1928b,Landsberg1928b,Landsberg1928a,Landsberg1928}, while establishing a unified and intuitive description across temporal regimes.

\begin{figure*}
\centering
\includegraphics{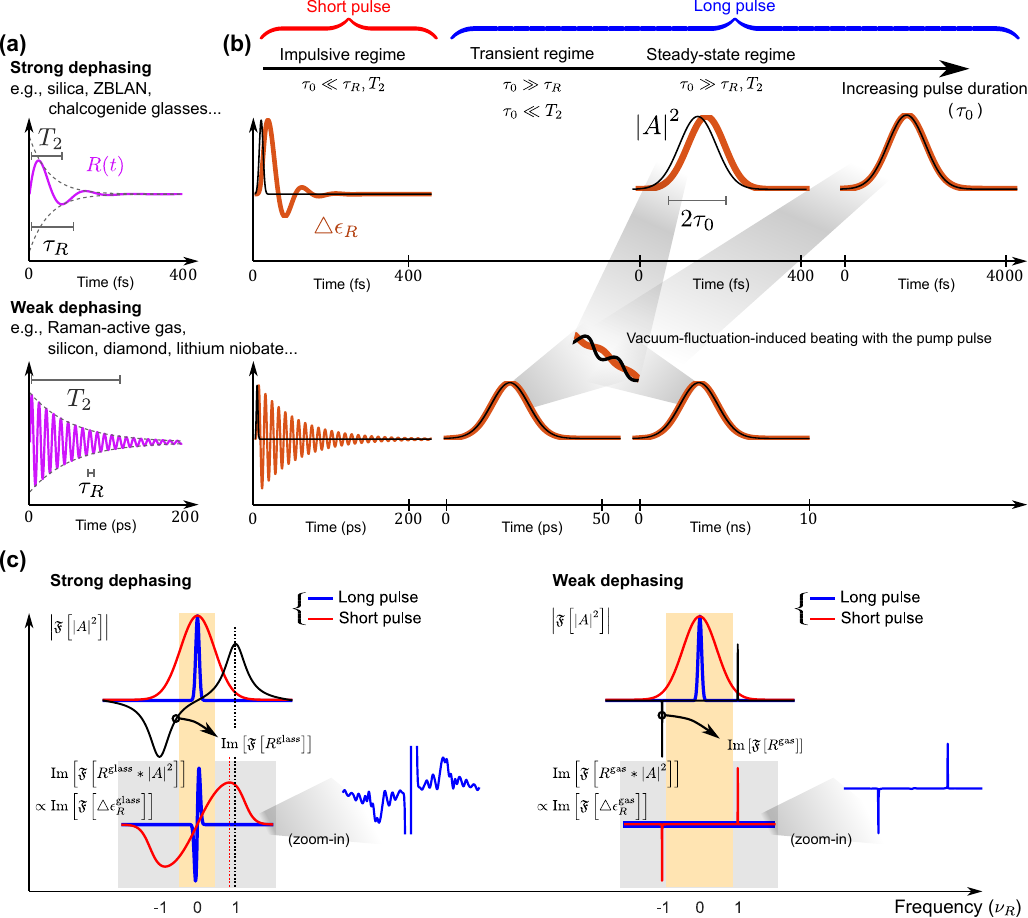}
\caption{Physical pictures of Raman scattering in time and frequency domains. (a) and (b) are Raman dynamics in the time domain while (c) is in the frequency domain. (a) Raman temporal response $R(t)$ with high (top) and low (bottom) dephasing rates, respectively. $\tau_R=1/\nu_R$: Raman period, $\nu_R$: Raman transition frequency, and $T_2$: Raman dephasing time. (b) Raman regimes. $\triangle\epsilon_R$: Raman-induced index modulation, $\tau_0$: duration parameter of a Gaussian or soliton pulse [which is approximately half the FWHM duration], $A(t)$: electric field. Inset shows the weak Raman-induced oscillatory waves, \ie phonon waves \cite{Bauerschmidt2015a,Chen2024}, superposed on the pulse-following index in the transient and steady-state regimes, due to beating between the pump and vacuum fluctuations. This leads to spontaneous Raman scattering. Without vacuum fluctuations and the subsequent beating, there will be no discrete-frequency Raman generation of Stokes signals far from the pump frequency. (c) Top: the relationship between the Raman gain spectrum (black) and the pulse (colored lines) in strong-dephasing (left) and weak-dephasing (right) media. Bottom: spectral profiles of the imaginary part of the Fourier transform ($\mathfrak{F}$) of the Raman convolution integral, which follows ${\Im\left[\mathfrak{F}\left[R\right]\mathfrak{F}\left[\abs{A}^2\right]\right]}$. Zoom-in views in (c) of the long-pulse convolution-integral profile are displayed to visualize the beating-induced index modulation in frequency due to vacuum fluctuations [as shown in the inset in (b)].}
\label{fig:Raman_regimes}
\end{figure*}

\section{Raman temporal regimes}\label{sec:Raman_temporal_regimes}
Analysis of Raman scattering across temporal regimes requires the use of a unidirectional pulse propagation equation that considers the Raman response in the time domain \cite{Chen2024}. The nonlinear terms in the equation are
\begin{align}
& \left[\partial_zA(z,t)\right]_{\text{NL}} \nonumber \\
& =
\begin{cases}
i\omega\kappa\left[\kappa_e\abs{A}^2A+\left(R\ast\abs{A}^2\right)A\right] \\
i\gamma\left[\left(1-f_R\right)\abs{A}^2A+f_R\left(h_R\ast\abs{A}^2\right)A\right],
\end{cases}
\label{eq:UPPE}
\end{align}
which can be formulated in terms of separate Raman and electronic contributions, or as a total nonlinear coefficient with a Raman fraction identified. They are related as follows. $\gamma=\omega\kappa\left(\kappa_e+\int_0^{\infty}R(t)\diff t\right)$ is the total nonlinear coefficient, $f_R=\int_0^{\infty}R(t)\diff t/\left(\kappa_e+\int_0^{\infty}R(t)\diff t\right)$ is the Raman fraction, $h_R(t)=R(t)/\left(\int_0^{\infty}R(t)\diff t\right)$ is the normalized Raman response function, $\kappa_e$ represents the electronic nonlinearity, $\kappa=1/\left(\epsilon_0^2\left[n_{\text{eff}}(\omega)\right]^2c^2A_{\text{eff}}(\omega)\right)$, $n_{\text{eff}}(\omega)$ is the effective refractive index of the propagating mode, $A_{\text{eff}}(\omega)$ is its effective mode area, and ${A(z,t)}$ is the electric field. We simplify the analysis by approximating $\omega(t)$, the instantaneous frequency, as a constant; a more general formulation should rely on the Fourier transform \cite{Chen2024}. Recent derivations \cite{Chen2024} from first-principle vibrational and rotational quantum dynamics have shown that their Raman responses $R(t)$ can be represented, in a quantitatively-accurate and physically-realistic manner, as a sum of damped sinusoidal waves $R(t)=\sum_jR_j(t)$ with $R_j(t)={\Theta(t)R^{\text{coeff}}_j\Exp^{-\gamma_{2,j}t}\sin(\omega_{R_j}t)}$, $\Theta(t)$ the Heaviside step function, and $\gamma_2={1/T_2}$. This form has also been used to accurately model the Raman response of glasses \cite{Lin2006}. Investigations rely on numerical modeling that accounts for realistic Raman response functions of solids and gases (Supplementary Sec.~2) \cite{Chen2024}. Liquids show analogous delayed responses, altered by intermolecular interactions \cite{Reichert2014,Zhao2018a} but governed by the same underlying physics.

Based on the relationship of the excitation pulse duration to the material response time, Raman scattering can be classified into impulsive, transient, and steady-state regimes [Figs.~\ref{fig:Raman_regimes}(a,b)] \cite{Carman1970,Chen2024}, which indicate its spectrotemporal nature. In the impulsive regime, the pulse duration is much shorter than the characteristic Raman period, the index modulation rises monotonically within the pulse, and this induces a continuous redshift ${\triangle\omega}=-{\diff\phi_R/\diff t}<0$ \cite{Korn1998,Chen2024}, with $\phi_R(t)$ the Raman-induced nonlinear phase proportional to $\triangle\epsilon_R(t)$. In the transient and steady-state regimes, the index modulation closely follows the intensity profile, which is the origin of Raman-enhanced SPM but yields minimal overall frequency shift [see Eqs.~(12a) and (13a) in Ref.~\cite{Chen2024}]. In this regime, optical beating with vacuum fluctuations (where the field has a nonzero background at all frequencies) or existing Raman signals drives phonon waves that subsequently amplify the Raman signal and produces a discrete frequency shift. Unlike steady-state Raman growth, which depends on instantaneous intensity, transient Raman growth depends on integrated pulse energy and exhibits a delay due to the buildup of phonons \cite{Carman1970,Chen2024}. Fundamentally, all Raman phenomena originate from the optical interaction with the (real-valued) temporally-varying Raman-induced index variation $\triangle\epsilon_R(t)\propto{R(t)\ast\abs{A(t)}^2}$ [Eq.~(\ref{eq:UPPE}) and see Supplementary Sec.~4], and the frequency-domain description of its action on the field results from the Fourier transform of $\triangle\epsilon_R(t)$. The traditional description based solely on the Raman gain spectrum ${\Im\left[\mathfrak{F}\left[R\right]\right]}$ relies on the ``slow-field'' approximation, which reduces the convolution to a Fourier integral at $\triangle\omega$: ${\triangle\epsilon_RA^S}\sim{\mathfrak{F}\left[R\right](\triangle\omega)\times\abs{A^P}^2A^S}$ with $A^{P/S}$ denoting the pump (P) or Stokes (S) field and $\triangle\omega=\omega^S-\omega^P$ their frequency difference, which may be detuned from $\left(-\omega_R\right)$ with a detuned seeded Stokes field \cite{Agrawal2013}. This approximation is not valid beyond the steady-state regime. In addition, if the interacting fields are not sufficiently narrowband relative to the Raman gain bandwidth, the reduction to a single $\triangle\omega$ becomes invalid.

\begin{figure*}
\centering
\includegraphics{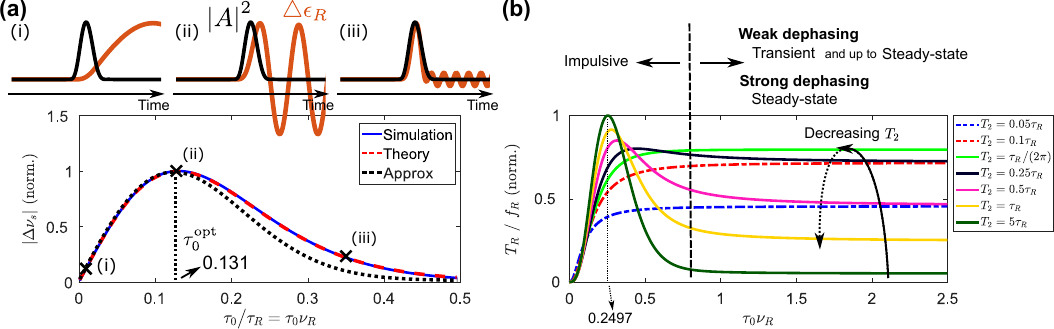}
\caption{Intrapulse continuous redshifting. (a) Simulated magnitude of impulsive Raman redshift (blue) with varying Raman period $\tau_R$ normalized by $\tau_0$. Propagation distance is minimized to exclude dispersion and electronic effects, for comparison to Eq.~(\ref{eq:SSFS_im}). $\tau_0^{\text{opt}}={0.131\tau_R}$ (equivalently, $\tau_{\text{FWHM}}^{\text{opt}}={0.231\tau_R}$). Red and black dashed lines are the results of Eqs.~(\ref{eq:SSFS_im}) and (\ref{eq:SSFS_im2}), respectively. Although Eq.~(\ref{eq:SSFS_im}) is an exact analytical formula of the impulsive redshifting, it does not provide a physically-intuitive formulation due to the $\csch^2(\cdot)$ term, so an approximate relation Eq.~(\ref{eq:SSFS_im2}) is derived. It shows a linear dependence on $\omega_R$ for small $(\omega_R\tau_0)$ and decreasing redshift with increasing $(\omega_R\tau_0)$ due to the exponential term. Top figures show the temporal profiles of the pulse $\abs{A}^2$ (black) and the Raman-induced index modulations $\triangle\epsilon_R$ (orange) for the values indicated on the plot below. (b) ${T_R/f_R}$ evaluated as a function of pulse duration $\tau_0$ normalized by $\nu_R$, for the indicated dephasing times $T_2$. Dividing by $f_R$ removes the contribution from the electronic nonlinearity. We keep the steady-state convention of $T_R$ in Eq.~(\ref{eq:T_R}) for consistency with prior studies; however, to avoid conflation with the electronic nonlinearity, $\gamma f_R$ should be chosen to represent the strength of the Raman nonlinearity [Eq.~(\ref{eq:UPPE})], instead of putting $f_R$ within $T_R$ [Eq.~(\ref{eq:T_R})].}
\label{fig:impulsive_redshift_TR}
\end{figure*}

\section{Intrapulse Raman frequency shifting}
The frequency-domain framework of intrapulse (stimulated) Raman scattering (SRS) in glasses, starting from Gordon's theory \cite{Mitschke1986,Gordon1986}, relies on strong dephasing of the Raman transitions. In silica, for example, the dephasing time is \SI{32}{\fs} and the Raman period is \SI{77}{\fs} \cite{Lin2006}. The strong dephasing creates a region of linear (in frequency) Raman gain near zero frequency [Fig.~\ref{fig:Raman_regimes}(c) left] \cite{Mitschke1986}, so SRS can induce a redshift regardless of the pulse bandwidth. The redshift reaches its maximum when the pulse bandwidth becomes comparable to the Raman transition frequency. In the time domain, SRS induces an index modulation that is slightly delayed with respect to the pulse envelope due to dephasing [Fig.~\ref{fig:Raman_regimes}(b)], and the resulting nonlinear phase modulation induces the continuous frequency redshift. The delay, and thus the redshift, diminishes when the pulse becomes too long, or reaches a maximum when the pulse is sufficiently short to drive SRS impulsively.

The situation is quite different in media with weak Raman dephasing. Long dephasing times translate to narrowband Raman gain spectra, and pulses longer than the Raman period do not experience any intrapulse Raman shift [blue lines in Fig.~\ref{fig:Raman_regimes}(c) right]. A femtosecond-duration pulse can be in the transient or impulsive Raman regime. In the transient regime, Raman scattering plays out as Raman-enhanced SPM or interpulse amplification of discrete Stokes signals through the Raman gain. Only when the pulse bandwidth becomes comparable to the Raman transition frequency does it undergo intrapulse continuous Raman shifting [red lines in Fig.~\ref{fig:Raman_regimes}(c) right]. In the time domain, only pulses shorter than the Raman period can impulsively induce the delayed index variation that leads to continuous redshift.

An analytic expression for intrapulse redshift can be derived following the process used by \etal{Z.\ Chen} \cite{Chen2010} (see Supplementary Sec.~1 for details):
\begin{equation}
\dod{\omega_s}{z}=-\dfrac{4\omega_s\kappa_sE_s}{15\tau_0^3}\displaystyle\int_0^{\infty}B(t,\tau_0)R(t)\diff t,
\label{eq:SSFS_Chen}
\end{equation}
where
\begin{align}
B(t,\tau_0) & =\frac{15}{8}\csch^4\left(\frac{t}{\tau_0}\right) \nonumber \\
& \hspace{1em} \times\Big[4t+2t\cosh\left(\frac{2t}{\tau_0}\right)-3\tau_0\sinh\left(\frac{2t}{\tau_0}\right)\Big].
\end{align}
$\omega_s$ represents the pulse's intensity-averaged frequency, such that $\gamma_s=\gamma(\omega_s)$ and $\kappa_s=\kappa(\omega_s)$. $E_s$ is the pulse energy. Further simplification into steady-state (ss), transient (tr), and impulsive (im) regimes leads to
\begingroup\allowdisplaybreaks
\begin{subequations}
\begin{align}
\dod{\omega_s}{z}\Big\rvert_{\text{ss}} & =-\dfrac{4\omega_s\kappa_sE_s}{15\tau_0^3}\displaystyle\int_0^{\infty}tR(t)\diff t \label{eq:SSFS_ss} \\
\dod{\omega_s}{z}\Big\rvert_{\text{tr}} & =-\dfrac{\omega_s\kappa_sE_s}{6\pi\tau_0}\displaystyle\sum_jR^{\text{coeff}}_j\left(\gamma_{2,j}\tau_{R,j}\right)^2 \label{eq:SSFS_tr} \\
\dod{\omega_s}{z}\Big\rvert_{\text{im}} & =-\frac{\pi^2\tau_0^2\omega_s\kappa_sE_s}{8}\sum_jR^{\text{coeff}}_j\omega_{R_j}^3\csch^2(\frac{\pi\omega_{R_j}\tau_0}{2}) \label{eq:SSFS_im} \\
& \approx-\frac{1}{2}\omega_s\kappa_s E_s\sum_jR^{\text{coeff}}_j\omega_{R_j}\Exp^{-\frac{\omega_{R_j}^2\tau_0^2}{12/\pi^2}} \label{eq:SSFS_im2}
\end{align} \label{eq:SSFS_all}
\end{subequations}
\endgroup
Eq.~(\ref{eq:SSFS_ss}) is the traditional steady-state Raman shifting formula that has been widely studied \cite{Santhanam2003}. We derived the continuous-shifting formulae for the transient and impulsive regimes [Eqs.~(\ref{eq:SSFS_tr}) and (\ref{eq:SSFS_im})] from the duration-dependent $B(t,\tau_0)$, which applies beyond the steady-state Raman regime.

Eqs.~(\ref{eq:SSFS_all}) quantify the qualitative description above. In the steady-state regime, redshifting monotonically diminishes with the inverse-cubic dependence on the pulse duration. In the transient regime, which exists only with weak dephasing (${T_2\gg\tau_R}$), the delay in the Raman-induced index modulation is minimal [Fig.~\ref{fig:Raman_regimes}(b)], which yields weak continuous redshifting because ${\gamma_2\tau_R\ll1}$ in Eq.~(\ref{eq:SSFS_tr}). Therefore, Raman dephasing is the origin of long-pulse continuous redshifting, through the linear Raman gain around zero frequency, or equivalently the temporally-delayed index (Fig.~\ref{fig:Raman_regimes}). Mathematically, the dephasing term disrupts the balanced sinusoidal cancellation in the convolution integral of $\triangle\epsilon_R(t)$, slowing the index rise and delaying the index peak in time. On the other hand, impulsive SRS always induces a delayed index with respect to the ultrashort pulse, and produces significant continuous redshift. Fig.~\ref{fig:impulsive_redshift_TR}(a) shows that the impulsive redshift becomes significant when the pulse duration $\tau_0$ is smaller than half the Raman period. Moreover, the magnitude of the shift diminishes for pulses longer than $\tau_0^{\text{opt}}$, which approaches the transient regime. Only the steady-state and impulsive regimes produce efficient intrapulse redshifting.

To visualize intrapulse redshifting across different Raman regimes, it is helpful to define the Raman time $T_R$ as
\begin{align}
T_R(\tau_0) & =\int_0^{\infty} B(t,\tau_0)\left[f_Rh_R(t)\right]\diff t \nonumber \\
& =\int_0^{\infty}B(t,\tau_0)\left[\frac{\omega_s\kappa_s}{\gamma_s}R(t)\right]\diff t, \label{eq:T_R}
\end{align}
which is duration-dependent. With ${\gamma_sf_Rh_R(t)}={\omega_s\kappa_sR(t)}$, Eq.~(\ref{eq:SSFS_Chen}) is rewritten as ${\diff\omega_s/\diff z}={-\left(4\gamma_sT_RE_s\right)/\left(15\tau_0^3\right)}$. In the steady-state regime (\eg with a large $\tau_0$), $B(t,\tau_0)\approx t$; the Raman time becomes a material property independent of the pulse duration, and Eq.~(\ref{eq:SSFS_Chen}) reverts to the traditional SSFS formula [Eq.~(\ref{eq:SSFS_ss})] \cite{Santhanam2003}. Fig.~\ref{fig:impulsive_redshift_TR}(b) shows that, under relatively-weak Raman dephasing, impulsive redshifting is always more significant than transient and steady-state redshifting, at least until the dephasing is so fast that a rising index is not induced [dashed lines in Fig.~\ref{fig:impulsive_redshift_TR}(b)]. We also see that in the long-pulse regime, a significant level of dephasing facilitates the intrapulse redshifting: $T_2^{\max(T_R/f_R)}\approx{\tau_R/\left(2\pi\right)}$ (see derivation in Supplementary Sec.~1D).

We now turn to the relationship between Raman redshift and the SSFS. Because the redshift is due to the index rising within the pulse envelope, intrapulse continuous redshifting is independent of the dispersion. However, in the presence of anomalous dispersion, solitons retain their temporal compactness during propagation and that enables sustained redshifting. This is particularly important for impulsive redshifting, because it relies on ultrashort pulses. Without soliton dynamics, a short pulse quickly broadens in time owing to dispersion, which causes it to transition from the impulsive regime to the transient regime, where minimal redshifting occurs if the Raman dephasing is weak [Fig.~\ref{fig:impulsive_redshift_TR}(a)]. For a sufficiently-short propagation distance where the change of the pulse shape is negligible, intrapulse redshifting follows Eq.~(\ref{eq:SSFS_Chen}), which is independent of dispersion and electronic nonlinearity. In the case of extended propagation with anomalous dispersion of magnitude $\beta_2$, a fundamental soliton can form and sustain the Raman redshifting: the SSFS. A soliton number of $N=\sqrt{\gamma_{\text{soliton}}E_s\tau_0/\left(2\abs{\beta_2}\right)}=1$ implies that $\left(E_s\propto\tau_0^{-1}\right)$. Inserting this into Eq.~(\ref{eq:SSFS_ss}) shows that the steady-state SSFS obeys the well-known $\left(\tau_0^{-4}\propto E_s^4\right)$ dependence \cite{Mitschke1986,Gordon1986}, while inserting it into Eq.~(\ref{eq:SSFS_tr}) shows that transient SSFS is proportional to $\left(\tau_0^{-2}\propto E_s^2\right)$. The impulsive SSFS is still approximately proportional to $E_s$ when $\omega_R\tau_0\ll1$ \cite{Herrmann1994}. These analytic conclusions are fully supported by numerical simulations.

\begin{figure*}
\centering
\includegraphics{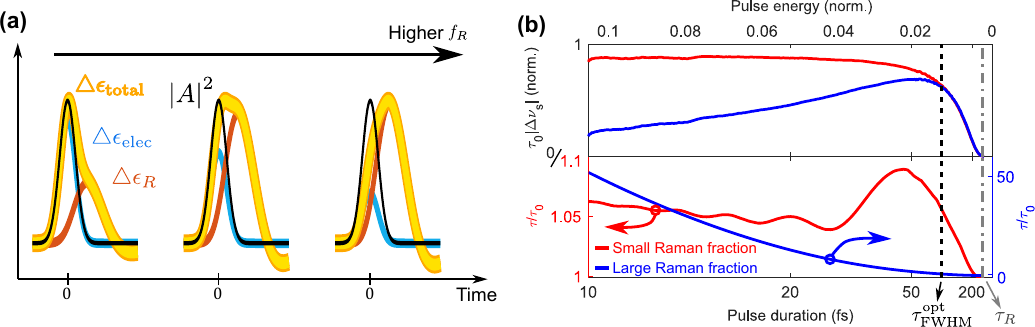}
\caption{Reduced SSFS due to Raman-induced temporal distortion at high Raman fractions. (a) Simulated induced index modulation with different Raman fractions from electronic (blue) and Raman (orange) nonlinearities, along with the total variation (yellow) and the pulse profile ($\abs{A}^2$; black). (b) Simulated magnitude of Raman impulsive redshift, in a medium with weak dephasing, and output pulse duration versus input duration (with the corresponding energy of the fundamental soliton), for small (red) or large (blue) Raman fraction. The pulse propagates over an extended distance to include all effects. For a pulse that propagates as a fundamental soliton (under $N=1$), the normalized impulsive redshift $\tau_0\abs{\triangle\nu_s}\propto{\tau_0E_s}\propto N^2$ should be constant [Eq.~(\ref{eq:SSFS_im2})], as shown in the red line at short durations.}
\label{fig:perturbative_redshift}
\end{figure*}

It is common to assume fundamental soliton propagation ($N=1$) in studies of the SSFS, but this must be applied appropriately to avoid incorrect conclusions. In most experiments, the amount of redshift is tuned by varying the injected pulse energy, which changes the soliton number of the pulse. The pulse dynamically evolves around the nearest integer soliton number, or can undergo fission if $N>1.5$. In more complicated environments such as a tapered fiber or a pressure gradient, pulses may not even have a fixed soliton number. Only under slow variations in the environment do pulses evolve adiabatically and strictly maintain the soliton number \cite{Gerome2007,Laegsgaard2009,Chao2014}. For these reasons, it is critical to analyze the SSFS with Eqs.~(\ref{eq:SSFS_Chen}) and (\ref{eq:SSFS_all}), without reformulation via $N=1$. In cases where $N=1$ does apply throughout the pulse evolution, the precise expression should be
\begin{equation}
\dod{\omega_s}{z}=-\frac{\omega_s\kappa_s\gamma_{\text{soliton}}^3E_s^4}{30\abs{\beta_2}^3}\int_0^{\infty}B(t,\tau_0)R(t)\diff t,
\end{equation}
which is proportional to $\left(\gamma_R\gamma_{\text{soliton}}^3\right)$, where $\gamma_R={\gamma_sf_R}={\omega_s\kappa_s\left(\int_0^{\infty}R(t)\diff t\right)}$ represents the strength of the Raman nonlinearity. Because the Raman-induced index is strongly delayed in the impulsive regime, the soliton dynamics are governed solely by the instantaneous electronic nonlinearity. Unlike steady-state and transient SSFS where $\gamma_{\text{soliton}}=\gamma_s$ is the total nonlinear coefficient [Eq.~(\ref{eq:UPPE})], the relevant coefficient in the impulsive regime reduces to $\gamma_{\text{soliton}}={\omega_s\kappa_s\kappa_e}$, reflecting only the electronic contribution. In the long-pulse regime, ${\gamma_R\gamma_{\text{soliton}}^3}={f_R\gamma_s^4}$ yields the fourth-power relation in $\gamma_s$, as in the traditional steady-state SSFS formula ${\diff\omega_s/\diff z}={-\left(\gamma_s^4T_RE_s^4\right)/\left(30\abs{\beta_2}^3\right)}$ \cite{Mitschke1986,Gordon1986,Santhanam2003}. On the other hand, in the impulsive regime, the frequency shift is proportional to $\left(\gamma_R\kappa_e^3\right)$, linear in the Raman nonlinearity and cubic in the electronic nonlinearity. Lastly, it is important to note that because the dependence on the electronic nonlinearity originates in $\gamma_{\text{soliton}}$ through the soliton relation, the SSFS no longer scales as $\left(\gamma_R\gamma_{\text{soliton}}^3\right)$ if the soliton relation is not consistently satisfied during propagation. More precisely, intrapulse Raman redshifting is intrinsically governed only by the Raman nonlinearity, as expressed in Eq.~(\ref{eq:SSFS_Chen}).

\section{Perturbative impulsive SSFS}
Stimulated Raman scattering can be enhanced dramatically by the use of a solid fiber or a gas-filled hollow waveguide \cite{Benabid2002}. The SSFS is an important application of SRS that underlies numerous frequency-tunable sources. It has been well-studied in silica \cite{Mitschke1986,Gordon1986,Wang2011}, tellurite \cite{Yan2010}, fluoride \cite{Yan2012}, and chalcogenide \cite{Alamgir2021} glass fibers, to produce Raman solitons in various wavelength ranges. Recent works have extended the process to Raman-active gases in hollow-core fibers, involving \ce{H2} \cite{Chen2022,Tani2022,Lippl2026} and \ce{N2} \cite{Eisenberg2025a}, for operation at higher peak powers.

\begin{figure*}
\centering
\includegraphics{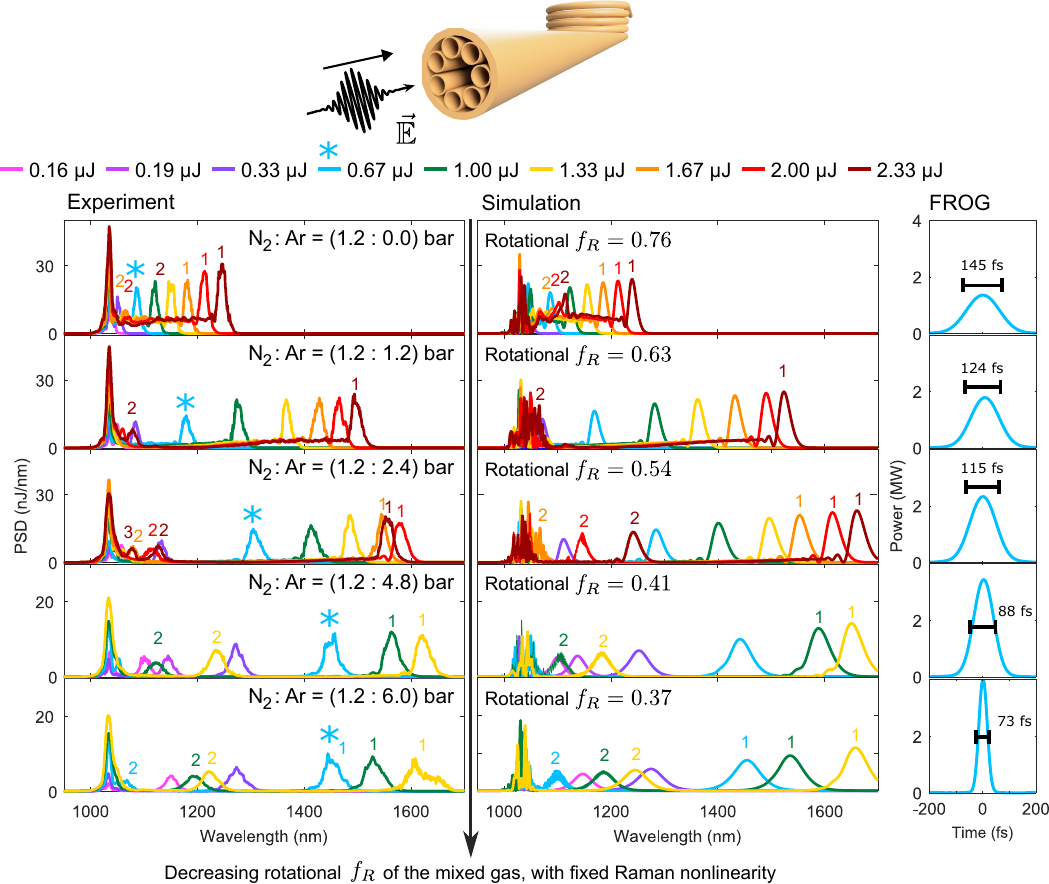}
\caption{Impulsive SSFS with different rotational Raman fractions. The rotational Raman fraction is tuned by introducing a gas mixture, with varying amounts of \ce{Ar} and a fixed \SI{1.2}{\bar} of \ce{N2}, into a hollow-core fiber. Measured and simulated output spectra produced by launching \SIadj{74}{\fs} pulses with energies indicated at the top are shown. Frequency-resolved optical gating (FROG) measurements of the intensity profiles of filtered pulses at an injected energy of \SI{0.67}{\micro\joule} (blue lines and labeled with stars) are shown in the rightmost column, for each value of the Raman fraction. In cases where multiple solitons are generated, their spectra are labeled with ``1'' for the first soliton and so on.}
\label{fig:Spectrum}
\end{figure*}

Raman-induced index variation can be excessive when its role in frequency-shifting is more than offset by its deleterious impact on the ensuing nonlinear evolution. In the impulsive regime where the Raman nonlinearity dominates over the electronic nonlinearity, the total nonlinear index modulation is governed primarily by the delayed Raman polarization and no longer follows the intensity profile [Fig.~\ref{fig:perturbative_redshift}(a)]. In the presence of anomalous dispersion, a soliton cannot form when the nonlinear phase modulation does not closely follow the intensity profile to balance the dispersive phase. The distorted pulse, therefore, does not follow the SSFS relation [Eq.~(\ref{eq:SSFS_im}) and Fig.~\ref{fig:perturbative_redshift}(b)]. Because Raman dephasing times in gases are typically long (\SIrange[range-phrase=--,range-units=single]{>10}{100}{\ps}), efficient SSFS must operate in the impulsive regime, but the ultrashort pulse becomes vulnerable to temporal distortions arising from excessive SRS. These distortions pose a fundamental challenge to energy scaling of SSFS-based frequency shifters. It is worth noting that there is no distortion from excessive SRS in the steady-state regime because the index delay scales with pulse duration [Fig.~\ref{fig:Raman_regimes}(c) left] and the Raman-induced index variation never significantly deviates from the pulse envelope. In the long-pulse regime, dephasing only slightly disrupts the pulse-following behavior of the Raman-induced index modulation, introducing only a modest temporal delay, rather than producing the inertial response characteristic of impulsive SRS. A significant consequence of this is that steady-state SSFS \cite{Dianov1985,Mitschke1986,Gordon1986} can in fact occur without staying in the perturbative Raman limit; \ie the Raman fraction can be large. Simulations of the SSFS in silica with either the actual Raman fraction \cite{Lin2006} or a hypothetical unity Raman fraction both generate high-quality Raman solitons (see Supplementary Sec.~5).

To alleviate the Raman distortion in the impulsive regime, the total nonlinear index variation must closely follow the pulse envelope, which can be achieved by reducing the Raman fraction $f_R$ (Fig.~\ref{fig:perturbative_redshift}), rendering SRS a weak perturbation to the dominant electronic nonlinearity. To investigate this approach, we measured the SSFS in an anti-resonant hollow-core fiber filled with mixtures of gases with varying Raman fractions. Linearly-polarized \SIadj{74}{\fs} pulses at \SI{1030}{\nm} were launched into a \SIadj{29}{\m} length of fiber with a \SIadj{34}{\micro\m} core and anomalous dispersion. With this pulse duration, vibrational SRS (with a Raman period of \SI{14}{\fs} in \ce{N2}) is always transient and contributes only to SPM. The use of linear polarization suppresses the polarization distortion during rotational SRS involving angular momentum \cite{Chen2024}. The optical input end was maintained under vacuum while the output end was filled with a gas mixture (Supplementary Sec.~6). This pressure gradient helps approximate a fundamental soliton: the low pressure at the input end inhibits the formation of a higher-order soliton, and the increasing pressure along the fiber preserves the short soliton duration by maintaining a constant value of ${\left(\gamma_{\text{soliton}}E\tau_0\right)}$ despite the gradual decrease of the pulse energy due to the quantum defect. The long fiber length also contributes to a reduction in the soliton number, as it lowers the required gas pressure for redshifting. The use of low gas pressures ensures that the dispersion is dominated by the anomalous waveguide contribution, effectively fixing the dispersion value across different pressures and gas species: $\beta_2=\num{-0.91}, \num{-0.80}\text{, and }\SI{-0.81}{\fs^2/\mm}$ for vacuum, \SIadj{7.2}{\bar} \ce{N2}, and {\num{1.2}:\SIadj{6}{bar}} \ce{N2}:\ce{Ar} mixture, respectively \cite{Marcatili1964,Bache2019}. Measurements of the output beam produced an invariant Gaussian spatial profile, which is evidence of purely single-mode propagation in the fiber. To focus on the nonlinear pulse propagation and avoid potential thermal effects \cite{Beetar2021,Truong2024}, experiments were performed with a low pulse repetition rate of \SI{7.5}{\kHz}. Numerical simulations of the experiments take into account the gradient pressure profile through continuous variation of the parameters (\eg dispersion, nonlinear response, and dephasing time) with propagation. No evidence of photoionization was observed in the experiments, consistent with simulations based on the Perelomov-Popov-Terent'ev model \cite{Perelomov1966,Couairon2007}.

Molecular gases commonly have strong Raman responses (Supplementary Table~S1). The rotational Raman nonlinearity of molecular nitrogen (\ce{N2}) is large as a result of contributions from many transitions around \SI{2}{\THz} \cite{Chen2024}, which yield a substantial Raman fraction of \num{0.76}. However, experiments aimed at observing the SSFS in the \ce{N2}-filled hollow-core fiber fail to generate isolated solitons (top row of Fig.~\ref{fig:Spectrum}) \cite{Eisenberg2025a}. Numerical simulations indicate that the strong rotational SRS disrupts the soliton formation. Raman solitons are accompanied by significant temporal and spectral pedestals, the latter of which are readily-visible in the top row of Fig.~\ref{fig:Spectrum} and are interpreted as Airy tails that precede the main soliton in time \cite{Gorbach2008}. As these pedestals form and broaden the pulse temporally, the system departs from the impulsive regime, which further weakens the Raman-shifting.

To enhance the SSFS in \ce{N2}, the contribution from the Raman nonlinearity should be reduced to facilitate soliton formation. Through controlled mixing of \ce{N2} with Raman-inactive \ce{Ar}, we systematically varied the rotational Raman fraction, which is $f_{R_{\text{rot}}}^{\ce{N2}}=I_{R_{\text{rot}}}^{\ce{N2}}/(\kappa_e^{\ce{Ar}}+\kappa_e^{\ce{N2}}+I_{R_{\text{rot}}}^{\ce{N2}}+I_{R_{\text{vib}}}^{\ce{N2}})\approx I_{R_{\text{rot}}}^{\ce{N2}}/(\kappa_e^{\ce{Ar}}+\kappa_e^{\ce{N2}}+I_{R_{\text{rot}}}^{\ce{N2}})$ [Eq.~(\ref{eq:UPPE})], where $I_R={\int_0^{\infty}R\diff t}\propto{1/\omega_R}$ is negligible for vibrational SRS due to its large transition frequency \cite{Chen2024}. The Raman fraction is reduced by fixing the \ce{N2} pressure and increasing the \ce{Ar} pressure. Several features are clear from the results, which agree very well with numerical simulations (Fig.~\ref{fig:Spectrum}). As the Raman fraction decreases from \num{0.76} to \num{0.41}, the power spectral density between the input spectrum and the longest-wavelength spectral lobe decreases, with the contrast between that lobe and the background reaching \SI{30}{\decibel} (Supplementary Fig.~S12). For fixed input pulse energy, the Raman shift increases markedly with decreasing Raman fraction. These observations are clear evidence of the benefits of reduced Raman-induced distortion. The clean and compact soliton temporal profile maintains the impulsive Raman interaction, and this underlies efficient frequency-shifting out to \SI{1650}{\nm}. For Raman fractions below \num{0.4}, the SSFS performance does not continue to improve. At that point, soliton formation is not corrupted by the Raman response; the increasing electronic nonlinearity increases the soliton number, which leads to soliton fission, observed as the appearance of a second soliton with decreasing Raman fraction. Fission adversely affects the SSFS directly by reducing the first soliton energy, and indirectly through degraded shifting owing to its lower energy. Reduction of Raman-induced distortion improves the performance until it is not the limiting factor, at a rotational Raman fraction near \num{0.4}, and below that, the perturbative Raman soliton dynamics, in which SRS perturbs the electronic soliton evolution, govern the process.

\begin{figure}[!b]
\centering
\includegraphics{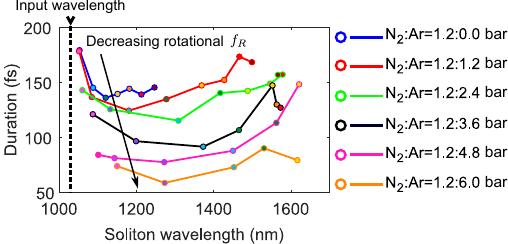}
\caption{Dependence of soliton duration on Raman fraction. FWHM duration of the reddest Raman soliton at the indicated \ce{N2}:\ce{Ar} mixing ratios of Fig.~\ref{fig:Spectrum}. Symbol colors follow the energy color sequence of Fig.~\ref{fig:Spectrum}.}
\label{fig:SSFS_duration}
\end{figure}

\begin{figure*}
\centering
\includegraphics{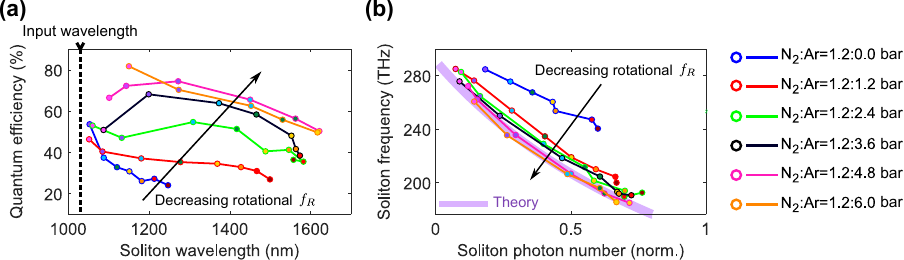}
\caption{Reduced Raman fraction enhances Raman soliton generation. (a) Quantum efficiency of the reddest Raman soliton at indicated \ce{N2}:\ce{Ar} mixing ratios of Fig.~\ref{fig:Spectrum}. (b) Measured (solid lines) and theoretical [colored region; Eq.~(\ref{eq:SSFS_im})] Raman soliton frequency with different soliton photon numbers in \ce{N2}:\ce{Ar} mixtures (Fig.~\ref{fig:Spectrum}). Photon numbers are retrieved only from the filtered reddest soliton at the output (Fig.~\ref{fig:Spectrum}) that remains a fundamental-like soliton. In practice, soliton duration varies during propagation due to energy loss from the quantum defect, dispersion and nonlinearity changes in a pressure gradient, and, most importantly, temporal pedestal formation from Raman-induced distortions, so the theoretical line calculated with fixed \SIadj{74}{\fs} duration represents a theoretical upper bound on the magnitude of the frequency shift. In both panels, symbol colors follow the energy color sequence of Fig.~\ref{fig:Spectrum}.}
\label{fig:SSFS_QE_Eq}
\end{figure*}

The soliton duration is also heavily-influenced by the Raman fraction. With a large Raman fraction, an upper bound on the soliton (FWHM) duration is approximately the duration of the first spike in the Raman temporal response [Fig.~\ref{fig:Raman_regimes}(a)] \cite{Nibbering1997,Chen2024}, which is a result of the coherent beating of phonon modes with distinct frequencies. When the pulse duration matches or exceeds the duration of the first spike, the redshift significantly decreases until the pulse enters the transient regime for all Raman transitions, at which point the shifting ceases. In \ce{N2}, the initial spike spans approximately \SI{200}{\fs} (Supplementary Fig.~S6) and the measured soliton duration stays below \SI{180}{\fs} (Fig.~\ref{fig:SSFS_duration}). A reduced Raman fraction decouples the soliton duration from the spike in the Raman response due to an increased electronically-induced index modulation. The Raman-induced index introduces a slight delay in the overall index modulation, which is dominated by the instantaneous electronic nonlinearity at low Raman fractions. This slightly-delayed response mimics the index profile associated with steady-state SRS [Fig.~\ref{fig:Raman_regimes}(b)], which redshifts without distortion. At high \ce{Ar} pressure, the soliton duration drops to \SI{60}{\fs} (Fig.~\ref{fig:SSFS_duration}). The combination of efficient conversion of microjoule-level pulses and short pulse duration yields peak powers that exceed \SI{1}{\MW} (Supplementary Fig.~S18). Understanding of the pulse propagation, which evolves from the short-pulse impulsive to transient regimes, is enabled by the time-domain framework.

The pedestal-free soliton formation and the short pulse duration that result from reducing the Raman fraction separately contribute to effective Raman-shifting. Pedestal-free soliton propagation maintains the pulse duration for sustained impulsive redshifting, and the shorter duration reduces the soliton number and suppresses subsequent fission. The two effects combine to yield frequency conversion with quantum efficiency that reaches \SI{80}{\percent} in the gas mixture with a $1{:}5$ mixing ratio and a \num{0.37} rotational Raman fraction [Fig.~\ref{fig:SSFS_QE_Eq}(a)]---a four-fold increase in efficiency over pure \ce{N2}. The measured frequency shifts with pure \ce{N2} fall well short of the theoretical predictions of Eq.~(\ref{eq:SSFS_im}), which assumes ideal soliton formation. Due to pedestal formation in pure \ce{N2}, the soliton requires a larger photon number to achieve the same frequency shift as an ideal soliton. With decreasing Raman fraction, the measured shifts approach the ideal theoretical results [Fig.~\ref{fig:SSFS_QE_Eq}(b)]. The physical intuition that accompanies these results is simple: a prerequisite for strong SSFS is formation of a high-quality soliton.

To further test the new impulsive redshifting relation [Eq.~(\ref{eq:SSFS_im})] and the perturbative Raman technique, we conducted analogous SSFS experiments in \ce{H2} (Supplementary Sec.~7). Because \ce{H2} has a single dominant rotational Raman transition, it exhibits a low rotational Raman fraction of \num{0.38} \cite{Chen2024}. When launched into the same fiber filled with \ce{H2}, the \SIadj{74}{\fs} pulses undergo rapid soliton compression and impulsively drive the Raman transition. The resulting evolution is very similar to that observed in a \ce{N2}:\ce{Ar} mixture with a $1{:}4$ ratio, which has the same Raman fraction. The quantum efficiency of frequency conversion reaches \SI{73}{\percent}. In this regime of small soliton distortion due to the low Raman fraction, the redshifting behavior matches the prediction of Eq.~(\ref{eq:SSFS_im}) (Supplementary Fig.~S16), consistent with the trend observed at high mixing ratios in Fig.~\ref{fig:SSFS_QE_Eq}(b).

It is worth briefly discussing the recent use of \SIadj{4.5}{\fs} Raman time $T_R$ to fit the SSFS trend in \ce{H2} reported in \cite{Lippl2026}. The fitting succeeds because it is performed using ${\diff\omega_s/\diff z}={-\left(4\gamma_sT_RE_s\right)/\left(15\tau_0^3\right)}$ which is fundamentally Eq.~(\ref{eq:SSFS_Chen}). In this form, the dependencies on nonlinearity, pulse energy, and duration remain physically correct and are not distorted by the soliton relation; consequently, fitting the frequency shift versus pulse energy is feasible. Had they instead used the expression ${\diff\omega_s/\diff z}={-\left(\gamma_s^4T_RE_s^4\right)/\left(30\abs{\beta_2}^3\right)}$, the fit would fail because the $\gamma_s^4$, $E_s^4$, and $\abs{\beta_2}^3$ scalings from consistent fundamental soliton propagation are not satisfied in their propagation. Their successful fitting therefore reflects the choice of formula rather than a physically-meaningful determination of a constant Raman time. The value of \SI{4.5}{\fs} is approximately the maximum Raman time in \ce{H2} (see Supplementary Fig.~S3).

\section{Discussion}
One might wonder if the measured improvement in Raman shifting is due to the increase in the total nonlinearity that accompanies increased \ce{Ar} pressure. This would be a natural question within the conventional framework of steady-state SSFS (\eg in glasses), which is proportional to $\left(f_R\gamma_s^4\right)$. As discussed above, steady-state analysis is invalid here; the frequency-domain picture says nothing about the Raman-induced distortion, nor the suppression of distortion with decreasing Raman fraction. In addition, pulses in these experiments span a range of soliton numbers and undergo nontrivial soliton dynamics in the presence of a pressure gradient. The observation of soliton fission indicates that the pressure increase along the waveguide is not gentle enough to support adiabatic soliton evolution, in which the pulse reshapes itself during propagation to maintain a fundamental soliton. As a result, the SSFS must be analyzed in the form of Eqs.~(\ref{eq:SSFS_Chen}) and (\ref{eq:SSFS_im}), which show that it is independent of the electronic nonlinearity.

One might also ask whether experiments should be performed at a fixed total pressure while varying the partial pressures to change the Raman fraction. This approach is only meaningful in the long-pulse transient or steady-state Raman regime, where the Raman and electronic responses both follow the pulse's temporal profile and the notion of a fixed ``total'' nonlinearity, or more precisely, the nonlinear index modulation which defines the soliton number, remains valid. In the impulsive regime, the concept of fixing the total nonlinearity becomes ambiguous because the Raman nonlinearity exhibits temporal delay. Moreover, because the soliton number in the impulsive regime is governed primarily by the electronic nonlinearity, changing the Raman fraction alters the soliton number and therefore the subsequent nonlinear evolution. In other words, the soliton number is not fixed even when the total pressure is held constant. By fixing the Raman nonlinearity, thereby defining the maximum possible redshift in our experiments, the temporal quality of the Raman soliton maps directly onto the observed redshift. This enables straightforward assessment of soliton quality and facilitates validation of the perturbative Raman theory.

Raman-induced soliton distortion has previously been attributed to the development of an asymmetric intensity profile \cite{Akhmediev1996,Gagnon1990}. These analyses, however, rely on assumptions appropriate to silica fibers: a weak, steady-state Raman response. Under such conditions, the Raman-induced index modulation follows the pulse envelope with only a slight temporal delay, and the resulting distortion arises solely from this quasi-instantaneous response. Consequently, these treatments do not capture the impulsive temporal distortion considered here, where the induced index modulation departs substantially from the pulse profile.

Our investigation of impulsive SRS reveals the decisive role of Raman-induced temporal index modulation in Raman-soliton formation, and identifies a clear path to highly-efficient and high-energy ultrashort-pulse generation with frequency shift that approaches the theoretical upper bound of an ideal soliton. The frequency-domain framework for the Raman response is inadequate to explain the distortion introduced by SRS, much less identify the remedy based on limiting the Raman response to a perturbation on the electronic response. Understanding derived from the analytic theory is consistent with the pulse propagation found in numerical simulations, which in turn agree with the experimental results.

Reduction of Raman-induced distortion is realized through direct control of the nonlinear optical response of the gas. There is only one prior report of mixing of Raman-active and -inactive gases to control the nonlinear response \cite{Imani2025}: in that work, \ce{N2} was added to \ce{Ar} to induce asymmetric spectral broadening, as expected based on the simplest considerations. Direct control of the nonlinear response function, as demonstrated here, should be distinguished from the strategy of introducing a noble gas to control Raman gain suppression through the \emph{linear} wave-vector mismatch ${\triangle\beta}=2\beta^P-\beta^S-\beta^{AS}$ with $\beta^k$ the wave vector of the $k$ (P: pump, S: Stokes, AS: anti-Stokes) field \cite{Hosseini2016,Hosseini2017a}. More broadly, reduction of Raman-induced distortion does not require mixing of gases. For example, it could also be realized by the use of input pulses much shorter than the Raman period, so that the Raman-induced index varies little across the pulse's temporal extent and the electronic nonlinearity dominates. The optimal Raman fraction will depend on pulse duration, as shorter pulses resist distortion from stronger Raman nonlinearity.

Controlled experiments aimed at understanding the Raman process are presented here. A long fiber was employed to allow the Raman soliton to fully evolve and reach its duration limit for each Raman fraction, to ensure that the measurements reflect the underlying physics (Fig.~\ref{fig:SSFS_duration}). Without attempts to optimize the performance, pulses with megawatt peak powers are generated (Supplementary Fig.~S18). As a practical issue, a shorter fiber will be desirable in a source intended for applications. As long as the Raman fraction is low, it will be possible to determine the minimum fiber length for efficient generation of acceptably-short pulses. The performance of this approach is limited only by technical issues; a fundamental soliton is theoretically predicted to shift with unity quantum efficiency. When combined with the customizable transmission bands of anti-resonant hollow-core fibers, impulsive SSFS with a reduced Raman-induced distortion offers an attractive approach to femtosecond pulse generation, with potential for scaling to higher energies (\SIadj{50}{\micro\joule} pulses are produced in simulations) and to other spectral regions (\eg the mid-wave infrared). Soliton dynamics have also been demonstrated in large-core capillaries \cite{Travers2019,Brahms2019,Travers2024}, so further scaling of this process in energy or peak power may be possible with a larger-core fiber. The combination of efficient generation of femtosecond pulses across wide frequency bands, implementation in a compact footprint with coilable fiber, and simple alignment may make SSFS an attractive alternative to existing crystal-based optical parametric amplification \cite{Cerullo2003,Manzoni2016} with further engineering. By maintaining the fundamental soliton propagation without fission, the coherent intrapulse Raman redshift should proceed in a phase-stable manner; if a Raman soliton preserves the carrier-envelope phase of the input pulse, it may find application in further frequency conversion. As a secondary point, this work illustrates how mixtures of innocuous gases can replace hazardous gases (\eg combustible \ce{H2} \cite{Chen2022,Tani2022,Lippl2026}), which are commonly forbidden for use outside research laboratories.

The benefits of controlling the nonlinear response are not limited to the case of Raman soliton formation. The time-domain framework also provides an improved treatment of discrete interpulse Raman frequency shifting, which generates Stokes sidebands far from the pump frequency. This process arises only in the long-pulse transient and steady-state regimes, and can be accurately described by a time-dependent Raman gain spectrum \cite{Chen2024}. Because continuous intrapulse redshifting also occurs in both the transient and steady-state regimes, the two processes compete. In the transient regime, the intrapulse redshift is weak, so discrete Raman generation always dominates. In contrast, in the steady-state regime, the dominant process depends on the Raman dephasing rate, pulse duration, and the presence or absence of an existing Stokes seed. Moreover, temporally stretching a pulse, \eg by introducing a chirp, facilitates discrete Raman generation; the Raman-induced index modulation ${\triangle\epsilon_R(t)\propto R(t)\ast\abs{A(t)}^2}$ is reduced because the stretched pulse has a more narrowband intensity envelope $\mathfrak{F}\left[\abs{A(t)}^2\right]$, which allows discrete spontaneous Raman generation from vacuum fluctuations to dominate [Fig.~\ref{fig:Raman_regimes}(c)]. Detailed discussion of the discrete-generation mechanism is provided in \cite{Chen2026arXiv}.

\section{Conclusion}
In conclusion, we have applied a general time-domain framework based on the delayed index modulation to the analysis of continuous Raman frequency-shifting. It complements and updates the widely-employed frequency-domain description based on the Raman gain spectrum, which fails to capture important spectrotemporal Raman dynamics. By resolving the Raman and electronic contributions to the nonlinear response of molecules in the impulsive Raman regime, we identify Raman-induced temporal distortion imposed on a propagating pulse and its detrimental impact on optical frequency shifting. Reducing the Raman fraction of the nonlinear response suppresses this distortion, which substantially enhances all aspects of frequency-shifting performance.

The time-domain treatment is based on a general expression that can be applied in any temporal Raman regime or at transitions between regimes, and to any Raman-active materials. The core of this framework is the Raman-induced index modulation, which may or may not be correlated with the traditional phonon concept, typically understood as a collective molecular oscillation. The fundamental definition of a phonon in this framework and its relationship to the index modulation, as well as the new physical phenomena that arise from them beyond what is discussed here, are the subjects of a separate publication \cite{Chen2026arXiv}.

\begin{backmatter}
\bmsection{Funding} National Institutes of Health (R01EB033179, U01NS128660), the Office of Naval Research (N00014-19-1-2592), the Army Research Office (W911NF2410008), and the Air Force Research Laboratory (FA8650-19-F-1025).

\bmsection{Acknowledgment} Y.-H.C. was partially supported by a Mong Fellowship from Cornell Neurotech.

\bmsection{Disclosures} Y.-H.C.\@ and F.W.\@ are inventors on a patent application. The other authors declare no competing interests.

\bmsection{Data availability} The code used in this work has been made publicly available at \url{https://github.com/AaHaHaa/gas_UPPE}.

\end{backmatter}

\bibliography{reference}

\bibliographyfullrefs{reference}

\end{document}


\begingroup
\hbadness=10000 
\maketitle
\endgroup

\newpage 
\section{Continuous Raman redshifting}
\label{sec:continuous_Raman_redshifting}
\subsubsection*{Foreword to impulsive redshift}
Before discussing intrapulse Raman redshifting, we briefly review the historical development of impulsive stimulated Raman scattering (SRS), in which temporal treatments were first introduced. However, the early emphasis on the impulsive regime limited the extension of temporal analyses to other regimes. More precisely, a clear classification of temporal regimes, together with a corresponding unified theoretical framework, has only been established in a recent year \cite{Chen2024}.

Impulsive redshifting has been traditionally studied in the context of nonlocal interactions with femtosecond pump-probe experiments \cite{Yan1985,Yan1987,Yan1987a,Ruhman1987,Ruhman1987a,Ruhman1987b,Ruhman1988,Weiner1991,Dhar1994,Bell1996,Korn1998,Nazarkin1998,Zhavoronkov2002,McCamant2003,Zheltikov2010,Saleh2015b,Belli2018} or controlling molecular alignment \cite{Normand1992,Pentlehner2013,Mullins2022,Wang2025}. However, their discussions involve several limitations. Most prior studies on pump-probe experiments have focused on vibrational Raman scattering, owing to its relatively simple scalar formulation, whereas rotational SRS generally entails non-negligible polarization coupling. On the other hand, studies on molecular alignment focused only on the rotational Raman scattering. A comprehensive Raman description incorporating polarization effects has only recently been established \cite{Chen2024}, enabling a single framework of impulsive Raman scattering to systematically cover both rotational and vibrational modes, beyond linear polarizations. Although redshifting phenomena have been extensively investigated, the observed frequency shifts remain small, as they are not essential for probing molecular dynamics or controlling the molecular alignment. Application of impulsive SRS to the generation of new frequencies was thus limited. Moreover, analyses have largely been confined to ultrashort durations within the impulsive regime. The transitional nonlinear response beyond this regime -- into the transient or steady-state domain -- remains insufficiently understood, despite its critical role in nonlinear pulse evolution. Traditional treatments of vibrational Raman scattering typically rely on the harmonic-oscillator model, which necessitates solving an additional coupled equation for the oscillator displacement; similarly, traditional treatments of rotational Raman scattering employ a quantum-mechanical framework to resolve its transient molecular alignment dynamics with a linearly-polarized driving field, which introduces complexity into the analysis and hinders physically-intuitive understanding. Recent advances in unified Raman theory \cite{Chen2024} have alleviated this situation by reformulating and unifying both responses as a summation of sinusoidal Raman temporal functions, even for gases [see Eq.~(31) and its derivation in Supplementary Secs.~5 and 6 in \cite{Chen2024}, as well as its Supplementary Sec.~3 to understand constraints introduced in various Raman models in prior works]. The Raman response is reduced to a convolution term of the form $\left[\left(\Exp^{-\gamma_2t}\sin(\omega_Rt)\right)\ast\abs{A(t)}^2\right]$ in the nonlinear pulse propagation equation. This formulation further directs the analysis toward the index variation to which the optical field directly responds. We employ this model throughout this article, rather than the vibrational oscillator displacement $\mathbb{Q}$ or rotational alignment angle $\left\langle\cos^2\left(\frac{\pi}{2}-\theta\right)\right\rangle$ from conventional Raman models.

\subsubsection*{Derivation of the redshifting formula}
In this section, we derive the continuous Raman redshifting formula due to the delayed Raman-induced index. Expressions for the frequency shift in various regimes are provided. The discussion here applies to both vibrational and rotational Raman processes. The comprehensive discussion is possible with the scalar-field unidirectional pulse propagation equation (UPPE) unified for arbitrary Raman temporal regimes and Raman types \cite{Chen2024}:
\begin{equation}
\left[\partial_zA(z,t)\right]_{\text{NL}}=
\begin{cases}
i\gamma\left[\left(1-f_R\right)\abs{A}^2A+f_R\left(h_R\ast\abs{A}^2\right)A\right] & \text{in the ($f_R$,$h_R(t)$) representation} \\
i\omega\kappa\left[\kappa_e\abs{A}^2A+\left(R\ast\abs{A}^2\right)A\right] & \text{in the $R(t)$ representation}
\end{cases},
\label{eq:UPPE}
\end{equation}
where
\begingroup\allowdisplaybreaks
\begin{subequations}
\begin{align}
\gamma & =\omega\kappa\left(\kappa_e+\int_0^{\infty}R(t)\diff t\right)\text{: (total) nonlinear coefficient} \label{eq:gamma_UPPE} \\
f_R & =\frac{\int_0^{\infty}R(t)\diff t}{\kappa_e+\int_0^{\infty}R(t)\diff t}\text{: Raman fraction} \label{eq:f_R} \\
h_R(t) & =\frac{R(t)}{\int_0^{\infty}R(t)\diff t}\text{: normalized Raman response function}
\end{align} \label{eq:transformation_hR_R}
\end{subequations}
\endgroup
and $\kappa_e=3\epsilon_0\chi^{(3)}_{\text{electronic}}/4$ represents the electronic nonlinearity with $\chi^{(3)}_{\text{electronic}}$ the third-order nonlinear susceptibility of the electronic response. $\kappa=1/\left(\epsilon_0^2\left[n_{\text{eff}}(\omega)\right]^2c^2A_{\text{eff}}(\omega)\right)$, where $n_{\text{eff}}(\omega)$ is the effective index and $A_{\text{eff}}(\omega)$ is the effective area, with both generally being frequency-dependent. Here, $\omega(t)$ is the instantaneous frequency such that $\kappa(t)=\kappa\left(\omega(t)\right)$ and $\gamma(t)=\gamma\left(\omega(t)\right)$. For a gas with dephasing time much longer than its Raman period, the Raman-induced phase modulation corresponds to $\int_0^{\infty}R(t)\diff t\approx R^{\text{coeff}}/\omega_R$, where $R(t)=\Theta(t)R^{\text{coeff}}\Exp^{-\gamma_2t}\sin(\omega_Rt)$ ($\Theta(t)$ is the Heaviside function, $\gamma_2$ is the dephasing rate, and $\omega_R$ is the Raman transition angular frequency \cite{Chen2024}).

Expressions for the redshifting $\od{\omega_s}{z}$ are derived in Part \ref{subsec:Exact}, and are summarized and discussed in Part \ref{subsec:summary_SSFS}. The results do not account for variation in the pulse with propagation. Extension to include pulse propagation is in Part \ref{subsec:freq_shift_extended}.

\subsection{Exact analytic formulation}
\label{subsec:Exact}
In this section, we present an analytic formulation of the intrapulse Raman redshift. The formulation follows the equation derived by \etal{Z.\ Chen} \cite{Chen2010}, which relies on the moment method. They derived an effective Raman time (see derivation detail in Appendix~\ref{sec:appendix_Raman_time}):
\begin{equation}
T'_R(\tau_0)=\int_0^{\infty}B(t,\tau_0)h_R(t)\diff t.
\label{eq:T_R_tau0}
\end{equation}
Unlike the traditional Raman time $T_R=f_R\int_0^{\infty} th_R\diff t$, this effective Raman time can be applied for timescales beyond the steady-state Raman regime ($\tau_0\gg\tau_R,T_2$), which helps generalize the intrapulse redshifting relation across various Raman temporal regimes. Note that the moment method here does not restrict the pulse to be a hyperbolic-sech soliton. It is generally applicable to propagations where the pulse temporal shape either remains unchanged or changes only slightly (\eg under a short propagation distance). Compared to the original version by \etal{Z.\ Chen}, here we remove the Raman fraction term, such that the commonly-used Raman time $T_R=f_RT'_R$, which we will see is useful when switching between the normalized and non-normalized Raman representations. Also,
\begin{equation}
B(t,\tau_0)=\frac{15}{8}\csch^4\left(\frac{t}{\tau_0}\right)\left[4t+2t\cosh\left(\frac{2t}{\tau_0}\right)-3\tau_0\sinh\left(\frac{2t}{\tau_0}\right)\right],
\end{equation}
where $\tau_0$ is the soliton duration [approximately half the full-width-at-half-maximum (FWHM) duration]. The evolution of the redshift is governed by
\begin{equation}
\dod{\omega_s}{z}=
\begin{cases}
-\dfrac{4\gamma_sf_RT'_RE_s}{15\tau_0^3} & \text{in the ($f_R$,$h_R(t)$) representation} \\
-\dfrac{4\omega_s\kappa_sE_s}{15\tau_0^3}\displaystyle\int_0^{\infty}B(t,\tau_0)R(t)\diff t & \text{in the $R(t)$ representation [with Eq.~(\ref{eq:transformation_hR_R})]}
\end{cases},
\label{eq:SSFS_Chen}
\end{equation}
$\omega_s$ represents the pulse's intensity-averaged frequency [Eq.~(\ref{eq:omega_s_Chenmoment})], or the center frequency if it has a symmetric spectrum, such that $\gamma_s=\gamma(\omega_s)$ and $\kappa_s=\kappa(\omega_s)$. $E_s$ is the soliton energy. We show Eq.~(\ref{eq:SSFS_Chen}) under representations with non-normalized and normalized Raman responses for different scenarios that we will discuss later. Note that $\gamma_sf_Rh_R(t)=\omega_s\kappa_sR(t)$ [Eq.~(\ref{eq:UPPE})].

To simplify the equation, we rewrite it in dimensionless form with
\begin{subequations}
\begin{align}
x & =\frac{t}{\tau_0} \\
\Omega_R & =\omega_R\tau_0.
\end{align}
\end{subequations}
By employing the single damped harmonic oscillator for the Raman response function, we then obtain
\begin{equation}
T'_R(\tau_0)=\tau_0^2\int_0^{\infty}B_d(x)\left[\frac{\gamma_2^2+\omega_R^2}{\omega_R}\Exp^{-\gamma_{2,d}x}\sin(\Omega_Rx)\right]\diff x,
\label{eq:T_R_t}
\end{equation}
where
\begin{subequations}
\begin{align}
\gamma_{2,d} & =\gamma_2\tau_0 \\
B_d(x) & =\frac{15}{8}\csch^4\left(x\right)\left[4x+2x\cosh\left(2x\right)-3\sinh\left(2x\right)\right]\quad\text{[see Fig.~\ref{fig:B_d}(a)]}. \label{eq:B_d}
\end{align}
\end{subequations}
Note that $B(t,\tau_0)=\tau_0B_d(x)\Big\rvert_{x=\frac{t}{\tau_0}}$.

\subsection*{Steady-state intrapulse redshifting}
In this regime, we consider only small $x$. $B_d(x)$ remains significantly nonzero within a limited interval $x=0\sim6$ [Fig.~\ref{fig:B_d}(a)]. In particular, it exhibits a linear regime when $x<1$, such that the dephasing contribution relevant to the Raman-time integral is $\gamma_{2,d}x<\gamma_{2,d}=\gamma_2\tau_0=\tau_0/T_2$. Therefore, as long as the dephasing time $T_2$ is much smaller than $\tau_0$, the integrand [Eq.~(\ref{eq:T_R_t})] remains significantly nonzero only at small $x$. Using
\begin{subequations}
\begin{align}
\sinh(x) & \approx x+\frac{x^3}{3!}+\frac{x^5}{5!}+\cdots \\
\cosh(x) & \approx 1+\frac{x^2}{2!}+\frac{x^4}{4!}+\cdots,
\end{align}
\end{subequations}
we find that
\begin{align}
B_d(x)\big\rvert_{x\ll1} & \approx\frac{15}{8}\frac{1}{x^4}\left[4x+2x\left(1+2x^2+\frac{2x^4}{3}\right)-3\left(2x+\frac{4x^3}{3}+\frac{4x^5}{15}\right)\right] \nonumber \\
& =\frac{15}{8}\frac{1}{x^4}\frac{8x^5}{15}=x,
\label{eq:B_d_ss}
\end{align}
so $B(t,\tau_0)=t$ when $T_2\ll\tau_0$, and we recover the traditional Raman time $T_R=f_R\int_0^{\infty} th_R(t)\diff t$, which is simply a constant and becomes an intrinsic material property.

In the steady-state regime, intrapulse redshifting follows 
\begin{equation}
\dod{\omega_s}{z}\Big\rvert_{\text{steady-state}}=
\begin{cases}
-\dfrac{4\gamma_sf_RE_s}{15\tau_0^3}\displaystyle\int_0^{\infty}th_R(t)\diff t & \text{in the ($f_R$,$h_R(t)$) representation} \\
-\dfrac{4\omega_s\kappa_sE_s}{15\tau_0^3}\displaystyle\int_0^{\infty}tR(t)\diff t & \text{in the $R(t)$ representation}
\end{cases},
\label{eq:SSFS_ss}
\end{equation}

It should be emphasized that the preceding derivation does not rely on the assumption of strong dephasing. The only requirement is that $T_2\ll\tau_0$. From a mathematical standpoint, weak (but not zero) dephasing continues to yield linear gain with frequency near zero frequency in the Raman spectral response [Fig.~1(c) in the article], albeit with a small magnitude. Consequently, the Raman-induced index modulation in the steady-state regime consistently exhibits a slight temporal delay relative to the driving pulse and produces a continuous redshift regardless of the pulse duration. It is still worth noting that strong dephasing (but not too strong to suppress SRS) creates a significantly-stronger linear gain near zero frequency -- equivalently, it leads to a larger Raman time -- and boosts the redshifting speed [see Fig.~2(b) in the article].

\subsection*{Transient intrapulse redshifting}
In the transient regime, $\Omega_R\gg1$. The dephasing time $T_2$ needs to be much larger than $\tau_0$; otherwise, the evolution occurs in the steady-state regime discussed above.

With $\Omega_R\gg1$, we rewrite the effective Raman time [Eq.~(\ref{eq:T_R_t})] to
\begin{equation}
T'_R(\tau_0)\approx\tau_0^2\omega_R\int_0^{\infty}B_d(x)\left[\Exp^{-\gamma_{2,d}x}\sin(\Omega_Rx)\right]\diff x.
\end{equation}

To solve this equation, we use the following relation for an arbitrary slowly-varying smooth function $f(t)$ that vanishes at infinity:
\begin{equation}
\int_0^{\infty}\left[\Exp^{-\gamma_2t}\sin(\omega_Rt)\right]f(t)\diff t=\frac{\gamma_2^2\tau_R}{\omega_R}\int_0^{\infty}f(t)\Exp^{-\gamma_2t}\diff t+\frac{1-\gamma_2\tau_R}{\omega_R}f(0),
\label{eq:tmp3}
\end{equation}
with the Raman period $\tau_R=2\pi/\omega_R$. (See the derivation in Appendix~\ref{sec:appendix_Derivation_tmp3}.)

\begin{figure}[!b]
\centering
\includegraphics[width=.9\linewidth]{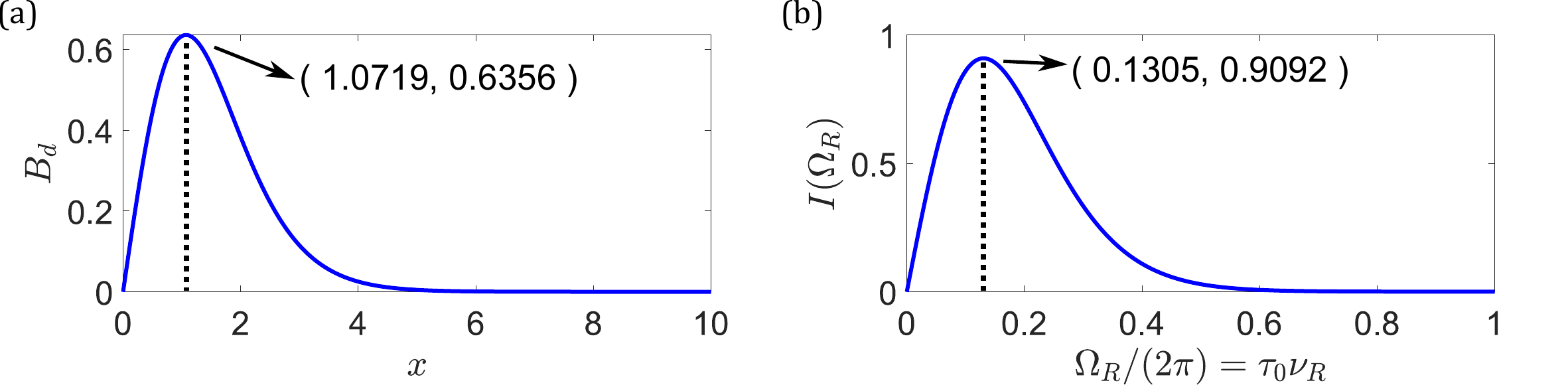}
\caption{(a) $B_d(x)$ and (b) $I(\Omega_R)$.}
\label{fig:B_d}
\end{figure}

Because $B_d(x)$ is slowly varying relative to $\Omega_R$ [Fig.~\ref{fig:B_d}(a)], we obtain
\begin{align}
T'_R(\tau_0) & =\tau_0^2\omega_R\left[\frac{\gamma_{2,d}^2\left(\tau_R/\tau_0\right)}{\Omega_R}\int_0^{\infty}B_d(x)\Exp^{-\gamma_{2,d}x}\diff x+\frac{1-\gamma_{2,d}\left(\tau_R/\tau_0\right)}{\Omega_R}B_d(0)\right] \nonumber \\
& =\tau_0^2\gamma_2^2\tau_R\int_0^{\infty}B_d(x)\Exp^{-\gamma_{2,d}x}\diff x\quad,\text{because }B_d(0)=0.
\label{eq:TRBd_tmp}
\end{align}
The integral relies on $\gamma_{2,d}=\gamma_2\tau_0$ which depends on pulse duration $\tau_0$.

In the transient regime, intrapulse shift is given by
\begin{equation}
\dod{\omega_s}{z}\Big\rvert_{\text{transient}}=
\begin{cases}
-\dfrac{4\gamma_sE_s}{15\tau_0}\displaystyle\sum_jf_{R,j}\gamma_{2,j}^2\tau_{R,j}\int_0^{\infty}B_d(x)\Exp^{-\gamma_{2,j}\tau_0x}\diff x & \text{in the ($f_R$,$h_R(t)$) representation} \\
-\dfrac{2\omega_s\kappa_sE_s}{15\pi\tau_0}\displaystyle\sum_jR^{\text{coeff}}_j\left(\gamma_{2,j}\tau_{R,j}\right)^2\int_0^{\infty}B_d(x)\Exp^{-\gamma_{2,j}\tau_0x}\diff x & \text{in the $R(t)$ representation}
\end{cases},
\label{eq:SSFS_transient}
\end{equation}
where we now generalize to include multiple Raman transitions. The $R(t)$ representation above is obtained using $\gamma_sf_R=\omega_s\kappa_s\left(\int_0^{\infty}R(t)\diff t\right)=\omega_s\kappa_sR^{\text{coeff}}\frac{\omega_R}{\omega_R^2+\gamma_s^2}\approx\omega_s\kappa_sR^{\text{coeff}}/\omega_R$ for each weakly-damped sinusoidal Raman transition.

Because $B_d(x)$ is significantly non-zero only around $x=0\sim6$, the integration range in Eq.~(\ref{eq:SSFS_transient}) is effectively limited to $x=0\sim6$. In situations where $\gamma_2\tau_0\ll1$ (\eg in a weakly-dephasing medium), the exponential term can be neglected ($\Exp^{-\gamma_2\tau_0x}\approx1$) and the $B_d$ integral becomes a constant $\int_0^{\infty}B_d(x)\diff x=5/4$ (see derivation in Appendix~\ref{sec:appendix_B_d}). These lead to
\begin{equation}
\dod{\omega_s}{z}\Big\rvert_{\text{transient}}\approx
\begin{cases}
-\dfrac{\gamma_sE_s}{3\tau_0}\displaystyle\sum_jf_{R,j}\gamma_{2,j}^2\tau_{R,j} & \text{in the ($f_R$,$h_R(t)$) representation} \\
-\dfrac{\omega_s\kappa_sE_s}{6\pi\tau_0}\displaystyle\sum_jR^{\text{coeff}}_j\left(\gamma_{2,j}\tau_{R,j}\right)^2 & \text{in the $R(t)$ representation}
\end{cases}.
\label{eq:SSFS_transient_approx}
\end{equation}

\subsection*{Impulsive intrapulse redshifting}
\label{subsec:Impulsive_intrapulse_redshifting}
\subsubsection{Derivation assuming small \texorpdfstring{$(\omega_R\tau_0)$}{ωRτ0}}
\label{subsubsec:Derivation assuming small}
Here, the Raman-induced index is impulsively driven [$\Omega_R<1$; Fig.~1(b) in the article], and the monotonic rise of the index within the pulse induces a continuous redshift.

The impulsive operation typically involves an ultrashort pulse. We assume that $\gamma_2\approx0$ to ignore dephasing during the duration of the pulse in the following derivation.
\begin{align}
\dod{\omega_s}{z}\big\rvert_{\gamma_{2,j}\approx0} & =-\frac{4\omega_s\kappa_sE_s}{15\tau_0}\int_0^{\infty}B_d(x)R_d(x)\diff x \nonumber \\
& =-\frac{4\omega_s\kappa_sE_s}{15\tau_0}\sum_jR^{\text{coeff}}_j\int_0^{\infty}B_d(x)\sin(\Omega_{R_j}x)\diff x=-\frac{4\omega_s\kappa_sE_s}{15\tau_0}\sum_jR^{\text{coeff}}_jI(\Omega_{R_j}),
\label{eq:tmp6}
\end{align}
where $R(t)=\Theta(t)\sum_jR^{\text{coeff}}_j\Exp^{-\gamma_{2,j}t}\sin(\omega_{R_j}t)$ and $R_d(x)=R(t)\rvert_{t=x\tau_0,\omega_R=\frac{\Omega_R}{\tau_0}}$. After lengthy algebra and using 
\begin{subequations}
\begin{align}
\sin(\Omega_Rx) & \approx\Omega_Rx-\frac{\Omega_R^3x^3}{6} \\
\csch^4(x) & =\frac{8}{3}\sum_{m=2}^{\infty}(m+1)m(m-1)\Exp^{-2mx} \\
\int_0^{\infty}x^n\Exp^{-ax}\diff x & =\frac{\Gamma(n+1)}{a^{n+1}},
\end{align}
\end{subequations}
we can show that, at small $\Omega_R$,
\begin{equation}
I(\Omega_R)\approx\frac{15}{8}\left(\Omega_R-\frac{\pi^2}{12}\Omega_R^3\right).
\label{eq:I_OmegaR_smallx}
\end{equation}
(Derivation details are in Appendix~\ref{sec:appendix_Derivation_Eq_I_Omega_R}). Hence,
\begin{align}
\dod{\omega_s}{z}\big\rvert_{\gamma_{2,j}\approx0} & \approx-\frac{\omega_s\kappa_sE_s}{2\tau_0}\sum_jR^{\text{coeff}}_j\left(\Omega_{R_j}-\frac{\pi^2}{12}\Omega_{R_j}^3\right) \nonumber \\
& =-\frac{\omega_s\kappa_sE_s}{2}\sum_jR^{\text{coeff}}_j\left(\omega_{R_j}-\frac{\pi^2}{12}\omega_{R_j}^3\tau_0^2\right)=-\frac{\omega_s\kappa_sE_s}{2}\sum_jR^{\text{coeff}}_j\omega_{R_j}\left[1-\frac{\pi^2}{12}\left(\omega_{R_j}\tau_0\right)^2\right].
\label{eq:dwdz_gas_Rcoeff}
\end{align}

Similarly, in terms of the representation with the Raman fraction $f_R$,
\begin{align}
\dod{\omega_s}{z}\big\rvert_{\gamma_2\approx0} & \approx-\frac{4\gamma_sf_RE_s}{15\tau_0}\omega_RI(\Omega_R) \nonumber \\
& \approx-\frac{\gamma_sf_RE_s}{2}\omega_R^2\left[1-\frac{\pi^2}{12}\left(\omega_R\tau_0\right)^2\right].
\label{eq:dwdz_gas_fr}
\end{align}
In gases, the notation using $R^{\text{coeff}}$ is preferred, to decouple the analysis from the electronic nonlinearity. The concept of Raman fraction couples electronic and Raman nonlinearities. Therefore, tuning components in a gas mixture can also change the Raman fraction. For example, the $\omega_R^2$ relation in Eq.~(\ref{eq:dwdz_gas_fr}) is potentially misleading, as varying $\omega_R$ also varies the Raman fraction $f_R$. Raman-induced phase modulation is proportional to $R^{\text{coeff}}/\omega_R$ with weak dephasing [Eq.~(\ref{eq:f_R})]. We can recover Eq.~(\ref{eq:dwdz_gas_Rcoeff}) from Eq.~(\ref{eq:dwdz_gas_fr}) with the following relation:
\begin{align}
\gamma_sf_R & =\omega_s\kappa_s\int_0^{\infty}R(t)\diff t\quad\text{[Eq.~(\ref{eq:UPPE})]} \nonumber \\
& =\omega_s\kappa_sR^{\text{coeff}}\frac{\omega_R}{\omega_R^2+\gamma_2^2}\approx\omega_s\kappa_s\frac{R^{\text{coeff}}}{\omega_R}\quad,\text{under weak dephasing [Eq.~(12a) in \cite{Chen2024}]}.
\end{align}
On the other hand, Eq.~(\ref{eq:dwdz_gas_Rcoeff}) accurately captures the linear $\omega_R$ relation in impulsive redshifting, as will be shown later in Fig.~\ref{fig:SSFS_freq_shift}. In addition, the normalization requirement on the Raman response $h_R(t)$ results in non-intuitive variation of the function because of the presence of multiple Raman transitions in gases.

\subsubsection{More-intuitive formulation through generalization under weak dephasing}
\label{subsubsec:Intuitive}
In the previous section, we derived the impulsive redshift under the condition $\omega_R\tau_0\ll1$. Here, we address the problem from a different perspective, to generalize it beyond the small $\left(\omega_R\tau_0\right)$ limit.

To begin, the Raman-induced phase modulation follows [Eq.~(\ref{eq:UPPE})]
\begin{equation}
\dpd{\phi_R}{z}(z,t)=\omega\kappa\left(R\ast\abs{A}^2\right)=\kappa R^{\text{coeff}}\omega\left[\Theta(t)\sin(\omega_Rt)\right]\ast\abs{A}^2.
\end{equation}
To simplify and limit our discussion within gases in this article, we assume weak dephasing where $\gamma_2\approx0$. This leads to a frequency redshift
\begin{align}
\dpd{\omega}{z}(z,t) & =-\dpd{}{z}\left(\dpd{\phi_R}{t}\right)=-\dpd{}{t}\left(\dpd{\phi_R}{z}\right) \nonumber \\
& =-\kappa R^{\text{coeff}}\omega\omega_R\int_{-\infty}^t\cos\left(\omega_R\left(t-\tau\right)\right)\abs{A(\tau)}^2\diff\tau.
\end{align}
The soliton $sech^2$ profile generally introduces difficulties in analytic derivations, so we assume that the temporal profile is Gaussian,
\begin{equation}
\abs{A(\tau)}^2=A_0^2\Exp^{-\nicefrac{\tau^2}{\tau_0^2}}.
\end{equation}
Thus, the frequency redshift follows
\begin{align}
\dpd{\omega}{z}(z,t) & =-\kappa R^{\text{coeff}}\omega\omega_RA_0^2\int_{-\infty}^t\cos\left(\omega_R\left(t-\tau\right)\right)\Exp^{-\nicefrac{\tau^2}{\tau_0^2}}\diff\tau \nonumber \\
& =-\kappa R^{\text{coeff}}\omega\omega_RA_0^2\int_{-\infty}^t\frac{\Exp^{i\omega_R\left(t-\tau\right)}+\Exp^{-i\omega_R\left(t-\tau\right)}}{2}\Exp^{-\nicefrac{\tau^2}{\tau_0^2}}\diff\tau \nonumber \\
& =-\frac{1}{2}\kappa R^{\text{coeff}}\omega\omega_RA_0^2\left[\Exp^{i\omega_Rt}\int_{-\infty}^t\Exp^{-i\omega_R\tau-\nicefrac{\tau^2}{\tau_0^2}}\diff\tau+\Exp^{-i\omega_Rt}\int_{-\infty}^t\Exp^{i\omega_R\tau-\nicefrac{\tau^2}{\tau_0^2}}\diff\tau\right].
\label{eq:tmp1}
\end{align}
By noting that 
\begin{equation}
\int_{-\infty}^t\Exp^{-\tau^2}\diff\tau=\frac{\sqrt{\pi}}{2}\left(1+\erf(t)\right),
\end{equation}
where $\erf(\cdot)$ is the error function, we can derive that
\begin{equation}
\int_{-\infty}^t\Exp^{i\omega_R\tau-\nicefrac{\tau^2}{\tau_0^2}}\diff\tau=\Exp^{-\frac{\omega_R^2\tau_0^2}{4}}\frac{\tau_0\sqrt{\pi}}{2}\left[1+\erf\left(\frac{t}{\tau_0}-\frac{i\omega_R\tau_0}{2}\right)\right].
\label{eq:erf_relation}
\end{equation}
With this relation [Eq.~(\ref{eq:erf_relation})], Eq.~(\ref{eq:tmp1}) becomes
\begin{align}
\dpd{\omega}{z}(z,t) & =-\frac{1}{2}\kappa R^{\text{coeff}}\omega\omega_RA_0^2\Exp^{-\frac{\omega_R^2\tau_0^2}{4}}\frac{\tau_0\sqrt{\pi}}{2}\left[2\cos(\omega_Rt)+\Exp^{i\omega_Rt}\erf\left(\frac{t}{\tau_0}+\frac{i\omega_R\tau_0}{2}\right)+\Exp^{-i\omega_Rt}\erf\left(\frac{t}{\tau_0}-\frac{i\omega_R\tau_0}{2}\right)\right] \nonumber \\
& =-\frac{1}{2}\kappa R^{\text{coeff}}\omega\omega_RA_0^2\Exp^{-\frac{\omega_R^2\tau_0^2}{4}}\tau_0\sqrt{\pi}\left[\cos(\omega_Rt)+\Re\left[\Exp^{i\omega_Rt}\erf\left(\frac{t}{\tau_0}+\frac{i\omega_R\tau_0}{2}\right)\right]\right].
\label{eq:tmp2}
\end{align}
The Gaussian pulse energy
\begin{equation}
E_s=\int_{-\infty}^{\infty}A_0^2\Exp^{-\nicefrac{\tau^2}{\tau_0^2}}\diff\tau=A_0^2\tau_0\sqrt{\pi},
\end{equation}
so we can rewrite Eq.~(\ref{eq:tmp2}) as
\begin{align}
\dpd{\omega}{z}(z,t) & =-\frac{1}{2}\kappa R^{\text{coeff}}\omega\omega_RE_s\Exp^{-\frac{\omega_R^2\tau_0^2}{4}}\left[\cos(\omega_Rt)+\Re\left[\Exp^{i\omega_Rt}\erf\left(\frac{t}{\tau_0}+\frac{i\omega_R\tau_0}{2}\right)\right]\right].
\label{eq:SSFS_redshift}
\end{align}
By noting that
\begin{equation}
i\erfi(x)=\erf(ix),
\end{equation}
where $\erfi(\cdot)$ is the imaginary error function and is an odd function, we see that at $t=0$,
\begin{align}
\dpd{\omega}{z}(z,t=0) & =-\frac{1}{2}\kappa R^{\text{coeff}}\omega\omega_RE_s\Exp^{-\frac{\omega_R^2\tau_0^2}{4}}.
\label{eq:SSFS_redshift_tis0}
\end{align}

\begin{figure}[!ht]
\centering
\includegraphics[width=.9\linewidth]{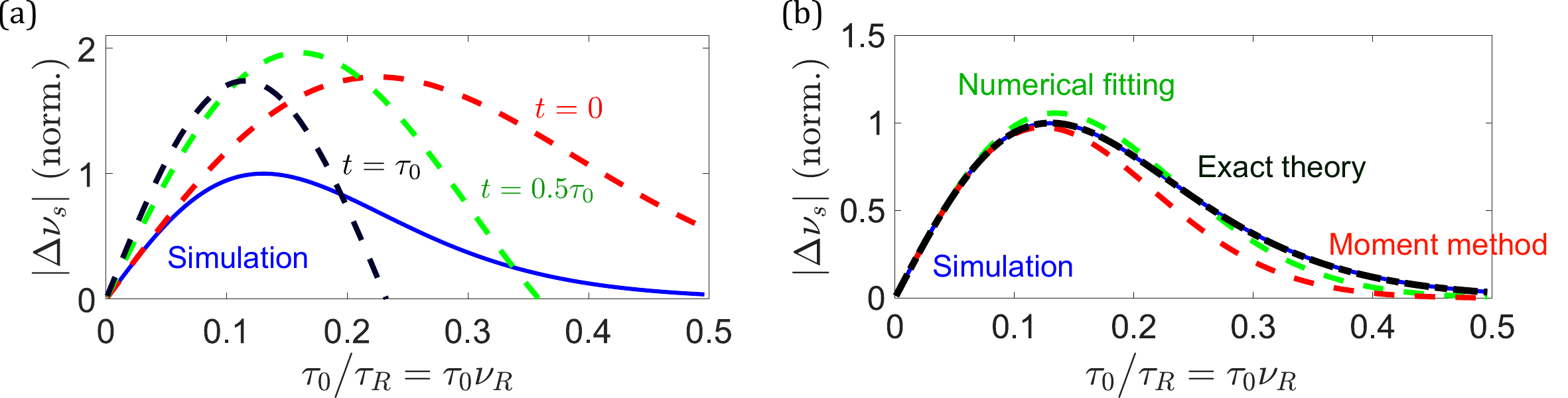}
\caption{Magnitude of impulsive intrapulse frequency redshift (blue) with different Raman transition frequencies $\nu_R$. The shift is determined by the difference between the field-averaged center frequencies of the input and output pulses. (a) also includes the theoretical redshifts at different points in time (note that the pulse is centered at $t=0)$, computed with Eq.~(\ref{eq:SSFS_redshift}). (b) includes the fitted line [green, Eq.~(\ref{eq:SSFS_redshift_fitting})], the generalization from the moment method [red, Eq.~(\ref{eq:SSFS_redshift_MM})], and the exact analytical solution [black, Eq.~(\ref{eq:impulsive_SSFS})]. We can see that the exact analytical solution perfectly matches the simulation result.}
\label{fig:SSFS_freq_shift}
\end{figure}

Apparently, the amount of redshifting is time-dependent and varies throughout the pulse [Eq.~(\ref{eq:SSFS_redshift}), and dashed lines in Fig.~\ref{fig:SSFS_freq_shift}(a)]. Due to the dynamic balance that occurs in soliton evolution, the entire pulse remains nearly chirp-free \cite{Kormokar2021} and the overall redshift is the effect of an integral over the pulse. Therefore, Eq.~(\ref{eq:SSFS_redshift}) cannot accurately predict the final pulse's redshift. As the pulse is assumed to be the strongest at $t=0$, it is reasonable to assume that overall redshift is dominated by the value at $t=0$. Through reasonable numerical least-squared fitting by modifying Eq.~(\ref{eq:SSFS_redshift_tis0}), we find that the following equation accurately predicts the pulse's overall redshift:
\begin{equation}
\dod{\omega_s}{z}(z)=-\frac{1}{2}\kappa_s R^{\text{coeff}}\omega_s\omega_RE_s\Exp^{-\frac{\omega_R^2\tau_0^2}{c_{\text{fit}}}}.
\label{eq:SSFS_redshift_fitting}
\end{equation}
where $c_{\text{fit}}=1.428$ [green line in Fig.~\ref{fig:SSFS_freq_shift}(b)]. $\omega_s$ represents the pulse's center frequency, and $\kappa_s=\kappa(\omega_s)$. Eq.~(\ref{eq:SSFS_redshift_fitting}) indicates the possible generalization of the previously-derived soliton redshifting based on the moment method [Eq.~(\ref{eq:dwdz_gas_Rcoeff})] with $1+x\approx\exp(x)$:
\begin{equation}
\dod{\omega_s}{z}(z)=-\frac{1}{2}\kappa_s R^{\text{coeff}}\omega_s\omega_RE_s\Exp^{-\frac{\omega_R^2\tau_0^2}{12/\pi^2}}.
\label{eq:SSFS_redshift_MM}
\end{equation}
We find good agreement between the fitting value $c_{\text{fit}}=1.428$ and $12/\pi^2\approx1.2159$. Compared to the intricate Eq.~(\ref{eq:SSFS_Chen}) or Eq.~(\ref{eq:dwdz_gas_Rcoeff}) under limited conditions, Eq.~(\ref{eq:SSFS_redshift_MM}) can provide a more physically-intuitive understanding of the redshifting relation across the entire impulsive regime [Fig.~\ref{fig:SSFS_freq_shift}(b)].

\subsubsection{Exact analytic formulation}
\label{subsubsec:Exact_analytic_formulation}
In the previous sections (Secs.~\ref{sec:continuous_Raman_redshifting}'s \ref{subsubsec:Derivation assuming small} and \ref{subsubsec:Intuitive}), Eq.~(\ref{eq:tmp6}) is solved with approximations, starting from small $\Omega_R$ and then generalized with the exponential function. In fact, Eq.~(\ref{eq:tmp6}) can be solved exactly analytically without approximations.

To tackle this problem, we need to introduce an autocorrelation function of the $\sech^2(\cdot)$ function:
\begin{align}
A(x)=\int_{-\infty}^{\infty}\sech^2(t)\sech^2(t-x)\diff t,
\end{align}
which, as derived in Appendix~\ref{sec:appendix_Derivation_tmp7}, is
\begin{align}
A(x)=\frac{4\left[x\cosh(x)-\sinh(x)\right]}{\sinh^3(x)}.
\label{eq:tmp7}
\end{align}
Its derivative is
\begin{align}
\dod{A}{x} & =4\frac{\left[\cosh(x)+x\sinh(x)-\cosh(x)\right]\sinh^3(x)-\left[x\cosh(x)-\sinh(x)\right]3\sinh^2(x)\cosh(x)}{\sinh^6(x)} \nonumber \\
& =4\frac{x\sinh^2(x)-3x\cosh^2(x)+3\cosh(x)\sinh(x)}{\sinh^4(x)}=4\frac{-x-2x\cosh^2(x)+3\cosh(x)\sinh(x)}{\sinh^4(x)} \nonumber \\
& =4\frac{-2x-x\cosh(2x)+\frac{3}{2}\sinh(2x)}{\sinh^4(x)}=-2\csch^4(x)\left[4x+2x\cosh(2x)-3\sinh(2x)\right].
\end{align}
Therefore, Eq.~(\ref{eq:B_d}) becomes
\begin{equation}
B_d(x)=-\frac{15}{16}\dod{A}{x}
\end{equation}
and the integral in Eq.~(\ref{eq:tmp6}) can be solved with integration by parts:
\begin{align}
\int_0^{\infty}B_d(x)\sin(\Omega_Rx)\diff x & =-\frac{15}{16}A(x)\sin(\Omega_Rx)\Big|_0^{\infty}+\frac{15}{16}\Omega_R\int_0^{\infty}A(x)\cos(\Omega_Rx)\diff x \nonumber \\
& =\frac{15}{16}\Omega_R\int_0^{\infty}A(x)\cos(\Omega_Rx)\diff x,\quad\because\text{boundary terms are zero} \nonumber \\
& =\frac{15}{32}\Omega_R\int_{-\infty}^{\infty}A(x)\cos(\Omega_Rx)\diff x,\quad\because A(x)\text{ is even} \nonumber \\
& =\frac{15}{32}\Omega_R\int_{-\infty}^{\infty}A(x)\left[\cos(\Omega_Rx)+ic_s\sin(\Omega_Rx)\right]\diff x,\quad\because A(x)\sin(\Omega_Rx)\text{ is odd} \nonumber \\
& =\frac{15}{32}\Omega_R\cdot\frac{1}{C_{\mathfrak{F}}}\mathfrak{F}\left[A\right](\Omega_R),
\end{align}
where $C_{\mathfrak{F}}$ is the Fourier-transform constant and $c_s$ determines the Fourier-transform convention ($c_s=\pm1$) \cite{Chen2025arXiv}. From the Wiener-Khinchin theorem,
\begin{align}
\mathfrak{F}\left[A\right] & =\frac{1}{C_{\mathfrak{F}}}\abs{\mathfrak{F}\left[\sech^2(x)\right]}^2 \nonumber \\
& =\frac{1}{C_{\mathfrak{F}}}\abs{C_{\mathfrak{F}}\frac{\pi\Omega_R}{\sinh(\frac{\pi\Omega_R}{2})}}^2=C_{\mathfrak{F}}\frac{\pi^2\Omega_R^2}{\sinh^2(\frac{\pi\Omega_R}{2})},
\end{align}
so
\begin{equation}
\int_0^{\infty}B_d(x)\sin(\Omega_Rx)\diff x=\frac{15}{32}\pi^2\Omega_R^3\csch^2(\frac{\pi\Omega_R}{2}).
\label{eq:intBdsin}
\end{equation}
Finally, Eq.~(\ref{eq:tmp6}) becomes
\begin{align}
\dod{\omega_s}{z}\big\rvert_{\gamma_{2,j}\approx0} & =-\frac{4\omega_s\kappa_sE_s}{15\tau_0}\sum_jR^{\text{coeff}}_j\frac{15}{32}\pi^2\Omega_{R_j}^3\csch^2(\frac{\pi\Omega_{R_j}}{2}) \nonumber \\
& =-\frac{\pi^2\omega_s\kappa_sE_s}{8\tau_0}\sum_jR^{\text{coeff}}_j\Omega_{R_j}^3\csch^2(\frac{\pi\Omega_{R_j}}{2})=-\frac{\pi^2\tau_0^2\omega_s\kappa_sE_s}{8}\sum_jR^{\text{coeff}}_j\omega_{R_j}^3\csch^2(\frac{\pi\omega_{R_j}\tau_0}{2}).
\label{eq:impulsive_SSFS}
\end{align}

Although Eq.~(\ref{eq:impulsive_SSFS}) solves the impulsive SSFS exactly analytically, we find it useful to keep the derivation of Eq.~(\ref{eq:SSFS_redshift_MM}) here in the Supplement, as its $(\omega_R\tau_0)$ relation in the exponential function is more mathematically intuitive in showing the reducing impulsive redshifting with an increased pulse duration, or even the Raman frequency if adjustable, where the Raman-induced index modulation starts to follow the pulse's temporal profile. In fact, we can recover Eq.~(\ref{eq:SSFS_redshift_MM}) from Eq.~(\ref{eq:impulsive_SSFS}) by having
\begin{align}
\csch^2(x) & \approx\frac{1}{x^2}-\frac{1}{3}+\frac{x^2}{15}-\frac{2x^4}{189}+\cdots \nonumber \\
& =\frac{1}{x^2}\left(1-\frac{x^2}{3}+\frac{x^4}{15}-\frac{2x^6}{189}+\cdots\right)\approx\frac{1}{x^2}\Exp^{-\frac{x^2}{3}},\quad\text{if }\abs{x}\ll1,
\end{align}
which leads to
\begin{align}
\dod{\omega_s}{z}\big\rvert_{\gamma_{2,j}\approx0} & \approx-\frac{1}{2}\omega_s\kappa_sE_s\sum_jR^{\text{coeff}}_j\omega_{R_j}\Exp^{-\frac{\omega_{R_j}^2\tau_0^2}{12/\pi^2}}.
\end{align}

\clearpage
\subsection{Summary of the intrapulse redshifting relations}
\label{subsec:summary_SSFS}
In summary, intrapulse redshifting follows:
\begin{subequations}
\begin{align}
\dod{\omega_s}{z}\Big\rvert_{\text{steady-state}} & =
\begin{cases}
-\dfrac{4\gamma_sf_RE_s}{15\tau_0^3}\displaystyle\int_0^{\infty}th_R(t)\diff t & \text{in the ($f_R$,$h_R(t)$) representation} \\
-\dfrac{4\omega_s\kappa_sE_s}{15\tau_0^3}\displaystyle\int_0^{\infty}tR(t)\diff t & \text{in the $R(t)$ representation}
\end{cases} \label{eq:SSFS_ss_summary} \\
\dod{\omega_s}{z}\Big\rvert_{\text{transient}} & =
\begin{cases}
-\dfrac{4\gamma_sE_s}{15\tau_0}\displaystyle\sum_jf_{R,j}\gamma_{2,j}^2\tau_{R,j}\int_0^{\infty}B_d(x)\Exp^{-\gamma_{2,j}\tau_0x}\diff x & \text{in the ($f_R$,$h_R(t)$) representation} \\
-\dfrac{2\omega_s\kappa_sE_s}{15\pi\tau_0}\displaystyle\sum_jR^{\text{coeff}}_j\left(\gamma_{2,j}\tau_{R,j}\right)^2\int_0^{\infty}B_d(x)\Exp^{-\gamma_{2,j}\tau_0x}\diff x & \text{in the $R(t)$ representation}
\end{cases} \nonumber \\
& \approx
\begin{cases}
-\dfrac{\gamma_sE_s}{3\tau_0}\displaystyle\sum_jf_{R,j}\gamma_{2,j}^2\tau_{R,j} & \text{in the ($f_R$,$h_R(t)$) representation} \\
-\dfrac{\omega_s\kappa_sE_s}{6\pi\tau_0}\displaystyle\sum_jR^{\text{coeff}}_j\left(\gamma_{2,j}\tau_{R,j}\right)^2 & \text{in the $R(t)$ representation}
\end{cases}\quad,\text{if }\gamma_2\tau_0\ll1 \label{eq:SSFS_transient_summary} \\
& \hspace{13 em} \left[\text{Note that }\int_0^{\infty}B_d(x)\diff x=\frac{5}{4}\text{ (see Appendix~\ref{sec:appendix_B_d}).}\right] \nonumber \\
\dod{\omega_s}{z}\Big\rvert_{\text{impulsive}} & =-\frac{\pi^2\tau_0^2\omega_s\kappa_sE_s}{8}\sum_jR^{\text{coeff}}_j\omega_{R_j}^3\csch^2(\frac{\pi\omega_{R_j}\tau_0}{2}) \nonumber \\
& \approx-\frac{1}{2}\kappa_s \omega_sE_s\sum_jR^{\text{coeff}}_j\omega_{R_j}\Exp^{-\frac{\omega_{R_j}^2\tau_0^2}{12/\pi^2}}\approx-\frac{1}{2}\kappa_s \omega_sE_s\sum_jR^{\text{coeff}}_j\omega_{R_j}\quad,\text{if }\omega_{R_j}\tau_0\ll1. \label{eq:SSFS_impulsive_summary}
\end{align} \label{eq:SSFS_all}
\end{subequations}

The fundamental difference between continuous redshifting in the steady-state and transient regimes resides in the relation between the dephasing time $T_2$ and the pulse duration $\tau_0$: when $T_2\ll\tau_0$, it is in the steady-state regime; otherwise, it is in the transient regime (and $\tau_R$ is small in both situations). Hence, unlike what is discussed in the article, in general, dephasing does not need to be strong for continuous redshifting to simply occur (whether it is weak or strong), and $T_2$ and $\tau_R$ remain decoupled. Mathematically, nonzero dephasing induces a linear spectral gain near zero frequency [Fig.~1(c) left], which results in the temporally-delayed index modulation and continuous redshifting. However, if the linear gain is too weak, or equivalently the amount of temporal delay is too small, the pulse undergoes a slow and weak continuous redshift. Subsequently, discrete (Stokes/anti-Stokes) frequency generation, which typically dominates both transient and steady-state dynamics, emerges and overtakes continuous redshifting. Thus, for continuous redshifting to occur and prevail in the overall process, strong dephasing is required (\eg $T_2\approx\tau_R$) to induce a more significantly-delayed index and stronger linear gain near zero frequency. An alternative way to induce a strong delayed index is impulsive excitation, \ie the impulsive redshifting. \textbf{The overall conclusion is that effective continuous redshifting can occur either in the steady state with strong dephasing, or with impulsive excitation [Fig.~2(b)].}

\begin{figure}[!ht]
\centering
\includegraphics[width=0.9\linewidth]{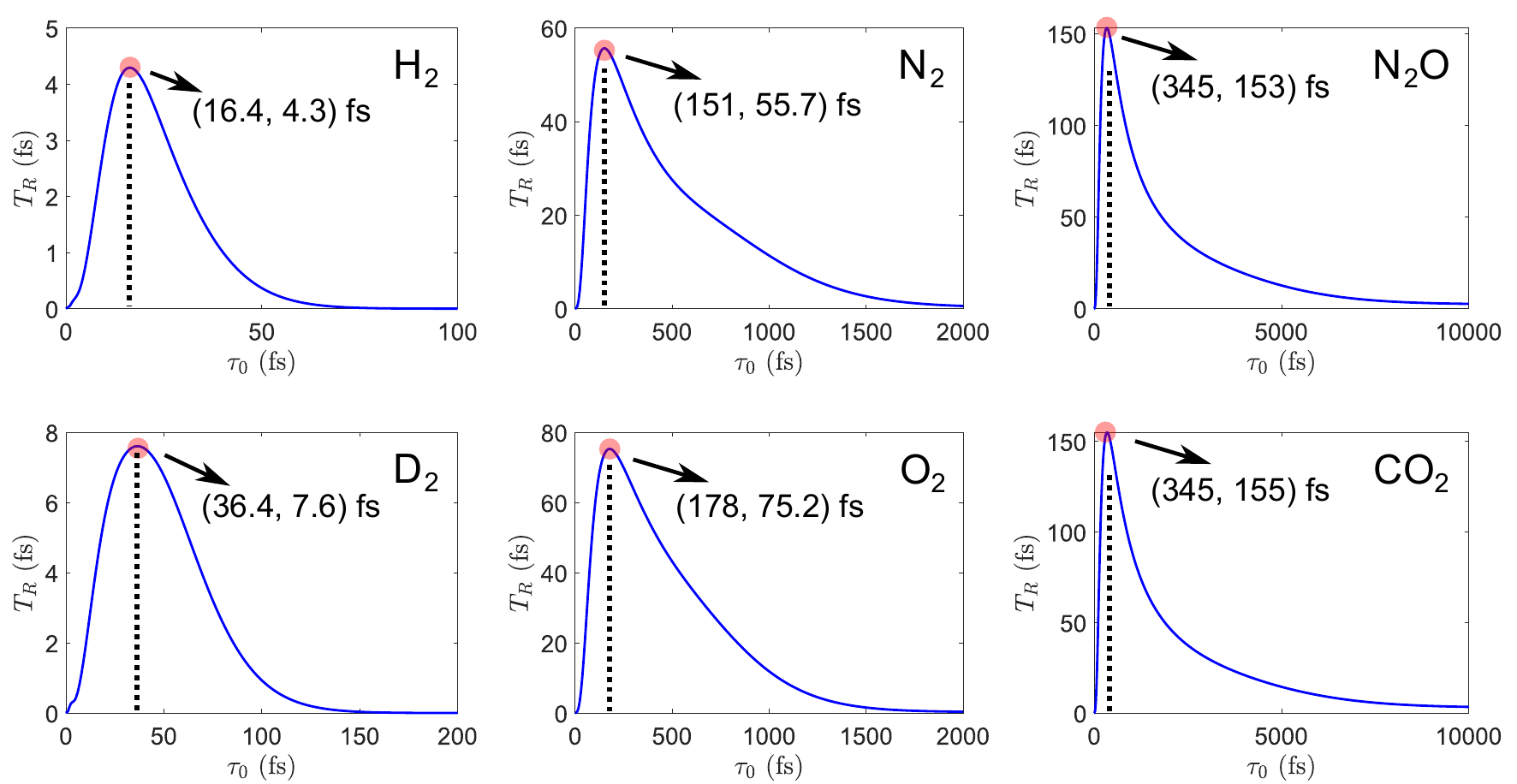}
\caption{Raman time $T_R(\tau_0)=f_RT'_R(\tau_0)$ [Eq.~(\ref{eq:T_R_tau0})] evaluated as a function of the pulse duration $\tau_0$, in different gases under \SIadj{1}{\bar} pressure. Number shown in each figure is the duration and the corresponding Raman time, where $T_R$ is maximum (red dot). Pressure is crucial here, as it determines the dephasing time that can change the overall shape [see Fig.~2(b)].}
\label{fig:effective_Raman_time}
\end{figure}

\begin{figure}[!ht]
\centering
\includegraphics[width=0.9\linewidth]{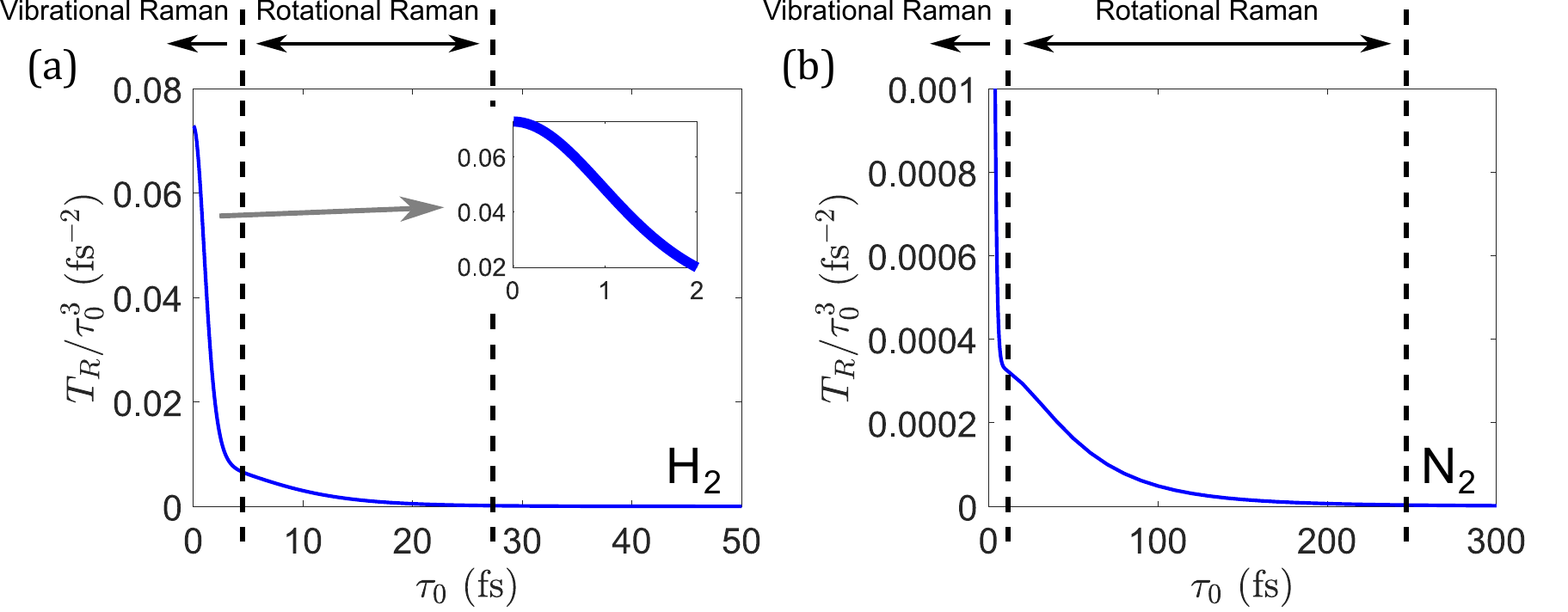}
\caption{$T_R(\tau_0)/\tau_0^3$ evaluated as a function of the pulse duration $\tau_0$, in \SIadj{1}{\bar} (a) \ce{H2} and (b) \ce{N2}, respectively. Pulse energy is fixed. This is an accurate metric to determine the Raman redshifting rate under a fixed energy. A flat curve (with zero slope) represents the independence in redshifting of the pulse duration.}
\label{fig:effective_Raman_time_over_tau03}
\end{figure}

The transition duration at which the impulsive redshift begins to diminish is governed not only by the individual Raman transition frequencies but also by the collective Raman distribution. This duration corresponds to the regime where the Raman-induced index change starts to track the pulse's temporal profile, a behavior strongly influenced by the beating among all Raman modes. Ultimately, the resulting first temporal spike determines the effective transition duration (Fig.~\ref{fig:Raman_response_gas}). This transition can be quantitatively characterized by computing the effective Raman time across different pulse durations (Fig.~\ref{fig:effective_Raman_time}). The integral $T_R=f_R\int_0^{\infty}B(t,\tau_0)h_R(t)\diff t$ of effective Raman time explicitly incorporates the influence of the beating and therefore captures the resulting temporal spike. It is worth noting that the duration of maximum redshift here does not rely only on the effective Raman time due to the $\left(T_RE_s/\tau_0^3\right)$ relation in the shifting formula, which involves the duration term. The pulse can vary with or without following the fundamental soliton relation. If pulse duration is reduced under a fixed energy, the overall redshifting rate follows $\left(T_R/\tau_0^3\right)$. Taking \ce{H2} as an example [Fig.~\ref{fig:effective_Raman_time_over_tau03}(a)], it initially increases to a pulse duration around \SI{20}{\fs} (as in Fig.~\ref{fig:effective_Raman_time}) due to the rotational Raman response. Prior to reaching a plateau (\eg independent of $\tau_0$ in the impulsive regime of the rotational SRS), it increases again due to the vibrational Raman response with \SIadj{8}{\fs} period. At ultrashort durations, where both Raman processes are impulsive, the curve ultimately flattens. In \ce{N2} [Fig.~\ref{fig:effective_Raman_time_over_tau03}(b)], with a reduced duration, it starts to rise fast at \SI{150}{\fs} (Fig.~\ref{fig:effective_Raman_time}). More Raman transitions contribute to impulsive redshifting as the duration is reduced. Nevertheless, \textbf{the duration of the maximum effective Raman time corresponds to the duration where the pulse begins to redshift significantly}.

In Fig.~2(a) of the article, the computation varies the Raman transition frequency but not the pulse duration or energy, so only $T_R$ is varied and affects intrapulse Raman redshifting [Eq.~(\ref{eq:SSFS_Chen})]. More accurately speaking, redshifting follows $\int_0^{\infty}B_d(x)\Exp^{-\gamma_{2,d}x}\diff x\sim T_R/\tau_0^2$ [Eq.~(\ref{eq:TRBd_tmp})], leading to a different relation from Fig.~\ref{fig:effective_Raman_time_over_tau03} having a reduced redshift at small Raman frequencies $\tau_0\nu_R$ (normalized to the unchanged $\tau_0$).

\textbf{The bounding effect on the soliton duration during the impulsive intrapulse redshifting can be accurately characterized in Fig.~\ref{fig:effective_Raman_time_over_tau03}, which, as previously discussed, corresponds to the duration of maximum effective Raman time.} In practical experimental conditions where energy variation is affected by the quantum defect and is thus reasonably small, its intrapulse redshifting closely follows $T_R/\tau_0^3$, whose value approaches zero beyond a threshold duration determined by the first temporal spike of the Raman response. 

\clearpage
\subsection{Frequency shift with extended propagation}
\label{subsec:freq_shift_extended}
In the previous sections, we derived only the Raman shifting rate. Here, we will derive the frequency shift of a fundamental soliton after extended propagation.

We assume that the propagation is lossless except for the quantum defect from the Raman shifting; that is to say, the photon number is fixed. To simplify the problem, we also assume that the amount of redshift is small, such that the energy variation is small and the variation of the pulse duration can be neglected in calculation if necessary.

With these assumptions, we first derive the steady-state SSFS shift. We rewrite Eq.~(\ref{eq:SSFS_ss_summary}) by using the fundamental soliton number relation:
\begin{align}
\dod{\omega_s}{z} & =-\frac{4\kappa_s\omega_sE_s}{15\tau_0^3}\int_0^{\infty}tR(t)\diff t=-\frac{\kappa_s^4\omega_s^4E_s^4}{30\abs{\beta_2}^3}\left(\int_0^{\infty}tR(t)\diff t\right)\left(\kappa_e+\int_0^{\infty}R(t)\diff t\right)^3, \nonumber \\
& \hspace{30em}\begin{cases}
N=\sqrt{\dfrac{\gamma_{\text{soliton}}E_s\tau_0}{2\abs{\beta_2}}}=1 \\
\gamma_{\text{soliton}}=\gamma_s\text{ in the steady-state regime}
\end{cases} \nonumber \\
& =-\frac{\kappa_s^4\omega_s^4\left(N^{\text{photon}}\hbar\omega_s\right)^4}{30\abs{\beta_2}^3}\left(\int_0^{\infty}tR(t)\diff t\right)\left(\kappa_e+\int_0^{\infty}R(t)\diff t\right)^3 \nonumber \\
& =-\frac{\kappa_s^4\left(N^{\text{photon}}\right)^4\hbar^4}{30\abs{\beta_2}^3}\left(\int_0^{\infty}tR(t)\diff t\right)\left(\kappa_e+\int_0^{\infty}R(t)\diff t\right)^3\omega_s^8.
\label{eq:tmp5}
\end{align}
$N^{\text{photon}}$ is the photon number of the soliton. We then obtain the shift after propagation through a distance $z$:
\begin{equation}
\text{Steady-state SSFS: }\frac{1}{\left[\omega_s(z)\right]^7}-\frac{1}{\left[\omega_s(0)\right]^7}=\frac{7\kappa_s^4\left(N^{\text{photon}}\right)^4\hbar^4}{30\abs{\beta_2}^3}\left(\int_0^{\infty}tR(t)\diff t\right)\left(\kappa_e+\int_0^{\infty}R(t)\diff t\right)^3z.
\label{eq:ss_SSFS_extended}
\end{equation}

For the transient regime, we consider only a simplified situation where $\gamma_2\tau_0\ll1$ to avoid the duration dependence of the $B_d$ integral [Eq.~(\ref{eq:SSFS_transient_approx})]. Here, $\gamma_{\text{soliton}}=\gamma_s$ as in the steady-state regime because of the pulse-following Raman-induced index change.
\begin{align}
\dod{\omega_s}{z} & =-\dfrac{\omega_s^2\kappa_s^2\left(N^{\text{photon}}\hbar\omega_s\right)^2}{12\pi\abs{\beta_2}}\left(\kappa_e+\int_0^{\infty}R(t)\diff t\right)\sum_jR^{\text{coeff}}_j\left(\gamma_{2,j}\tau_{R,j}\right)^2 \nonumber \\
& =-\dfrac{\kappa_s^2\left(N^{\text{photon}}\right)^2\hbar^2}{12\pi\abs{\beta_2}}\left(\kappa_e+\int_0^{\infty}R(t)\diff t\right)\sum_jR^{\text{coeff}}_j\left(\gamma_{2,j}\tau_{R,j}\right)^2\omega_s^4.
\end{align}
Thus, after propagation over a distance $z$:
\begin{align}
\text{Transient SSFS: }\frac{1}{\left[\omega_s(z)\right]^3}-\frac{1}{\left[\omega_s(0)\right]^3}=\dfrac{\kappa_s^2\left(N^{\text{photon}}\right)^2\hbar^2}{4\pi\abs{\beta_2}}\left(\kappa_e+\int_0^{\infty}R(t)\diff t\right)\sum_jR^{\text{coeff}}_j\left(\gamma_{2,j}\tau_{R,j}\right)^2z.
\label{eq:transient_SSFS_extended}
\end{align}

Similarly, for the impulsive SSFS,
\begin{equation}
\dod{\omega_s}{z}=-\frac{\pi^2\tau_0^2\omega_s\kappa_sE_s}{8}\sum_jR^{\text{coeff}}_j\omega_{R_j}^3\csch^2(\frac{\pi\omega_{R_j}\tau_0}{2})=-\frac{\pi^2\tau_0^2\kappa_sN^{\text{photon}}\hbar}{8}\sum_jR^{\text{coeff}}_j\omega_{R_j}^3\csch^2(\frac{\pi\omega_{R_j}\tau_0}{2})\omega_s^2.
\end{equation}
We assume that $\tau_0$ is approximately fixed during the propagation. After propagation through a distance $z$:
\begin{equation}
\text{Impulsive SSFS: }\frac{1}{\omega_s(z)}-\frac{1}{\omega_s(0)}=\frac{\pi^2\tau_0^2\kappa_sN^{\text{photon}}\hbar}{8}\sum_jR^{\text{coeff}}_j\omega_{R_j}^3\csch^2(\frac{\pi\omega_{R_j}\tau_0}{2})z.
\label{eq:impulsive_SSFS_extended}
\end{equation}

The above derivations assume that the nonlinearity is constant during propagation, \eg under a constant pressure of a Raman-active gas. In the experiments described in the article, a pressure gradient was employed. Here we re-derive the impulsive SSFS relation with the gradient pressure profile. To derive this relation, we need to use the gradient pressure formula in Eq.~(\ref{eq:gradient_p}). Because of the vacuum setting in one gas cell $p_1=0$, the pressure varies as $p(z)=\frac{p_2}{\sqrt{L}}\sqrt{z}$, which leads us to
\begin{align}
& \begin{aligned}
\dod{\omega_s}{z} & =-\frac{\pi^2\tau_0^2\left(p(z)\kappa_s^{\text{\SI{1}{\bar}}} \right)N^{\text{photon}}\hbar}{8}\sum_jR^{\text{coeff}}_j\omega_{R_j}^3\csch^2(\frac{\pi\omega_{R_j}\tau_0}{2})\omega_s^2 \\
& =-\frac{\pi^2\tau_0^2p_2\kappa_s^{\text{\SI{1}{\bar}}} N^{\text{photon}}\hbar}{8\sqrt{L}}\sum_jR^{\text{coeff}}_j\omega_{R_j}^3\csch^2(\frac{\pi\omega_{R_j}\tau_0}{2})\omega_s^2\sqrt{z}
\end{aligned} \nonumber \\
& \begin{aligned}
\Rightarrow \frac{1}{\omega_s(L)}-\frac{1}{\omega_s(0)} & =\frac{\pi^2\tau_0^2\kappa_s^{\text{\SI{1}{\bar}}} N^{\text{photon}}\hbar}{8}\sum_jR^{\text{coeff}}_j\omega_{R_j}^3\csch^2(\frac{\pi\omega_{R_j}\tau_0}{2})\frac{p_2}{\sqrt{L}}\frac{2}{3}L^{3/2} \\
& =\frac{\pi^2\tau_0^2\left(\frac{2}{3}p_2\kappa_s^{\text{\SI{1}{\bar}}}\right)N^{\text{photon}}\hbar}{8}\sum_jR^{\text{coeff}}_j\omega_{R_j}^3\csch^2(\frac{\pi\omega_{R_j}\tau_0}{2})L.
\end{aligned}
\end{align}
Therefore, the gradient pressure profile effectively reduces the pressure by a factor of $2/3$ compared to constant pressure [Eq.~(\ref{eq:impulsive_SSFS_extended})].

\clearpage
\subsection{Maximum Raman time}
\label{subsec:Maximum_effective_Raman_time}
In this section, we analytically solve for the maximum Raman time in different regimes [Fig.~2(b) in the article].

With the simplest single damped sinusoidal Raman response function, the Raman time is
\begin{equation}
T_R(\tau_0)=f_R\int_0^{\infty}B(t,\tau_0)\left[\frac{\gamma_2^2+\omega_R^2}{\omega_R}\Exp^{-\gamma_2t}\sin(\omega_Rt)\right]\diff t.
\end{equation}

First we solve for the long-pulse transient or steady-state regime where ${\tau_0\nu_R}\to\infty$. In this regime, the integrand can be approximated with a large $\tau_0$ under a limited range of $t$ because the exponential decay term $\gamma_2$ defines the effective $t$ range of the integration. As solved in Eq.~(\ref{eq:B_d_ss}), $B(t,\tau_0)\to t$ under a large $\tau_0$, leading to
\begin{align}
T_R & \approx f_R\frac{\gamma_2^2+\omega_R^2}{\omega_R}\int_0^{\infty}t\Exp^{-\gamma_2t}\sin(\omega_Rt)\diff t \nonumber \\
& =f_R\frac{\gamma_2^2+\omega_R^2}{\omega_R}\frac{2\gamma_2\omega_R}{\left(\gamma_2^2+\omega_R^2\right)^2} \nonumber \\
& =f_R\frac{2\gamma_2}{\gamma_2^2+\omega_R^2}.
\end{align}
Then, we rewrite the Raman fraction as a function of $\gamma_2$:
\begin{align}
f_R(\gamma_2) & =\frac{R^{\text{coeff}}\frac{\omega_R}{\gamma_2^2+\omega_R^2}}{\kappa_e+R^{\text{coeff}}\frac{\omega_R}{\gamma_2^2+\omega_R^2}} \nonumber \\
& =\frac{R^{\text{coeff}}\omega_R}{\kappa_e\left(\gamma_2^2+\omega_R^2\right)+R^{\text{coeff}}\omega_R}.
\label{eq:f_R(gamma2)}
\end{align}
From it,
\begin{equation}
f_R^0=f_R(0)=\frac{R^{\text{coeff}}\omega_R}{\kappa_e\omega_R^2+R^{\text{coeff}}\omega_R},
\end{equation}
so
\begin{equation}
\begin{cases}
R^{\text{coeff}}\omega_R=f_R^0\left(\kappa_e\omega_R^2+R^{\text{coeff}}\omega_R\right)=f_R^0C \\
\kappa_e\omega_R^2=\left(1-f_R^0\right)\left(\kappa_e\omega_R^2+R^{\text{coeff}}\omega_R\right)=\left(1-f_R^0\right)C
\end{cases},
\end{equation}
leading to
\begin{equation}
f_R(\gamma_2)=\frac{f_R^0C}{\left(1-f_R^0\right)C\frac{\gamma_2^2}{\omega_R^2}+C}=\frac{f_R^0\omega_R^2}{\omega_R^2+\left(1-f_R^0\right)\gamma_2^2}.
\end{equation}
Hence,
\begin{align}
T_R & =\frac{f_R^0\omega_R^2}{\omega_R^2+\left(1-f_R^0\right)\gamma_2^2}\frac{2\gamma_2}{\gamma_2^2+\omega_R^2} \nonumber \\
& =2\frac{f_R^0}{\omega_R}\frac{1}{1+f_e^0u^2}\frac{u}{u^2+1},\quad \text{let }f_e^0=1-f_R^0\text{ and }u=\frac{\gamma_2}{\omega_R}.
\end{align}
Setting $\dod{\ln T_R}{u}=0$ and simplifying ($s=u^2$) gives a quadratic formula:
\begin{equation}
3f_e^0s^2+\left(1+f_e^0\right)s-1=0\quad\Rightarrow\quad s=\frac{-\left(1+f_e^0\right)+\sqrt{\left((f_e^0\right)^2+14f_e^0+1}}{6f_e^0}.
\end{equation}
We obtain
\begin{equation}
\gamma_2^{\max(T_2)}=\omega_R\sqrt{\frac{-\left(1+f_e^0\right)+\sqrt{\left(f_e^0\right)^2+14f_e^0+1}}{6f_e^0}}.
\end{equation}
This gives the optimal $\gamma_2$ at two limits:
\begin{equation}
\begin{cases}
\kappa_e\to\infty\quad(f_e^0\to1): & \gamma_2^{\max(T_2)}=\omega_R\sqrt{\frac{-2+4}{6}}=\dfrac{\omega_R}{\sqrt{3}} \\
\kappa_e=0\quad(f_e^0=0): & \gamma_2^{\max(T_2)}\approx\omega_R\sqrt{\frac{-\left(1+f_e^0\right)+\left(1+7f_e^0\right)}{6f_e^0}}=\omega_R,\quad\text{using }\sqrt{1+x}\approx1+\frac{x}{2}\text{, as }x\ll1.
\end{cases}.
\end{equation}
That is to say, the optimal dephasing for maximum long-pulse Raman time varies with Raman fraction. This is because of the changing Raman fraction when Raman dephasing varies under the existence of electronic nonlinearity that is not affected by the varying dephasing [Eq.~(\ref{eq:f_R(gamma2)})]. Fig.~2(b) in the article solves $T_R/f_R$ to exclude the effect of Raman fraction, \ie the effect of electronic nonlinearity. Solving this is the same as setting $f_R=1$ here.

Second, we solve for the short-pulse impulsive regime where $\gamma_2\to0$.
\begin{equation}
T_R(\tau_0)=f_R\omega_R\int_0^{\infty}B(t,\tau_0)\sin(\omega_Rt)\diff t.
\end{equation}
With Eq.~(\ref{eq:intBdsin}),
\begin{equation}
T_R(\tau_0)=f_R\omega_R^4\tau_0^5\frac{15}{32}\pi^2\csch^2(\frac{\pi\omega_R\tau_0}{2}).
\end{equation}
Setting $\dod{\ln T_R}{\omega_R}=0$ and simplifying ($v=\tau_0\nu_R$) gives
\begin{equation}
\tanh(y)=\frac{2y}{5},
\end{equation}
where $y=\pi^2v$. This transcendental equation has the nontrivial root at \num{2.46406}, so
\begin{equation}
v^{\max(T_2)}=\frac{2.46406}{\pi^2}=0.24966.
\end{equation}

\clearpage
\section{Raman fraction and Raman gain}
\label{sec:fr_g}
This section aims to clarify the relationship between Raman fraction and Raman gain. Raman fraction is sometimes roughly defined as the ratio of Raman nonlinearity to the total nonlinearity. However, what does ``the nonlinearity'' mean specifically? In general, the nonlinear susceptibility $\chi^{(3)}$ in the spectral domain has real and imaginary parts. Because of real-valued electronic nonlinearity, people might think that the Raman fraction is determined by real parts of different nonlinearities (still, at which frequency?). On the other hand, the Raman gain is known to depend on the imaginary part at the Stokes or anti-Stokes frequency. Therefore, it seems reasonable to also calculate using the imaginary part of the Raman nonlinearity. Here, we clarify the potential confusion.

To begin, it is helpful to introduce the Raman integrals \cite{Chen2024}:
\begin{subequations}
\begin{align}
\mathcal{R}_1 & =R(t)\ast\abs{A(t)}^2=\int_{-\infty}^tR(t-\tau)\abs{A(\tau)}^2\diff\tau \\
\mathcal{R}_{2;i,j,k} & =A^i\int_{-\infty}^tR(t-\tau)A^j(\tau)A^k(\tau)\Exp^{-i\omega_R(t-\tau)}\diff\tau, \label{eq:R2}
\end{align}
\end{subequations}
where $i$, $j$, and $k$ can be either S (Stokes), P (pump), or AS (anti-Stokes); $i^*$, $j^*$, and $k^*$ represent complex conjugate of the corresponding fields. $\mathcal{R}_1$ and $\mathcal{R}_2$ govern Raman-induced index change and Raman gain, respectively. For details, please see \cite{Chen2024}. In the steady-state regime, they can be simplified to
\begin{subequations}
\begin{align}
\mathcal{R}_1 & \approx I_R\abs{A^P(t)}^2 \label{eq:I_R_ss} \\
\mathcal{R}_{2;i,j,k} & \approx\mathfrak{R}_{\text{ss}}A^iA^jA^k, \label{eq:R_ss_ss}
\end{align}
\end{subequations}
where
\begin{subequations}
\begin{align}
I_R & =\int_{-\infty}^tR(t-\tau)\diff\tau=\int_0^{\infty}R(\tau)\diff\tau \\
\mathfrak{R}_{\text{ss}} & =\int_{-\infty}^tR(t-\tau)\Exp^{-i\omega_R(t-\tau)}\diff\tau=\int_0^{\infty}R(\tau)\Exp^{-i\omega_R\tau}\diff\tau.
\end{align}
\end{subequations}

Before we enter the details and explanation, let me show the conclusion first: \textbf{Raman fraction is determined by the real part of $\mathfrak{F}\left[R\right]$ at zero frequency, which is $I_R$ [Eq.~(\ref{eq:I_R_ss})], and Raman gain is determined by its imaginary part at Stokes frequency $\mathfrak{R}_{\text{ss}}$ [Eq.~(\ref{eq:R_ss_ss})]}.

From Eq.~(\ref{eq:UPPE}), we can see that the nonlinear phase modulations induced by electronic and Raman nonlinearities are, respectively, $\kappa_e$ and $I_R$. Therefore as shown in Eq.~(\ref{eq:f_R}), the definition of Raman fraction depends only on the nonlinear phase modulation when SRS induces an index change that follows the pulse's temporal profile as the electronic one:
\begin{equation}
f_R=\frac{I_R}{\kappa_e+I_R}.
\end{equation}
This condition occurs not only in the steady-state regime shown here but also in the transient regime, whose derivation is more involved and can be found in \cite{Chen2024}. In the transient regime, Raman-induced index also follows the temporal profile and follows $I_R=\sum_jR_j^{\text{coeff}}/\omega_{R_j}$.

It is important to note that the Raman fraction here is an intrinsic material property, rather than a duration-dependent parameter \cite{Reichert2014,Zhao2018a,Chemnitz2017}. In \cite{Chemnitz2017}, \etal{Chemnitz} defines Raman fraction based on the effective nonlinearity:
\begin{align}
n_{2,\text{eff}} & =n_{2,\text{electronic}}+\frac{\int_{-\infty}^{\infty}\abs{A(t)}^2\int_0^{\infty}\epsilon_0^2n_{\text{eff}}^2cR(t-\tau)\abs{A(\tau)}^2\diff\tau\diff t}{\int_{-\infty}^{\infty}\abs{A(t)}^4\diff t} \nonumber \\
& =\epsilon_0^2n_{\text{eff}}^2c\left(\kappa_e+\frac{\int_{-\infty}^{\infty}\abs{A(t)}^2\int_0^{\infty}R(t-\tau)\abs{A(\tau)}^2\diff\tau\diff t}{\int_{-\infty}^{\infty}\abs{A(t)}^4\diff t}\right) \nonumber \\
& =\epsilon_0^2n_{\text{eff}}^2c\left(\kappa_e+R_{\text{avg}}\right).
\end{align}
$(\epsilon_0^2n_{\text{eff}}^2c)$ is added to make our Raman response function conform to the one used in \cite{Reichert2014} that has the same unit as $n_2$ [Eq.~(\ref{eq:kappa_e_n2})]. This formula can be derived from Eq~(\ref{eq:UPPE}):
\begin{align}
\int_{-\infty}^{\infty}A^*\left[\kappa_e\abs{A}^2A+\left(R\ast\abs{A}^2\right)A\right]\diff t=\int_{-\infty}^{\infty}\left[\kappa_e\abs{A(t)}^4+\abs{A(t)}^2\int_0^{\infty}R(t-\tau)\abs{A(\tau)}^2\diff\tau\right]\diff t,
\end{align}
which, in fact, is not the ``Raman fraction'' the nonlinear pulse propagation equation [Eq.~(\ref{eq:UPPE})] uses. As described in \cite{Reichert2014} [its Eqs.~(15) and (16)], this definition corresponds to the \emph{time-averaged} nonlinear index variation, whose derived Raman fraction $f_{R,\text{avg}}=R_{\text{avg}}/\left(\kappa_e+R_{\text{avg}}\right)$ approaches the true Raman fraction $f_R$ only in the long-pulse Raman regime due to pulse-following.

The Raman gain, on the other hand, depends only on $\mathfrak{R}_{\text{ss}}$ but not on $I_R$ \cite{Chen2024}:
\begin{alignat}{2}
g_{\text{ss}} & \mathrlap{=\kappa\omega_R\Im\left[\mathfrak{R}_{\text{ss}}\right]\abs{A^P}^2} \nonumber \\
& +\mybig|\Re\Bigg[\frac{1}{2} && \sqrt{2\kappa\left[\left(\sqrt{\omega^S\omega^{AS}}+\omega^P\right)\left(\kappa_e+\mathfrak{R}_{\text{ss}}\right)\right]\abs{A^P}^2-\triangle\beta}. \nonumber \\
&&\hspace{-1em} \times & \sqrt{2\kappa\left[\left(\sqrt{\omega^S\omega^{AS}}-\omega^P\right)\left(\kappa_e+\mathfrak{R}_{\text{ss}}\right)\right]\abs{A^P}^2+\triangle\beta}\Bigg]\mybig|,
\label{eq:Raman_gain_ss}
\end{alignat}
where $\triangle\beta$ is the wave-vector mismatch among the three fields. In the transient regime, $\left(\mathfrak{R}_{\text{ss}}\abs{A^P}^2\right)$ is replaced by $\mathfrak{R}_{\text{tr}}$ that depends on the time-integrated energy $E_{\text{int}}(t)=\int_{-\infty}^t\abs{A(\tau)}^2\diff\tau$.

\begin{figure}[!ht]
\centering
\includegraphics[width=.75\linewidth]{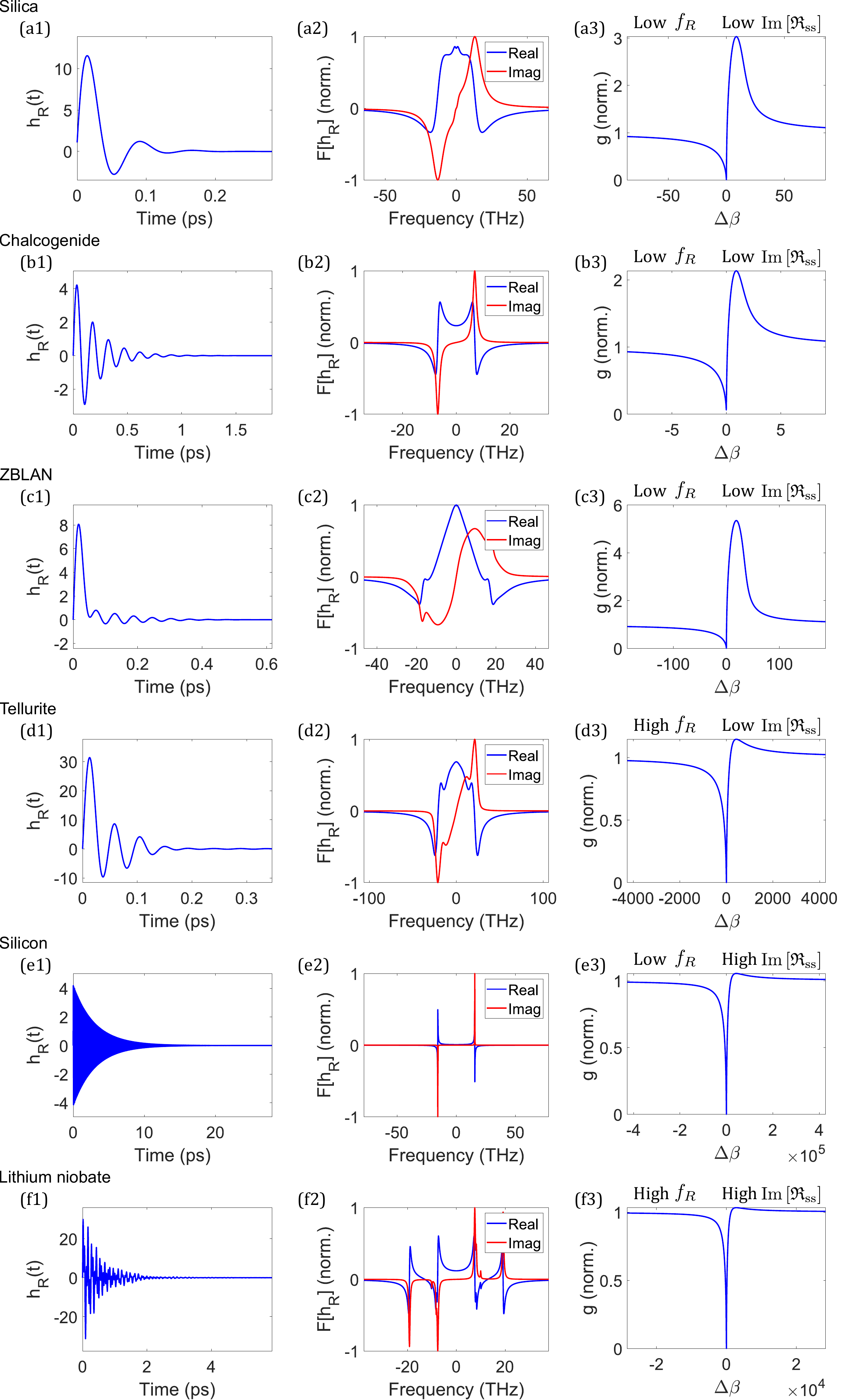}
\caption{Raman response function in time (a1--f1) and in frequency (a2--f2), as well as the Raman gain (a3--f3), for silica, chalcogenide, ZBLAN, tellurite, silicon, and lithium niobate. Note that values of wave-vector mismatch result from our choice of pulses and fiber parameters and will not change the gain shape. We assume a pulse of \SIadj{1}{\nano\joule} energy and \SIadj{1}{\ps} duration in a single-mode \SIadj{6}{\micro\m} fiber.}
\label{fig:Raman_response_solid}
\end{figure}

The shape of the Raman gain with respect to wave-vector mismatch is governed by the real and imaginary parts of $\left(\kappa_e+\mathfrak{R}_{\text{ss}}\right)$ \cite{Chen2024}. If a material has weak dephasing (as in Raman-active gases or silicon [Fig.~\ref{fig:Raman_response_solid}(e1--3)]), the imaginary part can be thousands of times larger than its real part. Therefore, even with a small Raman fraction so that $\kappa_e$ seems to dominate the term, its imaginary part can still be non-negligible.

Polarization is a crucial factor, as the Raman fraction will change with different polarizations. In general, the Raman nonlinearity is composed of an isotropic ($R_a$) and an anisotropic ($R_b$) parts \cite{Chen2024}:
\begin{subequations}
\begin{align}
R_a & =R^{\text{vib}}-2R^{\text{rot}} \\
R_b & =6R^{\text{rot}},
\end{align}
\end{subequations}
where $R^{\text{vib}}(t)$ and $R^{\text{rot}}(t)$ represent vibrational and rotational Raman response functions, respectively. Vibrational SRS is isotropic, while rotational SRS contributes to both the isotropic and anisotropic parts. Under different polarization bases, the nonlinear term in the vector UPPE follows, after expanding with $R^{\text{vib}}$ and $R^{\text{rot}}$:
\begin{subequations}
\begin{alignat}{2}
\hat{\mathcal{N}}A_+ & =i\omega\kappa && \Bigg\{\kappa_e\mathfrak{F}\left[\left(\frac{2}{3}\abs{A_+}^2+\frac{4}{3}\abs{A_-}^2\right)A_+\right] \nonumber \\
&&\hspace{-1em} + & \mathfrak{F}\left[\left[\left(R^{\text{vib}}+R^{\text{rot}}\right)\ast\left(\abs{A_+}^2+\abs{A_-}^2\right)\right]A_++6\left[R^{\text{rot}}\ast\left(A_+A_-^{\ast}\right)\right]A_-\right]\Bigg\} \label{eq:circular_NR} \allowdisplaybreaks \\
\hat{\mathcal{N}}A_x & =i\omega\kappa && \Bigg\{\kappa_e\mathfrak{F}\left[\left(\abs{A_x}^2+\frac{2}{3}\abs{A_y}^2\right)A_x+\frac{1}{3}A_y^2A_x^{\ast}\right] \nonumber \\
&&\hspace{-1em} + & \mathfrak{F}\left[\left[R^{\text{vib}}\ast\left(\abs{A_x}^2+\abs{A_y}^2\right)\right]A_x+\left[R^{\text{rot}}\ast\left(4\abs{A_x}^2-2\abs{A_y}^2\right)\right]A_x+3\left[R^{\text{rot}}\ast\left(A_xA_y^{\ast}+A_x^{\ast}A_y\right)\right]A_y\right]\Bigg\}, \label{eq:linear_NR}
\end{alignat} \label{eq:vector_NR}
\end{subequations}
where $(x,y)$ and $(+,-)$ represent linear and circular polarization bases, respectively. Under scalar scenarios where only one polarization is considered, the Raman nonlinearity is $\left(R_a+\frac{1}{2}R_b=R^{\text{vib}}+R^{\text{rot}}\right)$ under circular polarization but $\left(R_a+R_b=R^{\text{vib}}+4R^{\text{rot}}\right)$ under linear polarization. Similarly, the electronic nonlinearity also varies as $\left(\frac{2}{3}\kappa_e\right)$ under circular polarization but $\kappa_e$ under linear polarization. If the interacting field is not linearly polarized, the rotational SRS induces pronounced cross-polarized nonlinear coupling, because the cross-circularly-polarized rotational Raman gain inherently avoids any wave-vector-matching gain suppression and typically exhibits substantial gain (Fig.~14 in \cite{Chen2024}). Consequently, since the definition of the Raman fraction presupposes scalar-field operations, it is not applicable to media with strong rotational Raman nonlinearities under ``non-linearly-polarized'' excitation. Although it is possible to formulate the nonlinear pulse propagation with $\gamma(1-f_R)$, $\gamma f_Rf_a$, and $\gamma f_Rf_b$ to represent the electronic, isotropic, and anisotropic parts, respectively ($\gamma$: ``total'' nonlinear coefficient and $f_{a/b}$: fractional Raman contribution) \cite{Lin2006}, it might be more straightforward to formulate with those explicitly showing the strengths of each nonlinearity, such as using $\kappa_e$, $R^{\text{rot}}$, and $R^{\text{vib}}$ in Eq.~(\ref{eq:vector_NR}).

\begin{figure}[!ht]
\centering
\includegraphics[width=\linewidth]{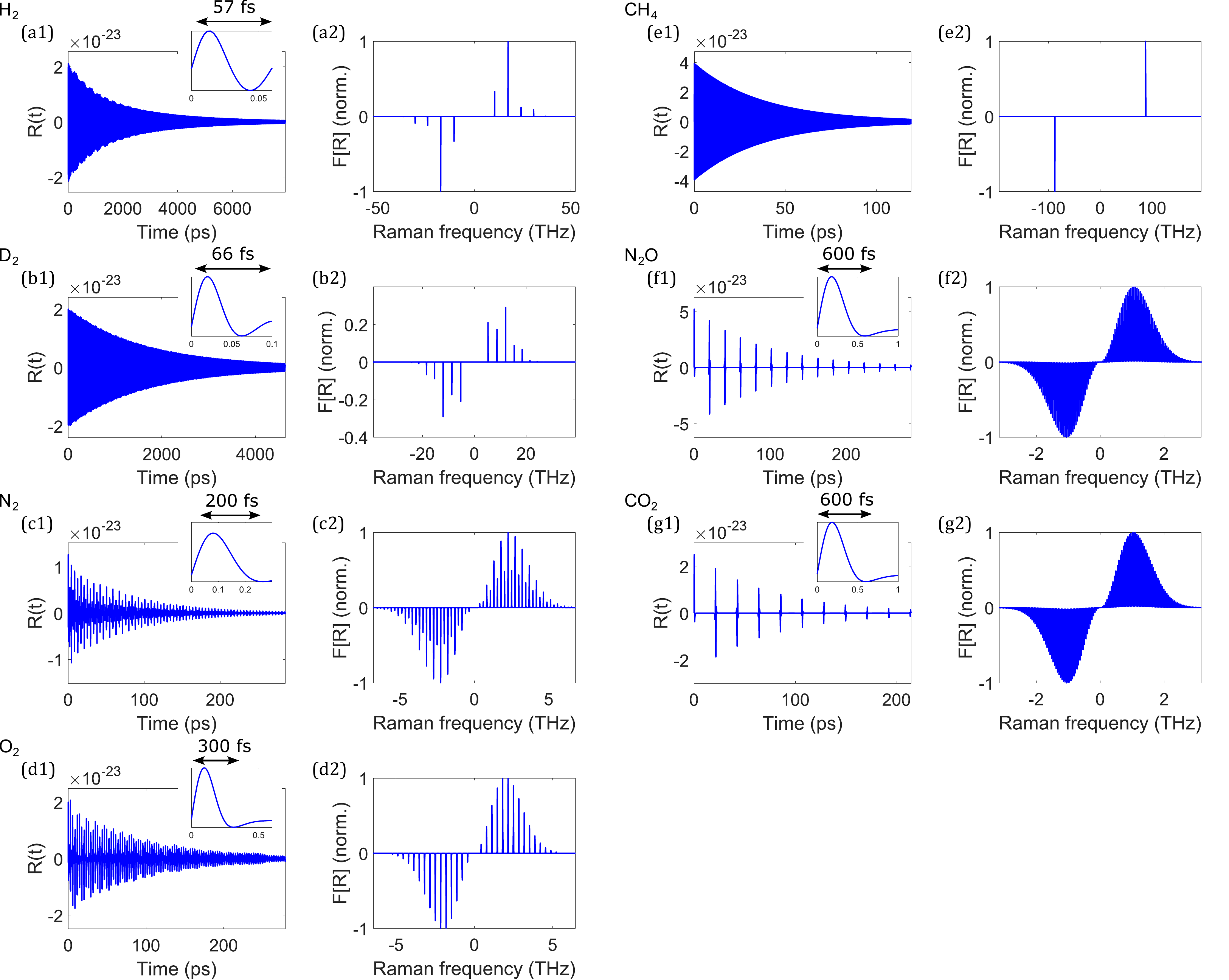}
\caption{Raman response function, involving rotational and vibrational parts, in time (a1--g1) and in frequency (a2--g2) for \ce{H2}, \ce{D2}, \ce{N2}, \ce{O2}, \ce{CH4}, \ce{N2O}, and \ce{CO2}. The gas pressure is \SI{1}{\bar}. Only rotational parts are shown, except for \ce{CH4} that exhibits only the vibrational Raman transition. Insets are the closeview of the first temporal spike of only the ``rotational'' Raman response.}
\label{fig:Raman_response_gas}
\end{figure}

\pagebreak
Below we show the Raman fractions of several materials for the case of linear polarization (Table~\ref{tab:fr}). Solids typically exhibit perturbative Raman scattering (having a small Raman fraction) with minimal Raman-induced distortions, whereas most gases exhibit non-perturbative Raman scattering. In gases with weak dephasing (Fig.~\ref{fig:Raman_response_gas}), $I_R\propto1/\omega_R$ for each Raman transition, so typically vibrational transitions with large transition frequencies do not significantly affect the Raman fraction; rotational Raman transitions with small frequencies dominate the Raman fraction. An exception is \ce{CH4}, which has no rotational Raman transition. Its vibrational transitions dominate the Raman-induced index change and result in a small Raman fraction of \num{0.082}. We do not list liquids here, as they contain more complicated Raman temporal patterns due to intermolecular interactions from collision, libration, and diffusive reorientation \cite{Reichert2014,Zhao2018a}.
\begin{table}[!ht]
\centering
\begin{tabular}{cc|cccc}
\toprule
& \textbf{Material} & \multicolumn{3}{c}{\textbf{Raman fraction}} & Reference \\
&& \textbf{From rotational Raman} & \textbf{From vibrational Raman} & \textbf{Total} \\
\midrule
\textbf{Solid} & Silica & \num{0} & \num{0.18} or \num{0.245} & \num{0.18} or \num{0.245} & \cite{Lin2006} \\
& Chalcogenide & \num{0} & \num{0.115} & \num{0.115} & \cite{Ou2016,Ung2010} \\
& ZBLAN & \num{0} & \num{0.24} & \num{0.24} & \cite{Yan2012} \\
& Tellurite & \num{0} & \num{0.51} & \num{0.51} & \cite{Yan2010} \\
& Silicon & \num{0} & \num{0.043} & \num{0.043} & \cite{Bristow2007,Lin2007} \\
& Lithium niobate & \num{0} & \num{0.58} & \num{0.58} & \cite{Bache2012} \\
\textbf{Gas} & \ce{H2} & \num{0.377} & \num{0.057} & \num{0.433} & \cite{Chen2024}\\
& \ce{D2} & \num{0.410} & \num{0.094} & \num{0.555} & \cite{Hanson1993,Russell1987,Kolos1967,Wahlstrand2015,Raj2018,Demtroeder2010,Minck1963} \\
& \ce{N2} & \num{0.759} & \num{0.017} & \num{0.777} & \cite{Chen2024,Hong2024,Rahn1986,Rahn1987,Lempert1990} \\
& \ce{O2} & \num{0.868} & \num{0.011} & \num{0.880} & \cite{Chen2024,Rahn1986,Rahn1987,Lempert1990} \\
& \ce{CH4} & 0 & \num{0.082} & \num{0.082} & \cite{Minck1963,Chen2017,Ottusch1988} \\
& \ce{N2O} & \num{0.953} & - & \num{0.953} & \cite{Truong2024,Beetar2021} \\
& \ce{CO2} & \num{0.941} & - & \num{0.941} & \cite{Truong2024,Jammu1966} \\
\bottomrule
\end{tabular}
\caption{Raman fractions (computed up to three digits from our model) of different materials under linear polarization. ``-'' means unknown or non-implemented in our model.\\
\textbf{Details about finding Raman fractions:} Since the concept of Raman fraction, along with the total $n_2$, is widely used in solids, we simply need to follow the derived models in prior works as referenced here. For gases, due to long pulse durations used in prior works, their measured $n_2$ involve both the electronic and (transient) Raman contributions, if not specified. Care must be taken to avoid double-counting the Raman contribution. In particular, their values, from our experience, require calibration with several experimental papers on Stokes generation. For \ce{H2} and \ce{N2}, please see Supplementary Sec.~11 in \cite{Chen2024}. Beyond those in \cite{Chen2024}, we have implemented the modified exponential gap (MEG) model for the collisional-narrowing effect of Q-branch vibrational Raman linewidth in \ce{N2} and \ce{O2} \cite{Rahn1986,Rahn1987,Lempert1990}, affecting Raman gain at high pressures. Vibrational SRS of \ce{N2} is calibrated against recent experimental nanosecond Stokes generation in \cite{Hong2024}. Raman parameters of \ce{O2} are calibrated with the same factor as those of \ce{N2} due to their common reference sources. The data of \ce{D2} is taken from the references above. It is currently calibrated only for its rotational SRS against \cite{Li2022b}. Raman parameter of \ce{CH4} is calibrated against \cite{Chen2017}. Rotational Raman parameters of \ce{N2O} are calibrated against \cite{Beetar2021}. Those of \ce{CO2} are taken from the same source as \ce{N2O} and calibrated according to results in \ce{N2O}. As our Raman model \cite{Chen2024} can already accurately predict the Raman phenomena with realistic Raman parameters (\eg polarizability, gas reduced mass, transitions between energy levels), the calibration factor is always close to one, as evidenced by these calibration results. Therefore, the lack of detailed calibrations for some media should not deviate the Raman fraction significantly. For details, please see the ``Raman\_model.m'' MATLAB function in the shared code \cite{github}.}
\label{tab:fr}
\end{table}

\clearpage
\section{Decoupling of rotational and vibrational Raman phenomena}
\label{sec:decoupling}
In this section, we would like to answer a simple question:
\begin{quote}
Can beating between vibrational Raman signals impulsively drive rotational Raman scattering?
\end{quote}
This problem is crucial to the generation of clean vibrational Raman signals. When a vibrational Raman signal is generated and amplified, it beats with the pump, inducing a highly-oscillatory temporal pattern in the total field. As a single pulse in the impulsive Raman regime can impulsively drive the index wave, this temporal profile with spikes at the vibrational Raman frequency might impulsively drive rotational Raman processes. The enhanced rotational Raman generation can significantly impair not only the efficiency but also the temporal fidelity of the vibrational Raman signals.

\begin{figure}[!ht]
\centering
\includegraphics[width=0.6\linewidth]{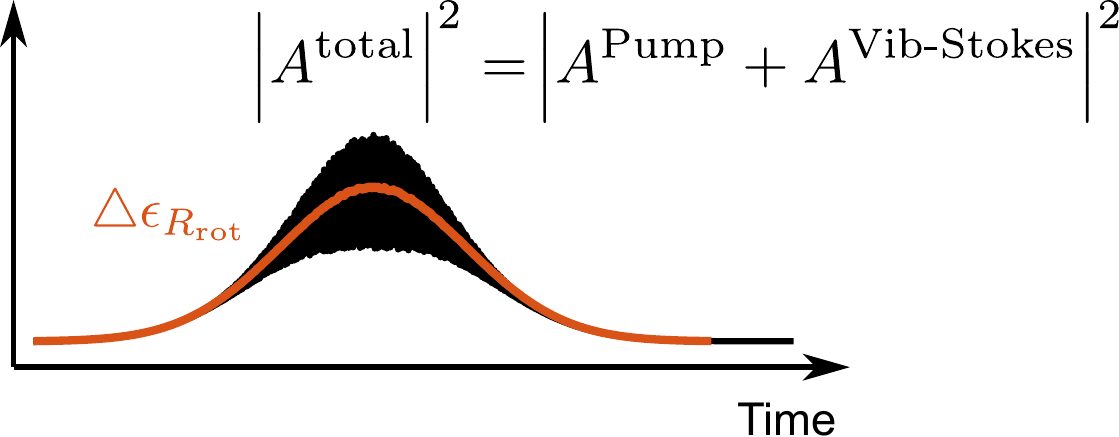}
\caption{Simulated Raman-induced permittivity change from rotational SRS (orange) and the total field (black). The total field is composed of the pump and the high-frequency vibrational Stokes components, which forms a train of temporal spikes at the vibrational Raman transition frequency. Here, vibrational Raman transition frequency is artificially made ten times larger than the rotational frequency.}
\label{fig:R_vib_rot}
\end{figure}

For simplicity, we assume that the field comprises a pump $A$ and a vibrational Stokes component:
\begin{equation}
A^{\text{total}}(t)=A(t)+c_{\text{vib}}\Exp^{-i\left(\omega_{R_{\text{vib}}}t+\phi\right)}A(t),
\end{equation}
where $c_{\text{vib}}\in\mathbb{R}$ and usually $0<c_{\text{vib}}<1$, and $\phi$ is the phase difference between two signals, so that its square magnitude is
\begin{equation}
\abs{A^{\text{total}}}^2=\abs{A}^2\left[1+c_{\text{vib}}^2+2c_{\text{vib}}\cos(\omega_{R_{\text{vib}}}t-\phi)\right].
\end{equation}
We introduce a useful relation, derived in \cite{Chen2024} for the analysis of transient Raman gain:
\begin{subequations}
\begin{alignat}{2}
& \sin(\omega_Rt)\ast f(t)\approx\frac{f(t)}{\omega_R} && ,\text{ as }f(t)\text{ is slowly varying compared to }\omega_R \label{eq:sinf} \\
& \cos(\omega_Rt)\ast f(t)\approx0 && ,\text{ as }f(t)\text{ is slowly varying compared to }\omega_R. \label{eq:cosf}
\end{alignat} \label{eq:sinfcosf}
\end{subequations}
There, only Eq.~(\ref{eq:sinf}) is derived but its cosine version [Eq.~(\ref{eq:cosf})] can be derived following the same process. In the transient Raman regime, the rotational Raman response of the material to this field is
\begingroup\allowdisplaybreaks
\begin{align}
\triangle\epsilon_{R_{\text{rot}}}^{\text{tr}} & \propto h_{R_{\text{rot}}}(t)\ast\abs{A^{\text{total}}}^2 \nonumber \\
& \propto\sin(\omega_{R_{\text{rot}}}t)\ast\abs{A^{\text{total}}}^2\quad,\gamma_2\approx0 \nonumber \\
& \approx\abs{A}^2\left(\frac{1+c_{\text{vib}}^2}{\omega_{R_{\text{rot}}}}\right)+2c_{\text{vib}}\int^t_{-\infty}\sin(\omega_{R_{\text{rot}}}(t-\tau))\abs{A(\tau)}^2\cos(\omega_{R_{\text{vib}}}\tau-\phi)\diff\tau \nonumber \\
& =\abs{A}^2\left(\frac{1+c_{\text{vib}}^2}{\omega_{R_{\text{rot}}}}\right)+2c_{\text{vib}}\int^t_{-\infty}\frac{1}{2}\big[\sin\left(\omega_{R_{\text{rot}}}t-(\omega_{R_{\text{rot}}}-\omega_{R_{\text{vib}}})\tau-\phi\right) \nonumber \\
& \hspace{14em}+\sin\left(\omega_{R_{\text{rot}}}t-(\omega_{R_{\text{rot}}}+\omega_{R_{\text{vib}}})\tau+\phi\right)\big]\abs{A(\tau)}^2\diff\tau\quad,\triangle\omega_R=\omega_{R_{\text{vib}}}-\omega_{R_{\text{rot}}}>0 \nonumber \\
& =\abs{A}^2\left(\frac{1+c_{\text{vib}}^2}{\omega_{R_{\text{rot}}}}\right)+c_{\text{vib}}\int^t_{-\infty}\big[\sin\left(-\triangle\omega_R(t-\tau)-\phi+\omega_{R_{\text{vib}}}t\right) \nonumber \\
& \hspace{13em}+\sin\left((\omega_{R_{\text{rot}}}+\omega_{R_{\text{vib}}})(t-\tau)+\phi-\omega_{R_{\text{vib}}}t\right)\big]\abs{A(\tau)}^2\diff\tau \nonumber \\
& \approx\abs{A}^2\left(\frac{1+c_{\text{vib}}^2}{\omega_{R_{\text{rot}}}}\right) \nonumber \\
& \hspace{1em}+c_{\text{vib}}\bigg[\cos(-\phi+\omega_{R_{\text{vib}}}t)\left(\sin(-\triangle\omega_Rt)\ast\abs{A}^2\right)+\sin(-\phi+\omega_{R_{\text{vib}}}t)\left(\cos(-\triangle\omega_Rt)\ast\abs{A}^2\right) \nonumber \\
& \hspace{5em}\cos(\phi-\omega_{R_{\text{vib}}}t)\left(\sin\left((\omega_{R_{\text{rot}}}+\omega_{R_{\text{vib}}})t\right)\ast\abs{A}^2\right)+\sin(\phi-\omega_{R_{\text{vib}}}t)\left(\cos\left((\omega_{R_{\text{rot}}}+\omega_{R_{\text{vib}}})t\right)\ast\abs{A}^2\right)\bigg], \nonumber \\
& \hspace{35em}\text{by assuming that }\omega_{R_{\text{vib}}}\gg\omega_{R_{\text{rot}}} \nonumber \\
& =\abs{A}^2\left[\frac{1+c_{\text{vib}}^2}{\omega_{R_{\text{rot}}}}+c_{\text{vib}}\left(\frac{\cos(-\phi+\omega_{R_{\text{vib}}}t)}{-\triangle\omega_R}+\frac{\cos(\phi-\omega_{R_{\text{vib}}}t)}{\omega_{R_{\text{rot}}}+\omega_{R_{\text{vib}}}}\right)\right] \nonumber \\
& =\abs{A}^2\left[\frac{1+c_{\text{vib}}^2}{\omega_{R_{\text{rot}}}}+c_{\text{vib}}\frac{2\omega_{R_{\text{rot}}}}{\omega_{R_{\text{rot}}}^2-\omega_{R_{\text{vib}}}^2}\cos(\omega_{R_{\text{vib}}}t-\phi)\right] \nonumber \\
& \approx\abs{A}^2\left[\frac{1+c_{\text{vib}}^2}{\omega_{R_{\text{rot}}}}+c_{\text{vib}}\frac{2\omega_{R_{\text{rot}}}}{\omega_{R_{\text{vib}}}^2}\cos(\omega_{R_{\text{vib}}}t-\phi)\right]\quad, \omega_{R_{\text{vib}}}\gg\omega_{R_{\text{rot}}} \nonumber \\
& \approx\abs{A}^2\frac{1+c_{\text{vib}}^2}{\omega_{R_{\text{rot}}}}\text{ is slowly varying.}
\label{eq:R_vib_rot_tr}
\end{align}
\endgroup
In the steady-state regime, the derivation is simpler because the field can be assumed instantaneous in the convolution. 
\begin{subequations}
\begin{alignat}{2}
& \mathrlap{\int_{-\infty}^t\Exp^{-\gamma_2(t-\tau)}\sin(\omega_R(t-\tau))\diff\tau=\int^{\infty}_0\Exp^{-\gamma_2\tau}\sin(\omega_R\tau)\diff\tau=\frac{\omega_R}{\gamma_2^2+\omega_R^2}} \nonumber \\
& \Rightarrow && \left(\Exp^{-\gamma_2t}\sin(\omega_Rt)\right)\ast f(t)\approx f(t)\frac{\omega_R}{\gamma_2^2+\omega_R^2}\quad,\gamma_2\gg1\text{ and }f(t)\text{ is slowly varying} \\
& \text{Also, } && \left(\Exp^{-\gamma_2t}\cos(\omega_Rt)\right)\ast f(t)\approx f(t)\frac{\gamma_2}{\gamma_2^2+\omega_R^2}\quad,\gamma_2\gg1\text{ and }f(t)\text{ is slowly varying}
\end{alignat}
\end{subequations}
Following the same procedure as in the transient version [Eq.~(\ref{eq:R_vib_rot_tr})],
\begin{align}
\triangle\epsilon_{R_{\text{rot}}}^{\text{ss}} & \propto\Exp^{-\gamma_{2,\text{rot}}t}\sin(\omega_{R_{\text{rot}}}t)\ast\abs{A^{\text{total}}}^2 \nonumber \\
& \approx\abs{A}^2\left(1+c_{\text{vib}}^2\right)\frac{\omega_{R_{\text{rot}}}}{\gamma_{2,\text{rot}}^2+\omega_{R_{\text{rot}}}^2}+2c_{\text{vib}}\int^t_{-\infty}\Exp^{-\gamma_{2,\text{rot}}(t-\tau)}\sin(\omega_{R_{\text{rot}}}(t-\tau))\abs{A(\tau)}^2\cos(\omega_{R_{\text{vib}}}\tau-\phi)\diff\tau \nonumber \\
& \approx\abs{A}^2\Bigg\{\left(1+c_{\text{vib}}^2\right)\frac{\omega_{R_{\text{rot}}}}{\gamma_{2,\text{rot}}^2+\omega_{R_{\text{rot}}}^2}+c_{\text{vib}}\Bigg[\left(\frac{-\triangle\omega_R}{\gamma_{2,\text{rot}}^2+(\triangle\omega_R)^2}+\frac{\omega_{R_{\text{rot}}}+\omega_{R_{\text{vib}}}}{\gamma_{2,\text{rot}}^2+(\omega_{R_{\text{rot}}}+\omega_{R_{\text{vib}}})^2}\right)\cos(\omega_{R_{\text{vib}}}t-\phi) \nonumber \\
&\hspace{17.5em} +\left(\frac{\gamma_{2,\text{rot}}}{\gamma_{2,\text{rot}}^2+(\triangle\omega_R)^2}-\frac{\gamma_{2,\text{rot}}}{\gamma_{2,\text{rot}}^2+(\omega_{R_{\text{rot}}}+\omega_{R_{\text{vib}}})^2}\right)\sin(\omega_{R_{\text{vib}}}t-\phi)\Bigg]\Bigg\} \nonumber \\
& =\abs{A}^2\Bigg\{\left(1+c_{\text{vib}}^2\right)\frac{\omega_{R_{\text{rot}}}}{\gamma_{2,\text{rot}}^2+\omega_{R_{\text{rot}}}^2}+c_{\text{vib}}\Bigg[\frac{2\omega_{R_{\text{rot}}}\left[\gamma_{2,\text{rot}}^2-\triangle\omega_R(\omega_{R_{\text{rot}}}+\omega_{R_{\text{vib}}})\right]}{\left[\gamma_{2,\text{rot}}^2+(\triangle\omega_R)^2\right]\left[\gamma_{2,\text{rot}}^2+(\omega_{R_{\text{rot}}}+\omega_{R_{\text{vib}}})^2\right]}\cos(\omega_{R_{\text{vib}}}t-\phi) \nonumber \\
&\hspace{17.5em} +\frac{4\gamma_{2,\text{rot}}\omega_{R_{\text{rot}}}\omega_{R_{\text{vib}}}}{\left[\gamma_{2,\text{rot}}^2+(\triangle\omega_R)^2\right]\left[\gamma_{2,\text{rot}}^2+(\omega_{R_{\text{rot}}}+\omega_{R_{\text{vib}}})^2\right]}\sin(\omega_{R_{\text{vib}}}t-\phi)\Bigg\} \nonumber \\
& \approx\abs{A}^2\left[\left(1+c_{\text{vib}}^2\right)\frac{\omega_{R_{\text{rot}}}}{\gamma_{2,\text{rot}}^2+\omega_{R_{\text{rot}}}^2}+c_{\text{vib}}\frac{2\omega_{R_{\text{rot}}}}{\omega_{R_{\text{vib}}}^2}\left(\cos(\omega_{R_{\text{vib}}}t-\phi)+\frac{2\gamma_{2,\text{rot}}}{\omega_{R_{\text{vib}}}}\sin(\omega_{R_{\text{vib}}}t-\phi)\right)\right], \nonumber \\
& \hspace{35em}\text{ with } \omega_{R_{\text{vib}}}\gg\omega_{R_{\text{rot}}}\approx\gamma_{2,\text{rot}} \nonumber \\
& \approx\abs{A}^2\left(1+c_{\text{vib}}^2\right)\frac{\omega_{R_{\text{rot}}}}{\gamma_{2,\text{rot}}^2+\omega_{R_{\text{rot}}}^2}\text{ is slowly varying.}
\label{eq:R_vib_rot_ss}
\end{align}
These two results [Eqs.~(\ref{eq:R_vib_rot_tr}) and (\ref{eq:R_vib_rot_ss})] show that a field with a train of temporal spikes at a much higher frequency than the Raman transition frequency does not impulsively drive the Raman scattering, in contrast to a single short temporal spike (Fig.~\ref{fig:R_vib_rot}).

\clearpage
\section{Nonlinear index change}
\label{sec:Nonlinear_index_change}
To understand the Raman perturbation, we must analyze the contributions from electronic and Raman nonlinearities. This can be visualized with the nonlinearly-induced index variation. Therefore, in this section, we will derive their equations, which underlie the analysis in the article.

To derive the equation for the nonlinear index change, we start with a different formulation of the UPPE:
\begin{equation}
\partial_zA(z,\Omega)=i\left[\beta(\omega)-\left(\beta_{(0)}+\beta_{(1)}\Omega\right)\right]A(z,\Omega)+\frac{i\omega}{4N^2(\omega)}P(z,\Omega),
\end{equation}
where the nonlinear polarization
\begin{equation}
\vec{P}(\vec{x},\Omega)=\frac{\vec{F}(\vec{r}_{\perp},\omega)}{N(\omega)}P(z,\Omega),
\end{equation}
$\vec{F}$ is the eigenmode profile and the normalization constant that converts the units of field to \si{\sqrt{\W}} is
\begin{equation}
N^2=\frac{\epsilon_0n_{\text{eff}}(\omega)c}{2}.
\end{equation}
$\Omega=\omega-\omega_0$. Also note that, from Eq.~(\ref{eq:UPPE}),
\begin{equation}
\omega\kappa=\frac{\omega}{4}\frac{1}{N^4}\frac{1}{A_{\text{eff}}}.
\end{equation}
Hence, we can obtain, again from Eq.~(\ref{eq:UPPE}),
\begin{align}
P(z,t) & =\triangle\epsilon(t)A(z,t) \nonumber \\
& =4N^2\kappa\left[\kappa_e\abs{A}^2A+\left(R\ast\abs{A}^2\right)A\right]=\frac{1}{N^2A_{\text{eff}}}\left[\kappa_e\abs{A}^2A+\left(R\ast\abs{A}^2\right)A\right],
\label{eq:P_tmp}
\end{align}
and thus
\begin{equation}
\triangle\epsilon(z,t)=\frac{1}{N^2A_{\text{eff}}}\left[\kappa_e\abs{A}^2+\left(R\ast\abs{A}^2\right)\right].
\end{equation}
An important concept for nonlinearities is that $\triangle\epsilon(t)$ is real-valued; its instantaneous and delayed features are the origin of all $\chi^{(3)}$ nonlinear phenomena. Any nonlinear analysis involving complex-valued susceptibility results from the spectral components. Now we have obtained the electronic and Raman contributions to the nonlinear index:
\begin{subequations}
\begin{align}
\triangle\epsilon_{\text{elec}}(z,t) & =\frac{1}{N^2A_{\text{eff}}}\left(\kappa_e\abs{A}^2\right) \label{eq:delta_n_t0_elec} \\
\triangle\epsilon_R(z,t) & =\frac{1}{N^2A_{\text{eff}}}\left(R\ast\abs{A}^2\right).
\end{align} \label{eq:delta_n_t0}
\end{subequations}

Since the field can contain widely-separated frequency components that also overlap in time, it is useful to rewrite Eq.~(\ref{eq:delta_n_t0}) as a general formula for any broadband case. By utilizing the Fourier transform ($\mathfrak{F}[\cdot]$), the multiplication can be done in the frequency domain such that the frequency-dependent coefficient can be correctly multiplied by the corresponding frequency component in the field part. Due to the convolution of the coefficient with the field $A$ in Eq.~(\ref{eq:P_tmp}): $\frac{1}{N^2A_{\text{eff}}}A$, we need to send it into the absolute squared to multiply with $A$ to produce the correct spectral effect. Finally, we obtain the following:
\begin{subequations}
\begin{align}
\triangle\epsilon_{\text{elec}} & =\abs{\mathfrak{F}^{-1}\left[\sqrt{\frac{1}{\left(N(\omega)\right)^2A_{\text{eff}}(\omega)}\kappa_e(\omega)}A(\Omega)\right]}^2 \\
\triangle\epsilon_R(z,t) & =R\ast\abs{\mathfrak{F}^{-1}\left[\sqrt{\frac{1}{\left(N(\omega)\right)^2A_{\text{eff}}(\omega)}}A(\Omega)\right]}^2. \label{eq:delta_n_t_broadband_Raman}
\end{align} \label{eq:delta_n_t_broadband}
\end{subequations}

\subsubsection*{A different view of Eq.~(\ref{eq:delta_n_t0_elec})}
First we rewrite the equation:
\begin{equation}
\triangle\epsilon_{\text{elec}}(z,t)=\frac{\kappa_e}{N^2}I,\quad I=\frac{\abs{A}^2}{A_{\text{eff}}}.
\end{equation}
With
\begin{align}
\kappa_e & =\frac{3}{4}\epsilon_0\chi^{(3)}_{\text{electronic}}=\frac{3}{4}\frac{4\epsilon_0^2n_{\text{eff}}^2c}{3}n_{2,\text{electronic}}\quad\text{\cite{Rau2008}} \nonumber \\
& =\epsilon_0^2n_{\text{eff}}^2cn_{2,\text{electronic}},
\label{eq:kappa_e_n2}
\end{align}
we obtain
\begin{align}
\triangle\epsilon_{\text{elec}} & =2\epsilon_0n_{\text{eff}}\left(n_{2,\text{electronic}}I\right) \nonumber \\
& =2\epsilon_0n_{\text{eff}}\triangle n_{\text{electronic}}. \label{eq:tmp8}
\end{align}

Because
\begin{align}
& \begin{aligned}
\epsilon_{\text{eff}}+\triangle\epsilon_{\text{electronic}} & =\epsilon_0\left(n_{\text{eff}}+\triangle n_{\text{electronic}}\right)^2 \\
& \approx\epsilon_0\left(n_{\text{eff}}^2+2n_{\text{eff}}\triangle n_{\text{electronic}}\right)
\end{aligned} \nonumber \\
& \Rightarrow\triangle\epsilon_{\text{electronic}}=2\epsilon_0n_{\text{eff}}\triangle n_{\text{electronic}},\quad\because\epsilon_{\text{eff}}=\epsilon_0n_{\text{eff}}^2,
\end{align}
which is consistent with Eq.~(\ref{eq:tmp8}).

\clearpage
\section{SRS does not need to be a perturbation for SSFS in the steady-state regime}
\label{sec:Steady-state_SSFS_doesn't need_perturbative_SRS}
It has been conventionally and widely accepted that steady-state SSFS, established by the seminal studies of Dianov, Mitschke, and Gordon \cite{Dianov1985,Mitschke1986,Gordon1986}, requires SRS to remain a weak, perturbative contribution. This perturbative assumption also underlies the standard derivation of the traditional steady-state SSFS formula \cite{Gordon1986,Santhanam2003}\footnote{Gordon is the first to study SSFS analytically \cite{Gordon1986}. Santhanam and Agrawal are the first to solve the steady-state frequency-shifting formula with the moment method \cite{Santhanam2003}, giving the widely-used formula to date.}, in which the soliton's $\sech^2(\cdot)$ profile is maintained and can be directly inserted into the frequency-shifting expression. In contrast, the present work shows that Raman-induced index modulation does not introduce appreciable temporal distortion in either the long-pulse transient regime or the steady-state regime, owing to the pulse-following nature of the Raman response in these two limits. Consequently, a small Raman fraction is not necessarily required to justify the use of a $\sech^2(\cdot)$ pulse shape. This overturns the widely-accepted notion that steady-state SSFS fundamentally relies on Raman scattering being restricted to the perturbative limit.

\begin{figure}[!ht]
\centering
\includegraphics[width=\linewidth]{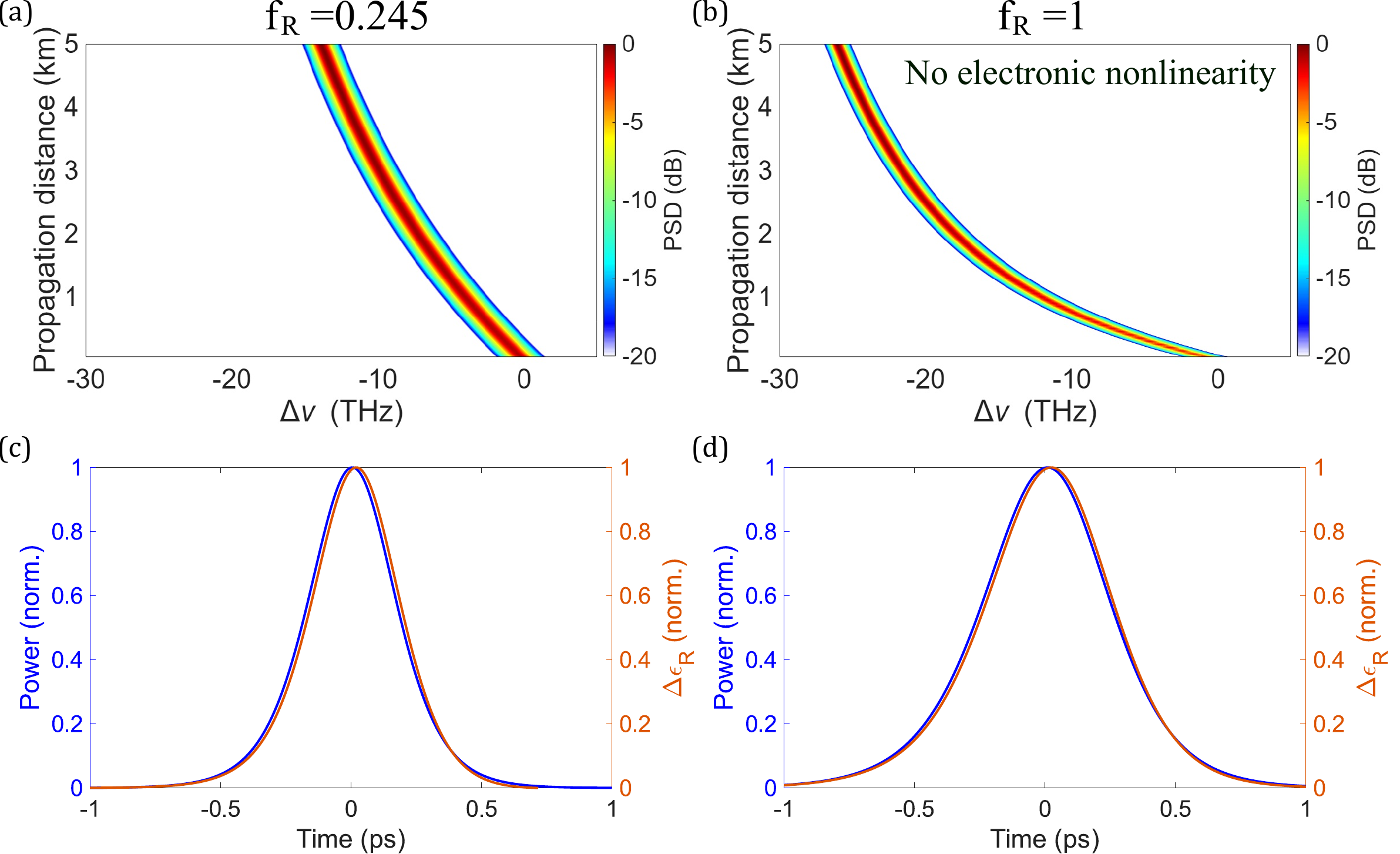}
\caption{Perturbative requirement of steady-state SSFS. Simulated SSFS when (a,c) $f_R=0.245$, the verified value in silica \cite{Lin2006}, and (b,d) $f_R=1$. Top figures are spectral evolutions with propagation and bottom figures are the output temporal profiles of the pulse and the Raman-induced permittivity change [Eq.~(\ref{eq:delta_n_t_broadband_Raman})]. A \SIadj{300}{\fs} fundamental soliton at \SI{1550}{\nm}, \ie with \SIadj{0.08}{\nano\joule} energy, is launched into \SIadj{5}{\km}-long \SIadj{6}{\micro\m}-core fiber. Only the Raman fraction is changed between the two cases. Dispersion $\beta$, (total) nonlinear coefficient $\gamma$, and Raman response function $h_R(t)$ aren't varied and follow silica's parameters \cite{Lin2006}. The smaller pulse duration, or the broader spectrum, under the smaller Raman fraction shows that electronic nonlinearity helps soliton preservation. However, the (steady-state) slightly-delayed Raman-induced index modulation, as seen in (b,d), is still able to maintain a good soliton. The soliton redshifts more under a larger Raman fraction simply due to a larger Raman nonlinearity.}
\label{fig:SS_SSFS}
\end{figure}

As shown in Fig.~\ref{fig:SS_SSFS}, a large Raman fraction can still produce a Raman soliton of high quality without appreciable pedestal formation. Although a smaller Raman fraction generally improves soliton preservation by enhancing the electronically-driven pulse-following index modulation, the Raman-induced index change in the steady-state regime remains sufficiently pulse-tracking to maintain a clean soliton. Thus, even with a relatively large Raman fraction, the steady-state Raman response supports a well-formed soliton without significant temporal distortion (compared to those of pure \ce{N2} in Fig.~4 of the article). Although in practice we cannot change the Raman fraction easily of a material, this comparison removes a long-standing misunderstanding about the perturbative requirement for the steady-state continuous Raman redshift.

Lastly, we emphasize that our findings do not imply that the traditional steady-state SSFS theory is incorrect. Rather, they demonstrate that one of its commonly assumed constraints --- namely, that Raman scattering must remain within the perturbative limit --- can be relaxed. Once this restriction is lifted, steady-state continuous Raman redshifting persists for arbitrary Raman fractions, including regimes far beyond the perturbative assumption embedded in the conventional SSFS framework. In this sense, our results extend the applicability of the traditional theory without contradicting its foundational structure.

\clearpage
\section{Gradient pressure}
\label{sec:gradient}
This section outlines the methodology for calculating the pressure profile along the fiber, information that is essential for accurate numerical simulations. Furthermore, understanding the transient dynamics of pressure-buildup enables precise determination of the time required to establish the target pressure distribution in experimental settings. Lastly, we investigate the effect of gradient pressure to the impulsive SSFS.

\subsection{Gradient pressure profile}
\label{subsec:gradient_pressure_profile}
\begin{figure}[!ht]
\centering
\includegraphics[width=0.4\linewidth]{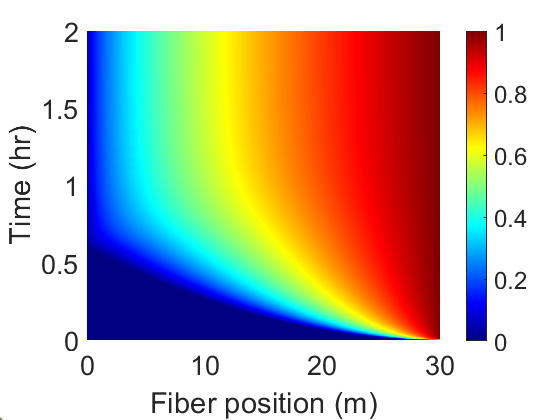}
\caption{Pressure transients from vacuum to a gradient pressure profile established by filling one fiber end with \SIadj{1}{\bar} \ce{N2} while the other end is under vacuum. \ce{N2} has dynamic viscosity of \SI{1.76e-5}{\Pa\cdot\s} \cite{eta}.}
\label{fig:1bar_gradientpressure}
\end{figure}

As we plan to operate around a few bars of pressure, the gas pressure gradient is characterized by the hydrodynamic process, where the evolution follows the Poiseuille flow \cite{Henningsen2008}. By applying the ideal gas law for compressible gases, we obtain the governing equation:
\begin{equation}
\dpd{p}{t}=\frac{d^2}{32\eta}\dpd{}{z}\left(p\dpd{p}{z}\right),
\label{eq:gradient_pde}
\end{equation}
which can be easily solved by MATLAB's PDE solver, ``pdepe()'' \cite{matlab_pdepe}. $p(z,t)$ is gas pressure, $t$ is time, $d$ is fiber core diameter, $\eta$ is dynamic viscosity, and $z$ is the coordinate along the fiber. As an example, we compute the time required to establish a gradient pressure profile, from an initial vacuum condition, by having vacuum and \SI{1}{\bar} \ce{N2} applied at the respective ends of the fiber (Fig.~\ref{fig:1bar_gradientpressure}). The result shows that it takes about \SI{1.1}{\hour} in a \SIadj{30}{\m}-long hollow-core fiber with \SIadj{30}{\micro\m} core diameter.

For simplicity, we also calculate the gradient pressure profile when it reaches the steady state, which is used in numerical modeling of nonlinear pulse propagation:
\begin{equation}
p(z)=\sqrt{p_1^2+\frac{z}{L}\left(p_2^2-p_1^2\right)},
\label{eq:gradient_p}
\end{equation}
where $p_1$ and $p_2$ are the pressures at two fiber ends, and $L$ is the fiber length. This calculation is done by having $\pd{p}{t}=0$ in Eq.~(\ref{eq:gradient_pde}), leading to $\left(p^2=C_1z+C_2\right)$, where $C_1$ and $C_2$ are found with boundary conditions.

\subsection{Effect of gradient pressure in impulsive SSFS}
\label{subsec:Effect_of_gradient_pressure_in_impulsive_SSFS}
\begin{figure}[!ht]
\centering
\includegraphics[width=0.7\linewidth]{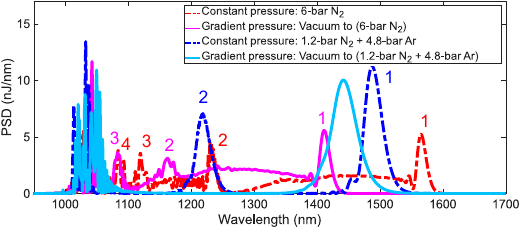}
\caption{Simulated spectra under different pressure conditions in a hollow-core fiber. The total pressure is fixed at \SI{6}{\bar}. The input energy is \SI{0.67}{\micro\joule}, to compare with the light-blue line in Fig.~4 in the article. In cases where multiple solitons are generated, their spectra are labeled with ``1'' for the first soliton, and so on.}
\label{fig:spectrum_gradient_effect}
\end{figure}

In the article, we mention that gradient pressure helps approximate a fundamental soliton during the propagation along the fiber. Here, we further compare it with constant-pressure propagations under pure \ce{N2} and a gas mixture (Fig.~\ref{fig:spectrum_gradient_effect}).

Because a pulse experiences more nonlinearity in constant \SIadj{6}{\bar} pressure than gradient pressure which goes from vacuum to \SI{6}{\bar}, it undergoes more redshifting (blue vs.\ light blue, and red vs.\ pink in Fig.~\ref{fig:spectrum_gradient_effect}). However, with the \SIadj{0.67}{\micro\joule} energy, the pulse, under a constant \SIadj{6}{\bar} pressure, has a soliton number of \num{3.6} and \num{3.3} under pure \ce{N2} and mixed \ce{N2}+\ce{Ar} gases, respectively, and undergoes soliton fission. On the other hand, the number of solitons, \ie ``effective'' soliton number, is minimized under the gradient pressure. At the beginning of gradient pressure, the pulse experiences no nonlinearity in a vacuum environment, \ie zero soliton number. As it propagates, the increasing gas pressure, and the nonlinearity, slowly increases the soliton number while the reducing energy in a redshifting soliton (due to quantum defect) decreases the soliton number. These two effects compete, facilitating and approximating soliton propagation of the minimum (soliton) order. In addition, we can see that Raman solitons from pure \ce{N2}, whether it's a constant or gradient pressure, always exhibit a huge spectral pedestal, reducing the achievable Raman-soliton energy, and thus the soliton generation efficiency.

\clearpage
\section{Supplementary experimental data}
\label{sec:data}
In this section, we provide supplementary data for experiments.

\subsection{Experimental setup}
\label{sbusec:Experimental_setup}
\begin{figure}[!ht]
\centering
\includegraphics[width=\linewidth]{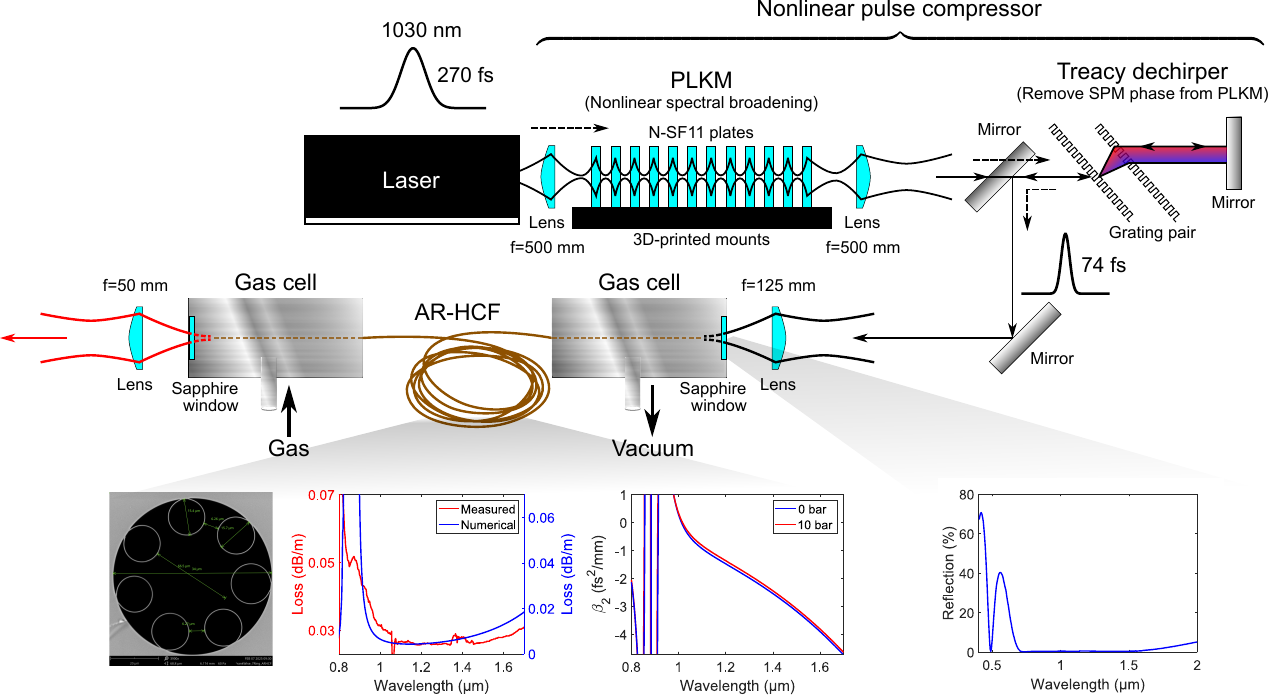}
\caption{Schematic of the experiment. The pulses from a fiber amplifier (Coherent Monaco) were nonlinearly compressed by a periodically-layered Kerr medium (PLKM) before being injected into the anti-resonant hollow-core fiber (AR-HCF). The fiber was filled with a gradient pressure profile. Gas was allowed into the cell at the optical output while the input end of the fiber was held under vacuum. Insets are the SEM image of the cross section, the loss \cite{Vincetti2016,Vincetti2019}, and the calculated dispersion curve \cite{Bache2019} of the AR-HCF, as well as the reflection curve of the AR-coated sapphire window [Edmund Optics \#36-576, NIR \RN{2} (\SIrange[range-phrase=--,range-units=single]{750}{1550}{\nm})].}
\label{fig:experiment_schematic}
\end{figure}

Here we show the schematics and details of the experiments (Fig.~\ref{fig:experiment_schematic}).

The \SIadj{29}{\m}-long anti-resonant hollow-core fiber has a core diameter of \SI{34}{\micro\m} and microstructure tubes with wall thickness around \SI{420}{\nm}. The fiber has a wide transmission band, from \num{850} to \SI{1700}{\nm} with loss less than \SI{0.05}{\dB/\m}. The fiber was coiled with \SIadj{60}{\cm} diameter and the ends were sealed in gas cells.

Gas mixing of \ce{N2} and \ce{Ar} requires extra care before being used for optical experiments. To precisely control the mixing ratio of a mixed gas, \ce{N2} and \ce{Ar} were first sent into a separate gas chamber. Here, gases were mixed with a higher total pressure than needed to reduce operational uncertainties. Most importantly, since different gases have distinct diffusion rates, pre-mixing ensures homogeneity of the gas mixture \cite{Hosseini2017a}. Otherwise, different diffusion rates can cause non-uniform gas concentrations along the fiber. Since the fiber is long and has a small core size, the gas flow rate is small. By having \SIadj{115}{\cm^3} and \SIadj{90}{\cm^3} inner volumes for our optical input and output gas cells, respectively, it takes more than a day for the cell pressures to vary by less than \SI{0.05}{\bar}, which is beneficial for long-term operations.

A fiber amplifier (Coherent Monaco) supplies \SIadj{270}{\fs} pulses at \SIadj{1030}{\nm} wavelength, with pulse energy as high as \SI{60}{\micro\J}. The pulses were compressed to \SI{74}{\fs} by a periodically-layered Kerr medium (PLKM) that consists of twelve N-SF11 glass plates with \SIadj{0.5}{mm} thickness \cite{Zhang2021,Wang2024b}, followed by a Treacy grating dechirper.

Input and output optical powers were measured with a thermal power meter that has a flat spectral response from \num{1} to \SI{2}{\micro\m} (Thorlabs S405C). The spectral and temporal profiles of input and output were measured with an optical spectral analyzer (Agilent 86143B) that covers \num{600} to \SI{1700}{\nm}, and frequency-resolved optical gating (Mesaphotonics FROG) (Figs.~\ref{fig:PLKM_compressed_input}--\ref{fig:FROG_H2}). The output Raman solitons were isolated by longpass filters (Thorlabs FELH series).

\clearpage
\subsection{Spectral measurements}
\label{subsec:Spectral_measurements}
Here, we show the measured spectra plotted on semi-logarithmic axes. This helps visualize the pedestal reduction at higher \ce{Ar} pressure, i.e., lower Raman fraction. The maximum intensity contrast between the Raman soliton peak and pedestal is \SI{30}{\decibel} in \ce{N2}:\ce{Ar} mixtures (black line at $1{:}5$ mixing ratio) and \SI{26}{\decibel} in \ce{H2} (almost all of them, especially those with large frequency shift). With increasing pressure of \ce{Ar} and fixed \ce{N2} pressure, the pedestal is suppressed. Note that the contrast value here is computed from a simple subtraction of the peak and the minimum value between the Raman soliton and the residual pump or the second soliton. We did not intend to claim that the gas mixture suppresses pedestal better than \ce{H2} here, as they have similar Raman fractions (see Table~\ref{tab:fr}).

\begin{figure}[!ht]
\centering
\includegraphics[width=\linewidth]{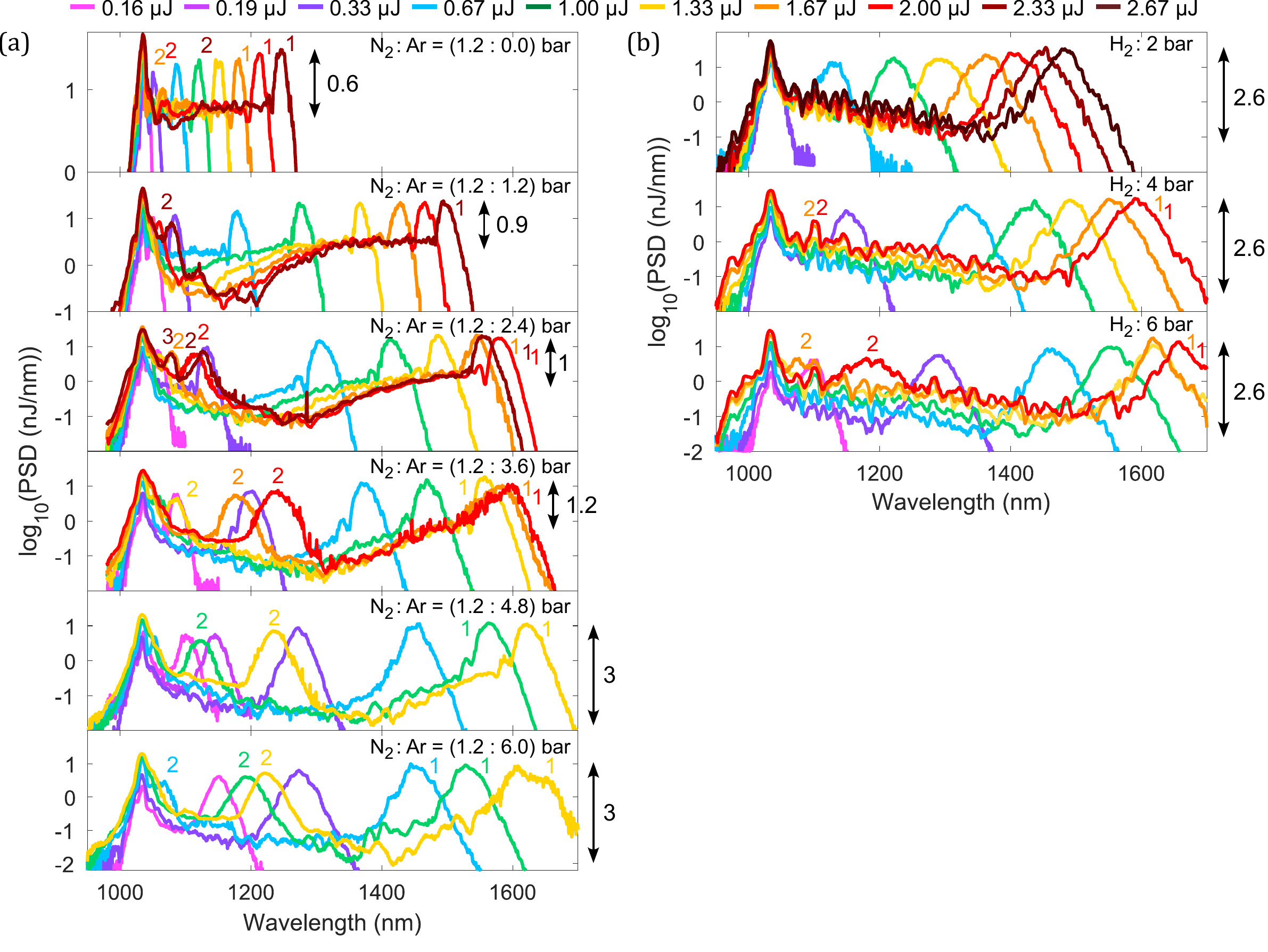}
\caption{Measured spectra on logarithmic y-axis. Spectra resulted from launching \SIadj{74}{\fs} pulses with varying energies. The contrast of the Raman soliton to its pedestal is indicated in each panel. In cases where multiple solitons are generated, their spectra are labeled with ``1'' for the first soliton, and so on.}
\label{fig:Spectrum_supplement}
\end{figure}

\begin{figure}[!ht]
\centering
\includegraphics[width=\linewidth]{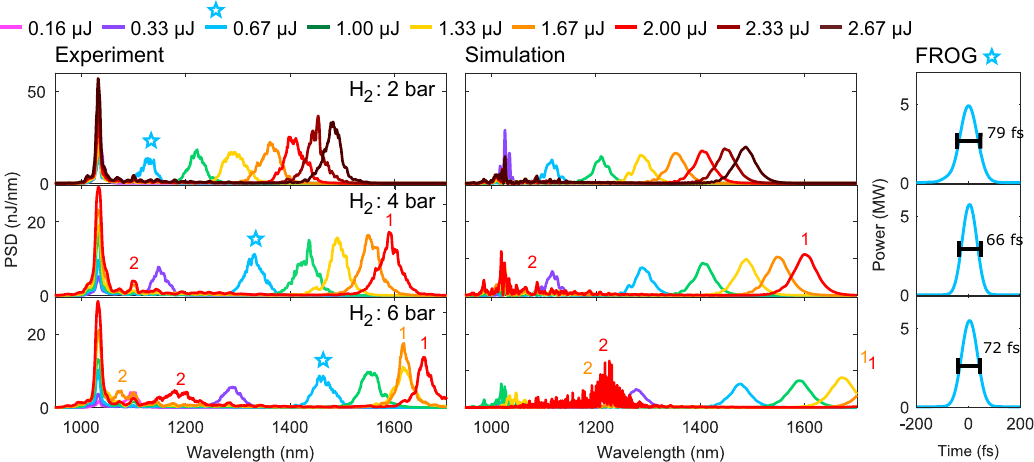}
\caption{Measured [linear scale of Fig.~\ref{fig:Spectrum_supplement}(b)] and simulated spectra, as well as FROG measurements of the filtered pulses at an injected energy of \SI{0.67}{\micro\joule} (blue lines and labeled with stars), are presented with different \ce{H2} pressures. In cases where multiple solitons are generated, their spectra are labeled with ``1'' for the first soliton, and so on. Complete FROG measurements for all solitons are shown in Fig.~\ref{fig:FROG_H2}.}
\label{fig:H2_spectrum}
\end{figure}

\clearpage
\subsection{Raman-soliton measurements}
\label{subsec:Soliton_measurements}
\begin{figure}[!ht]
\centering
\includegraphics[width=0.7\linewidth]{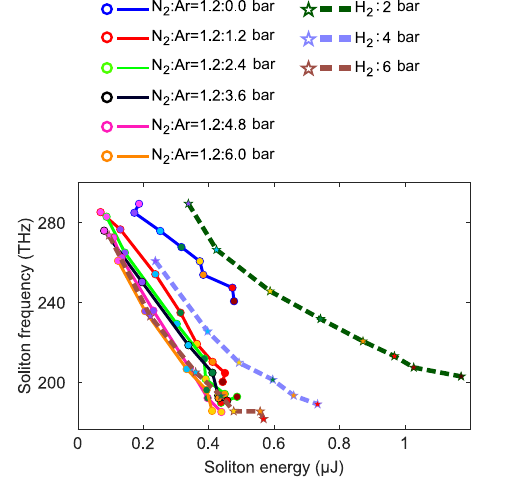}
\caption{Measured Raman soliton frequency with different soliton energies in \ce{N2}:\ce{Ar} mixtures (solid lines; filtered from Fig.~4 in the article) and in pure \ce{H2} (dashed lines; filtered from Fig.~\ref{fig:H2_spectrum}). Marker filled colors follow the energy color sequence in Fig.~\ref{fig:Spectrum_supplement}.}
\label{fig:exp_energy_freq}
\end{figure}

In the article, we show only the quantum efficiency of Raman-soliton generation in the \ce{N2}/\ce{Ar} mixtures. In Fig.~\ref{fig:exp_energy_freq}, we show the measured Raman soliton energies obtained by filtering out the longest-wavelength lobe of a spectrum and their frequencies. In addition, we show the measured quantum efficiency of Raman-soliton generation in \ce{H2} from Fig.~\ref{fig:H2_spectrum}.

From Fig.~\ref{fig:exp_energy_freq}, it may appear that \ce{H2} performs better than the \ce{N2}:\ce{Ar} mixtures due to its capability of generating higher energies (green dashed line). However, this is misleading, as we were doing a controlled experiment with the \ce{N2} fixed in the gas mixture. By reducing the total pressure of the gas mixture with a $1{:}5$ mixing ratio, which has almost the same, or even slightly lower, Raman fraction as \ce{H2}, we should also be able to produce microjoule-level Raman solitons with a large frequency redshift. The experiments show that the high-pressure \ce{H2} performs similarly to high-mixing-ratio (and thus high-pressure under a fixed \ce{N2} pressure) gas mixtures.

\begin{figure}[!ht]
\centering
\includegraphics[width=0.7\linewidth]{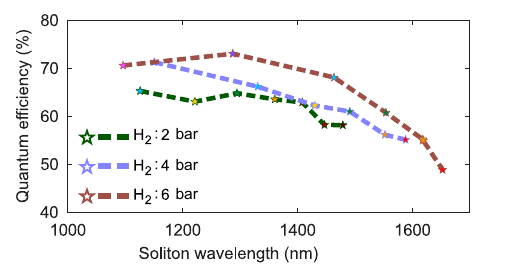}
\caption{Quantum efficiency of the reddest Raman soliton at different \ce{H2} pressures (filtered from Fig.~\ref{fig:H2_spectrum}). Marker filled colors follow the energy color sequence in Fig.~\ref{fig:H2_spectrum}.}
\label{fig:QE_H2}
\end{figure}

\clearpage
\subsection{Comparison of experiments and the theoretical redshifting equation}
\label{subsec:Comparison_shifting_equation}
\begin{figure}[!ht]
\centering
\includegraphics[width=0.7\linewidth]{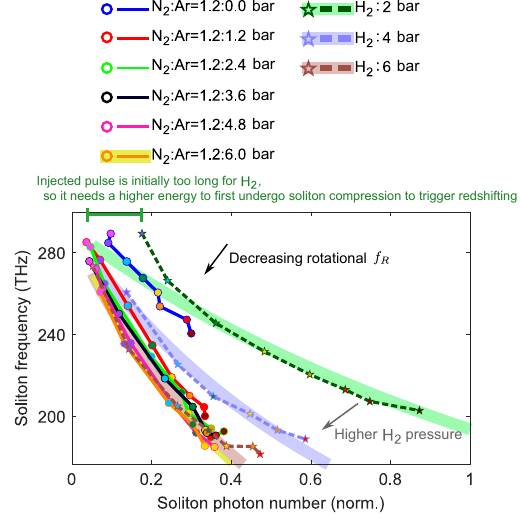}
\caption{Measured (solid and dashed lines) and calculated [colored regions; Eq.~(\ref{eq:impulsive_SSFS})] Raman soliton frequency with different soliton photon numbers in \ce{N2}:\ce{Ar} mixtures (Fig.~4 in the article) and in pure \ce{H2} (Fig.~\ref{fig:H2_spectrum}). We display the results with the photon number, rather than pulse energy, due to significant quantum defect when the redshift is large. Marker filled colors follow the energy color sequence in Fig.~\ref{fig:Spectrum_supplement}.}
\label{fig:SSFS_exp_theory_supplement}
\end{figure}

Analytical calculation of the impulsive redshift in Figs.~6(b) and \ref{fig:SSFS_exp_theory_supplement} is based on the redshifting rate in Eq.~(\ref{eq:impulsive_SSFS}), but the large quantum defects that accompany large shifts and the gradient pressure profile offset the amount of shifting from a simple relation proportional to the input soliton energy. A comparison is possible with calculations based on a fixed photon number in the soliton, leading to the frequency shift that follows Eq.~(\ref{eq:impulsive_SSFS_extended}) after an extended distance $z$ (see derivation in Sec.~\ref{sec:continuous_Raman_redshifting}\ref{subsec:freq_shift_extended}). To compare with our theory, the photon number in a soliton is retrieved only from the filtered reddest soliton at the output, which remains a fundamental-like soliton. Using the entire photon number at the input overestimates the redshift due to pedestals from Raman-induced distortion and soliton fission at high gas pressures. In addition, there are a few other details required for the calculation results before comparing with the measurements. Since the pulse can experience soliton compression during the evolution, the duration used in the equation can be smaller than the injected \SI{74}{\fs}. In particular, the pulse in \ce{H2} must be initially compressed below its \SIadj{57.8}{\fs} Raman period to trigger redshifting. Moreover, our fiber system exhibits loss so the pulse experiences redshifting under an effectively larger energy than the measured output energy. Because \ce{H2} has a small Raman fraction, its data serves as a reference point for initial parameter fitting, where we found approximately \SI{57}{\percent} loss to the soliton energy (similar to that in experiments caused by HCF-HCF splicing in our once-damaged fiber, fiber handling in gradient pressure sealing, etc.) and \SI{48}{\fs} for the effective soliton duration. Subsequently, \SIadj{74}{\fs} duration of our input pulse in experiments, along with the fitted loss, are directly applied to the analytical calculations for gases containing \SIadj{1.2}{\bar} \ce{N2}.

\clearpage
\subsection{FROG measurements}
\label{subsec:FROG}
\begin{figure}[!ht]
\centering
\includegraphics[width=.5\linewidth]{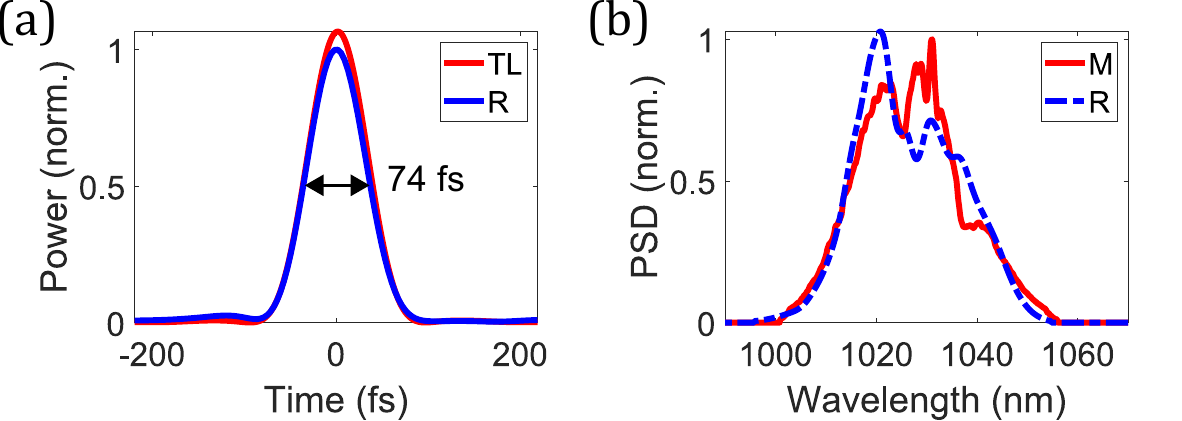}
\caption{FROG measurements of the PLKM-compressed beam. (a) Temporal profiles of the retrieved field (R), as well as its transform-limited (TL) counterpart. (b) Spectrum of the retrieved field (R), compared with an independently-measured spectrum (M) with an optical spectral analyzer (Agilent 86143B). This verifies the accuracy of the FROG measurement and retrieval.}
\label{fig:PLKM_compressed_input}
\end{figure}

\begin{figure}
\centering
\includegraphics[width=\linewidth]{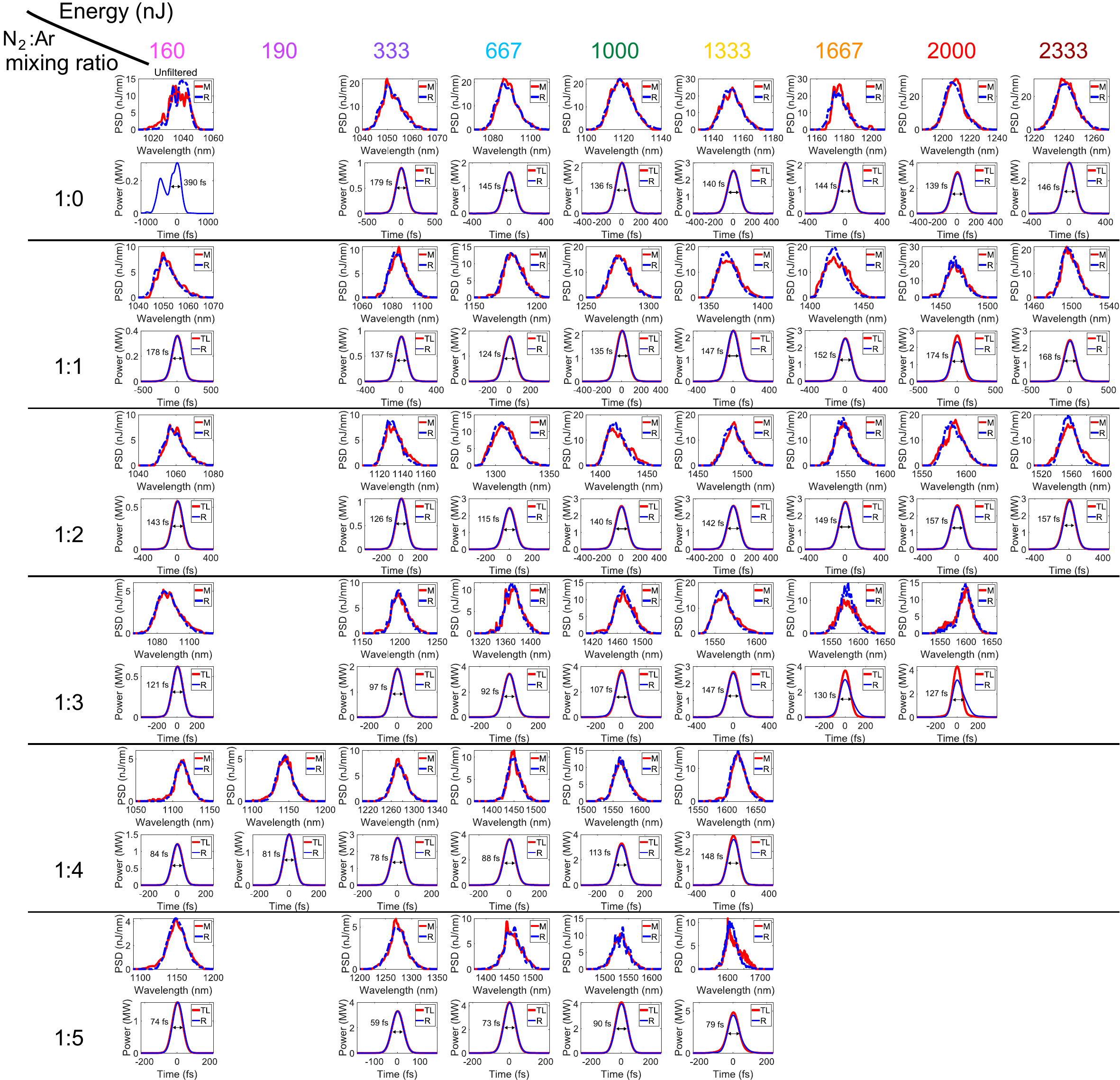}
\caption{FROG measurements of Raman solitons generated from the \ce{N2}:\ce{Ar} mixed gas at various mixing ratios and pulse energies. The \ce{N2} gas pressure is fixed at \SI{1.2}{\bar}. The energy shown here is the effective injected energy into the fiber after considering the coupling and splicing losses, as well as the loss from the superglue-induced distortion to seal the fiber with the gas cell. Bottom row of each mixing ratio are retrieved fields (R), as well as their corresponding transform-limited counterparts (TL). The good agreement in all cases confirms that they are chirp-free solitons. Spectra of the FROG-retrieved fields (R) are compared with independently-measured spectra (M) with an optical spectral analyzer (Agilent 86143B) to verify the accuracy of FROG measurements and retrievals.}
\label{fig:FROG_N2}
\end{figure}

\begin{figure}
\centering
\includegraphics[width=\linewidth]{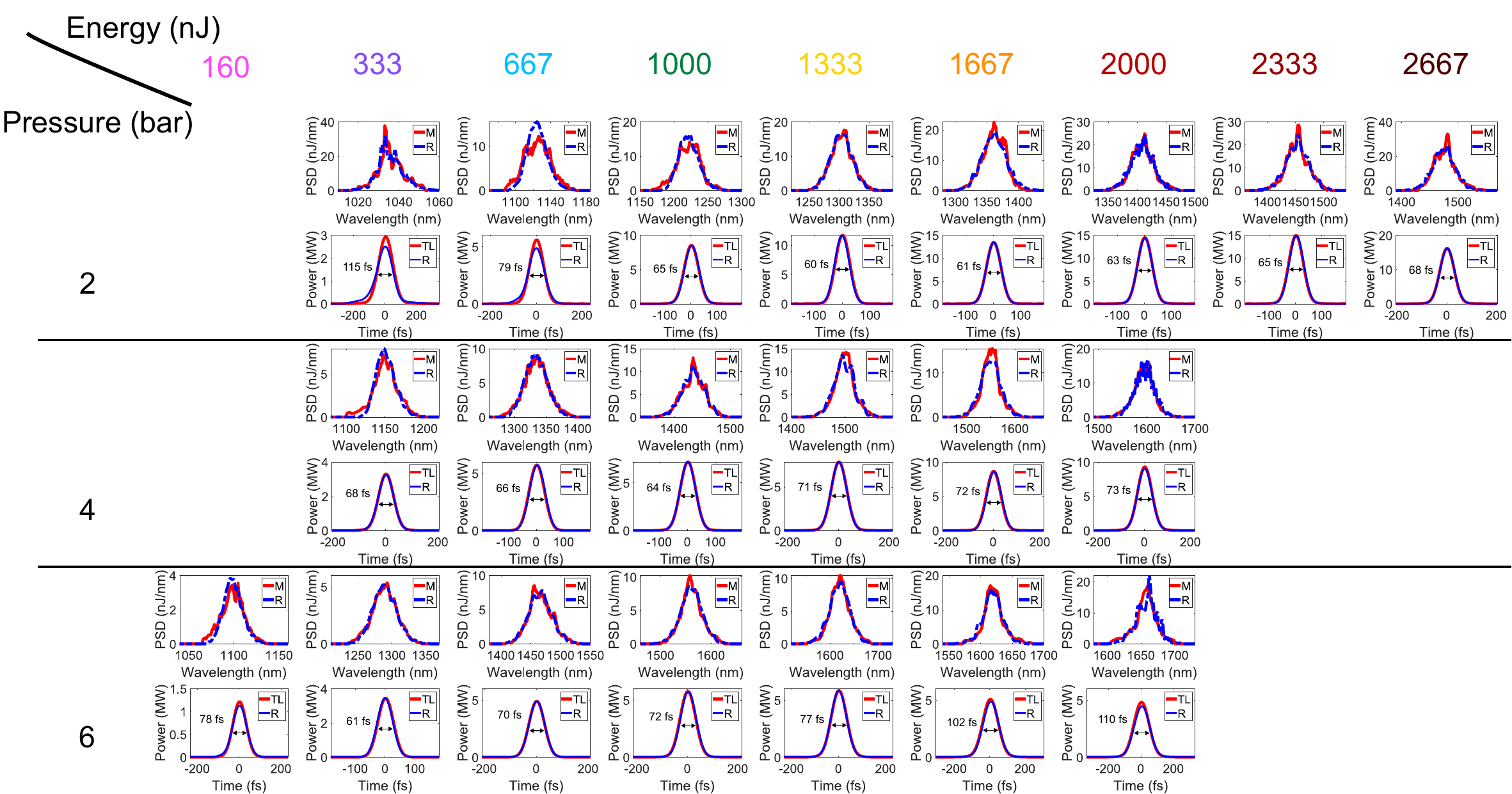}
\caption{FROG measurements of Raman solitons generated from the pure \ce{H2} gas at various pressures and interacting pulse energies. The energy shown here is the effective injected energy into the fiber after considering the coupling and splicing losses, as well as the loss from the superglue-induced distortion to seal the fiber with the gas cell. Bottom row of each pressure are retrieved fields (R), as well as their corresponding transform-limited counterpart (TL). The good agreement in all cases confirms that they are chirp-free solitons. Spectra of the FROG-retrieved fields (R) are compared with independently-measured spectra (M) with an optical spectral analyzer (Agilent 86143B) to verify the accuracy of FROG measurements and retrievals.}
\label{fig:FROG_H2}
\end{figure}

This section shows the experimental FROG measurements for the nonlinearly-compressed pulse by a PLKM for our gas experiments (Fig.~\ref{fig:PLKM_compressed_input}), along with the Raman solitons in mixed \ce{N2}:\ce{Ar} gases (Fig.~\ref{fig:FROG_N2}) and in pure \ce{H2} (Fig.~\ref{fig:FROG_H2}).

\clearpage
\subsection{PER measurements}
\label{subsec:PER}
Here we present measurements of the polarization extinction ratio (PER) of the filtered Raman soliton. The PER generally remains above \SI{10}{\dB} under linear polarization. A reduction is observed as the soliton approaches \SI{1600}{\nm}. Although the exact origin of this degradation is not fully understood, we suspect that it arises either from the soliton propagating close to the edge of the transmission band of our anti-resonant hollow-core fiber or from significant compression and fission of the higher-order soliton, whose elevated peak power may enhance undesirable polarization coupling. Recent work by \etal{Avrahamy} has experimentally shown that the propagating mode inside a hollow-core fiber exhibits polarization dependence in its spatial profile near the bandgap edge, and observed polarization dependence even in \ce{SF6} and \ce{CH4} with only ``isotropic'' vibrational Raman transitions \cite{Avrahamy2026}, which may align with our observations here. In the \ce{N2}/\ce{Ar} gas with high mixing ratios, we see significant drops in PER when there is a second or third soliton. This potentially shows that high peak power during soliton fission in a Raman-active gas can reduce the PER.

\begin{figure}[h!]
\centering
\includegraphics[width=\linewidth]{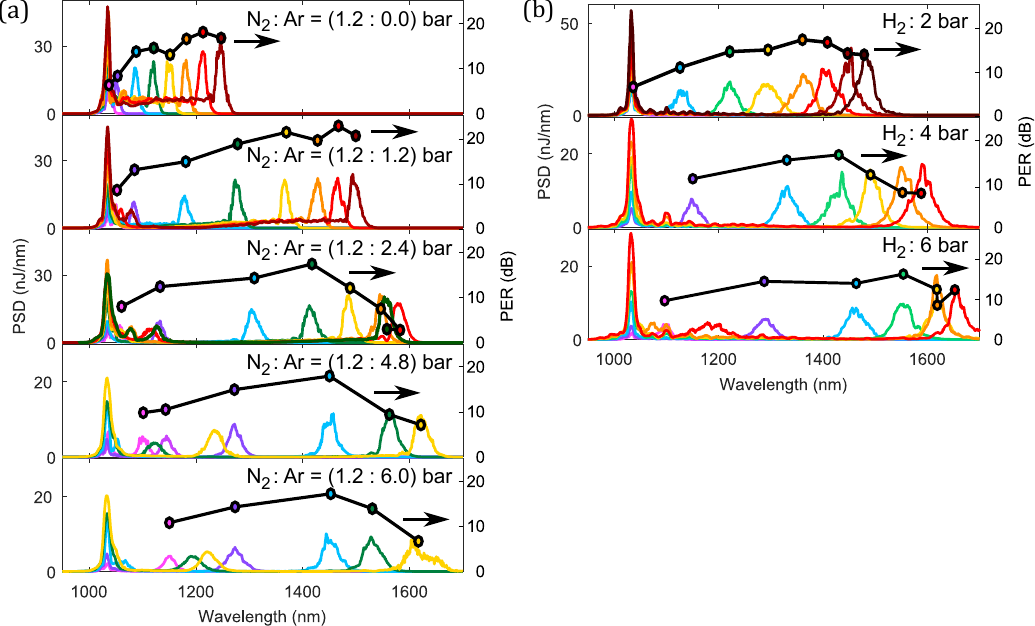}
\caption{PER measurements (black lines corresponding to the right y-axis) of Raman solitons under (a) mixed \ce{N2}/\ce{Ar} gas and (b) \ce{H2}, with spectra being the linear scale of those in Fig.~\ref{fig:Spectrum_supplement}.}
\label{fig:PER}
\end{figure}

\clearpage
\section{Author contributions}
\label{sec:Author_contributions}
Y.-H.C.\@ conceived the project, and performed theoretical calculations, numerical simulations, and experiments. W.W.\@ provided guidance on construction of the pulse compressor. S.H. derived the exact analytical formula of the impulsive SSFS equation beyond the first-order approximated formulation by Y.-H.C. J.E.A.-L.\@ and R.A.-C.\@ fabricated the hollow-core fiber. The paper was written by Y.-H.C.\@ and F.W., and reviewed by all authors. C.X.\@ and F.W.\@ secured the funding. F.W.\@ supervised the project.

\clearpage
\bibliography{reference_supplement}

\newpage
\appendix
\appendixpage

\section{Derivation of the effective Raman time [Eq.~(\ref{eq:T_R_tau0})]}
\label{sec:appendix_Raman_time}
The effective Raman time is a parameter that helps visualize the continuous intrapulse redshifting across Raman temporal regimes. It is derived by \etal{Z.\ Chen} \cite{Chen2010}, but they did not show a clear step-by-step derivation. Here, we derive it again, with an initial goal to ensure its correctness before application to our study.

The effective Raman time is obtained with the generalized nonlinear Schr\"{o}dinger equation (GNLSE), which is the narrowband version of the UPPE [Eq.~\ref{eq:UPPE}]:
\begin{equation}
\partial_zA(z,t)=-i\frac{\beta_2}{2}\partial_t^2A+i\gamma_s\left[\left(1-f_R\right)\abs{A}^2A+f_R\left(h_R\ast\abs{A}^2\right)A\right].
\label{eq:GNLSE}
\end{equation}
For simplicity and to focus on deriving the effective Raman time, we ignore higher-order dispersion and frequency-dependence of the nonlinear coefficient $\gamma_s$. For a complete discussion, please see the original derivation \cite{Chen2010}.

In the moment method, the soliton center frequency is described by
\begin{equation}
\omega_s=c_s\frac{i}{2E_s}\int_{-\infty}^{\infty}\left(A^{\ast}\partial_tA-A\partial_tA^{\ast}\right)\diff t,
\label{eq:omega_s_MM}
\end{equation}
where $c_s=1\text{ or }-1$ determines the Fourier-transform convention \cite{Chen2025arXiv}. We will prove it below:\\
\rule{\textwidth}{1pt}\\
The inverse Fourier transform of $A(\omega)$ follows
\begin{equation}
A(t)=C_{\mathfrak{IF}}\int_{-\infty}^{\infty}A(\omega)\Exp^{-ic_s\omega t}\diff\omega,
\end{equation}
where $C_{\mathfrak{F}}C_{\mathfrak{IF}}=\frac{1}{2\pi}$. So
\begin{align}
\int_{-\infty}^{\infty}A^{\ast}\partial_tA\diff t & =\int_{-\infty}^{\infty}\left(C_{\mathfrak{IF}}\int_{-\infty}^{\infty}\left[A(\omega)\right]^{\ast}\Exp^{ic_s\omega t}\diff\omega\right)\left(C_{\mathfrak{IF}}\int_{-\infty}^{\infty}\left(-ic_s\omega\right)A(\omega)\Exp^{-ic_s\omega t}\diff\omega\right)\diff t \nonumber \\
& =C_{\mathfrak{IF}}^2\int_{-\infty}^{\infty}\int_{-\infty}^{\infty}\left(-ic_s\omega'\right)\left[A(\omega)\right]^{\ast}A(\omega')\left[\int_{-\infty}^{\infty}\Exp^{-ic_s\left(-\omega+\omega'\right)t}\diff t\right]\diff\omega\diff\omega' \nonumber \\
& =C_{\mathfrak{IF}}^2\int_{-\infty}^{\infty}\int_{-\infty}^{\infty}\left(-ic_s\omega'\right)\left[A(\omega)\right]^{\ast}A(\omega')\left[2\pi\delta(\omega-\omega')\right]\diff\omega\diff\omega' \nonumber \\
& =-i2\pi C_{\mathfrak{IF}}^2c_s\int_{-\infty}^{\infty}\omega\abs{A(\omega)}^2\diff\omega
\end{align}
such that
\begin{equation}
\int_{-\infty}^{\infty}\left(A^{\ast}\partial_tA-A\partial_tA^{\ast}\right)\diff t=-i4\pi C_{\mathfrak{IF}}^2c_s\int_{-\infty}^{\infty}\omega\abs{A(\omega)}^2\diff\omega.
\end{equation}
With the generalized formulation of the Parseval's theorem \cite{Chen2025arXiv}:
\begin{equation}
\frac{1}{C_{\mathfrak{IF}}}\int_{-\infty}^{\infty}\abs{A(t)}^2\diff t=\frac{1}{C_{\mathfrak{F}}}\int_{-\infty}^{\infty}\abs{A(\omega)}^2\diff\omega,
\label{eq:Parseval}
\end{equation}
we obtain
\begin{align}
\omega_s & =\frac{\int_{-\infty}^{\infty}\omega\abs{A(\omega)}^2\diff\omega}{\int_{-\infty}^{\infty}\abs{A(\omega)}^2\diff\omega} \nonumber \\
& =\frac{\frac{i}{4\pi C_{\mathfrak{IF}}^2c_s}\int_{-\infty}^{\infty}\left(A^{\ast}\partial_tA-A\partial_tA^{\ast}\right)\diff t}{\frac{C_{\mathfrak{F}}}{C_{\mathfrak{IF}}}E_s}\quad,E_s=\int_{\infty}^{\infty}\abs{A(t)}^2\diff t \nonumber \\
& =c_s\frac{i}{2E_s}\int_{-\infty}^{\infty}\left(A^{\ast}\partial_tA-A\partial_tA^{\ast}\right)\diff t. \label{eq:omega_s_Chenmoment}
\end{align}
\rule{\textwidth}{1pt}

The presence of $c_s$ in Eq.~(\ref{eq:omega_s_MM}) means that the definition is contingent on the choice of the Fourier-transform convention. Here, we follow the physics definition $c_s=1$ so that $A(t)\sim \int A(\omega)\Exp^{i\left(kx-\omega t\right)}$. This is also the convention followed by the GNLSE in Eq.~(\ref{eq:GNLSE}). A different convention will lead to a different formulation of the GNLSE \cite{Chen2025arXiv}.

As derived by \etal{Z.\ Chen}, the energy variation is determined by the first-order term of the nonlinear coefficient. Therefore, the frequency-independence of $\gamma_s$ we assumed here also assumes that the energy remains constant during the propagation. Then,
\begin{align}
\dod{\omega_s}{z} & =\frac{i}{2E_s}\int_{-\infty}^{\infty}\left(\partial_zA^{\ast}\partial_tA-A^{\ast}\partial_z\partial_tA-\partial_zA\partial_tA^{\ast}-A\partial_z\partial_tA^{\ast}\right)\diff t \nonumber \\
& =\frac{i}{2E_s}\int_{-\infty}^{\infty}\left[\partial_t\left(A^{\ast}\partial_zA-A\partial_zA^{\ast}\right)+\left(\partial_zA^{\ast}\partial_tA-\partial_zA\partial_tA^{\ast}-\partial_tA^{\ast}\partial_zA+\partial_tA\partial_zA^{\ast}\right)\right]\diff t \nonumber \\
& =\frac{i}{E_s}\int_{-\infty}^{\infty}\left(\partial_zA^{\ast}\partial_tA-\partial_zA\partial_tA^{\ast}\right)\diff t\quad\because A(t)\rightarrow0\text{, as }t\rightarrow-\infty\text{ and }\infty.
\end{align}

We tackle the terms one by one after plugging in the GNLSE.
\subsubsection*{Dispersion}
\begin{align}
& \int_{-\infty}^{\infty}\left[\left(i\frac{\beta_2}{2}\partial_t^2A^{\ast}\right)\partial_tA-\left(-i\frac{\beta_2}{2}\partial_t^2A\right)\partial_tA^{\ast}\right]\diff t \nonumber \\
&\hspace{1em} =i\frac{\beta_2}{2}\left[\left(\partial_tA\partial_tA^{\ast}\bigg\rvert_{-\infty}^{\infty}-\int_{-\infty}^{\infty}\partial_t^2A\partial_tA^{\ast}\diff t\right)+\int_{-\infty}^{\infty}\partial_t^2A\partial_tA^{\ast}\diff t\right]\quad,\text{with integration by parts} \nonumber \\
&\hspace{1em} =0\quad\because A(t)\rightarrow0\text{, as }t\rightarrow-\infty\text{ and }\infty.
\end{align}
\subsubsection*{Electronic nonlinearity}
\begin{align}
& \int_{-\infty}^{\infty}\left[\left(-i\gamma_s\left(1-f_R\right)\abs{A}^2A\right)\partial_tA-\left(i\gamma_s\left(1-f_R\right)\abs{A}^2A\right)\partial_tA^{\ast}\right]\diff t \nonumber \\
&\hspace{1em} =-i\gamma_s\left(1-f_R\right)\int_{-\infty}^{\infty}\abs{A}^2\left(A\partial_tA+A\partial_tA^{\ast}\right)\diff t \nonumber \\
&\hspace{1em} =-i\gamma_s\left(1-f_R\right)\int_{-\infty}^{\infty}\abs{A}^2\partial_t\left(\abs{A}^2\right)\diff t \nonumber \\
&\hspace{1em} =-i\gamma_s\left(1-f_R\right)\frac{1}{2}\int_{-\infty}^{\infty}\partial_t\left(\abs{A}^4\right)\diff t\quad,\text{with integration by parts} \nonumber \\
&\hspace{1em} =0\quad\because A(t)\rightarrow0\text{, as }t\rightarrow-\infty\text{ and }\infty.
\end{align}
\subsubsection*{Raman nonlinearity}
\begin{align}
& \int_{-\infty}^{\infty}\left\{\left[-i\gamma_s f_R\left(h_R\ast\abs{A}^2\right)A^{\ast}\right]\partial_tA-\left[i\gamma_s f_R\left(h_R\ast\abs{A}^2\right)A\right]\partial_tA^{\ast}\right\}\diff t \nonumber \\
&\hspace{1em} =-i\gamma_s f_R\int_{-\infty}^{\infty}\left(h_R\ast\abs{A}^2\right)\left(A^{\ast}\partial_tA+A\partial_tA^{\ast}\right)\diff t \nonumber \\
&\hspace{1em} =-i\gamma_s f_R\int_{-\infty}^{\infty}\left(h_R\ast\abs{A}^2\right)\partial_t\left(\abs{A}^2\right)\diff t \nonumber \\
&\hspace{1em} =-i\gamma_s f_R\int_{-\infty}^{\infty}h_R(\tau)\int_{-\infty}^{\infty}\abs{A(t-\tau)}^2\partial_t\left(\abs{A(t)}^2\right)\diff t\diff\tau.
\end{align}
By assuming the $\sech^2$ shape of the pulse:
\begin{equation}
\abs{A(t)}^2=P_0\sech^2\left(\frac{t}{\tau_0}\right),
\end{equation}
and letting $x=t/\tau_0$ and $y=\tau/\tau_0$, we can rewrite the inner integral in the previous equation as
\begin{align}
\int_{-\infty}^{\infty}\abs{A(t-\tau)}^2\partial_t\left(\abs{A(t)}^2\right)\diff t & =\int_{-\infty}^{\infty}\left(P_0\sech^2(x-y)\right)\left(-\frac{2P_0}{\tau_0}\sech^2(x)\tanh(x)\right)\left(\tau_0\diff x\right) \nonumber \\
& =-2P_0^2\int_{-\infty}^{\infty}\sech^2(x-y)\sech^2(x)\tanh(x)\diff x.
\end{align}
Let $u=\tanh(x)$, then $\diff u=\sech^2(x)\diff x$. Also, let $v=\tanh(y)$, then
\begin{align}
\sech^2(x-y) & =1-\tanh^2(x-y)=1-\left(\frac{\tanh(x)-\tanh(y)}{1-\tanh(x)\tanh(y)}\right)^2 \nonumber \\
& =\frac{(1-uv)^2-(u-v)^2}{(1-uv)^2}=\frac{1+u^2v^2-u^2-v^2}{(1-uv)^2}=\frac{(1-u^2)(1-v^2)}{(1-uv)^2}.
\end{align}
Hence, we can continue to rewrite the integral as
\begin{align}
\int_{-\infty}^{\infty}\abs{A(t-\tau)}^2\partial_t\left(\abs{A(t)}^2\right)\diff t & =-2P_0^2(1-v^2)\int_{-1}^1\frac{1-u^2}{(1-uv)^2}u\diff u \nonumber \\
& =-2P_0^2(1-v^2)\dod{}{v}\int_{-1}^1\frac{1-u^2}{1-uv}\diff u \nonumber \\
& =-2P_0^2(1-v^2)\dod{}{v}\int_{-1}^1\left(\frac{1}{1-uv}-\frac{u^2}{1-uv}\right)\diff u.
\label{eq:tmp4}
\end{align}
The first integral is simply
\begin{align}
\int_{-1}^1\frac{1}{1-uv}\diff u & =-\frac{1}{v}\ln\frac{1-v}{1+v}=\frac{2}{v}\tanh^{-1}(v),
\end{align}
by using $\tanh^{-1}(x)=\frac{1}{2}\ln\frac{1+x}{1-x}$. The second integral can be solved by assuming $t=1-uv$:
\begin{align}
\int_{-1}^1\frac{u^2}{1-uv}\diff u & =\int_{1+v}^{1-v}\frac{\left(\frac{1-t}{v}\right)^2}{t}\left(\frac{\diff t}{-v}\right) \nonumber \\
& =-\frac{1}{v^3}\int_{1+v}^{1-v}\frac{(1-t)^2}{t}\diff t \nonumber \\
& =-\frac{1}{v^3}\left(\ln\frac{1-v}{1+v}+4v-2v\right) \nonumber \\
& =\frac{2}{v^3}\tanh^{-1}(v)-\frac{2}{v^2}.
\end{align}
Combined, 
\begin{align}
\int_{-1}^1\left(\frac{1}{1-uv}-\frac{u^2}{1-uv}\right)\diff u & =\frac{2v^2-2}{v^3}\tanh^{-1}(v)+\frac{2}{v^2}.
\end{align}
With the derivative relation: $\od{}{x}\tanh^{-1}(x)=\frac{1}{1-x^2}$, we can solve Eq.~(\ref{eq:tmp4}):
\begin{align}
\int_{-\infty}^{\infty}\abs{A(t-\tau)}^2\partial_t\left(\abs{A(t)}^2\right)\diff t & =-2P_0^2(1-v^2)\left(\frac{4v^4-6v^4+6v^2}{v^6}\tanh^{-1}(v)+\frac{2v^2-2}{v^3}\frac{1}{1-v^2}-\frac{4}{v^3}\right) \nonumber \\
& =-2P_0^2(1-v^2)\left[\frac{6-2v^2}{v^4}\tanh^{-1}(v)-\frac{6}{v^3}\right] \nonumber \\
& =-2P_0^2\left[\frac{6\cosh^2(y)-2\sinh^2(y)}{\sinh^4(y)}y-\frac{6\cosh(y)}{\sinh^3(y)}\right]\quad,1-\tanh^2(y)=\sech^2(y) \nonumber \\
& =-2P_0^2\left(\frac{2\cosh(2y)+4}{\sinh^4(y)}y-\frac{3\sinh(2y)}{\sinh^4(y)}\right)\quad,
\begin{cases}
\cosh^2(y)-\sinh^2(y)=1 \\
\sinh(2y)=2\sinh(y)\cosh(y) \\
\cosh(2y)=2\cosh^2(y)-1
\end{cases} \nonumber \\
& =-\frac{E_s^2}{2\tau_0^2}\csch^4\left(\frac{\tau}{\tau_0}\right)\left[\left(2\cosh\left(\frac{2\tau}{\tau_0}\right)+4\right)\frac{\tau}{\tau_0}-3\sinh\left(\frac{2\tau}{\tau_0}\right)\right]\quad,E_s=2P_0\tau_0.
\end{align}
Hence,
\begin{align}
\dod{\omega_s}{z} & =\frac{i}{E_s}\left(-i\gamma_s f_R\int_{-\infty}^{\infty}h_R(\tau)\left\{-\frac{E_s^2}{2\tau_0^2}\csch^4\left(\frac{\tau}{\tau_0}\right)\left[\left(2\cosh\left(\frac{2\tau}{\tau_0}\right)+4\right)\frac{\tau}{\tau_0}-3\sinh\left(\frac{2\tau}{\tau_0}\right)\right]\right\}\right)\diff\tau \nonumber \\
& =-\gamma_s f_R\frac{E_s}{2\tau_0^3}\int_{-\infty}^{\infty}h_R(\tau)\csch^4\left(\frac{\tau}{\tau_0}\right)\left[\left(2\cosh\left(\frac{2\tau}{\tau_0}\right)+4\right)\tau-3\tau_0\sinh\left(\frac{2\tau}{\tau_0}\right)\right]\diff\tau \nonumber \\
& =-\frac{4\gamma_s f_RE_s}{15\tau_0^3}T'_R,
\end{align}
where in the last line, the equation is written to conform with Gordon's formula for the steady-state SSFS and the traditional definition of the Raman time. The effective Raman time $T'_R$ is derived as
\begin{align}
T'_R(\tau_0) & =\int_{-\infty}^{\infty}h_R(\tau)B(\tau,\tau_0)\diff\tau \nonumber \\
 & =\int_0^{\infty}h_R(\tau)B(\tau,\tau_0)\diff\tau\quad\because h_R(t)\text{ is nonzero only when }t>0
\end{align}
with
\begin{equation}
B(\tau,\tau_0)=\frac{15}{8}\csch^4\left(\frac{\tau}{\tau_0}\right)\left[\left(2\cosh\left(\frac{2\tau}{\tau_0}\right)+4\right)\tau-3\tau_0\sinh\left(\frac{2\tau}{\tau_0}\right)\right],
\end{equation}
which is the same as that derived by \etal{Z.\ Chen}.

\clearpage
\section{Derivation of Eq.~(\ref{eq:tmp3})}
\label{sec:appendix_Derivation_tmp3}
Here, we aim to solve
\begin{equation}
\int_0^{\infty}\left[\Exp^{-\gamma_2t}\sin(\omega_Rt)\right]f(t)\diff t,
\end{equation}
where $f(t)$ is smooth and slowly varying relative to the sinusoidal variation determined by $\omega_R$. $f(t)$ also vanishes at infinity: $f(t\rightarrow\infty)=0$. The derivation procedure follows that in Supplementary Sec.~7 in \cite{Chen2024} for analyzing the transient Raman gain.

Due to the slowly-varying assumption of $f(t)$, we can solve the integral with a time spacing of Raman period $\tau_R=2\pi/\omega_R$ by treating $f(t)$ as ``stationary.'' With the Taylor series expansion, we rewrite the integral into summation:
\begin{align}
\int_0^{\infty}\left[\Exp^{-\gamma_2t}\sin(\omega_Rt)\right]f(t)\diff t & =\sum_{n=0}^{\infty}\int^{(n+1)\tau_R}_{n\tau_R}\left[\Exp^{-\gamma_2t}\sin(\omega_Rt)\right]f(t)\diff t \nonumber \\
& =\sum_{n=0}^{\infty}\int^{(n+1)\tau_R}_{n\tau_R}\left[\Exp^{-\gamma_2t}\sin(\omega_Rt)\right]\left[f(n\tau_R)+f'(n\tau_R)\left(t-n\tau_R\right)\right]\diff t.
\label{eq:sinf1tmp}
\end{align}
To solve this summation of integrals, we need to solve the following two integrals:
\begin{subequations}
\begin{align}
& \int^{(n+1)\tau_R}_{n\tau_R}\left[\Exp^{-\gamma_2t}\sin(\omega_Rt)\right]\diff t \label{eq:expsin1} \\
& \int^{(n+1)\tau_R}_{n\tau_R}\left[\Exp^{-\gamma_2t}\sin(\omega_Rt)\right]t\diff t. \label{eq:expsin2}
\end{align}
\end{subequations}

We can solve the first one [Eq.~(\ref{eq:expsin1})] with integration by parts:
\begin{align}
\int^{(n+1)\tau_R}_{n\tau_R}\left[\Exp^{-\gamma_2t}\sin(\omega_Rt)\right]\diff t & =\frac{1}{1+\frac{\gamma_2^2}{\omega_R^2}}\left(-\frac{1}{\omega_R}\Exp^{-\gamma_2t}\cos(\omega_Rt)-\frac{\gamma_2}{\omega_R^2}\Exp^{-\gamma_2t}\sin(\omega_Rt)\right)\Bigg\rvert_{n\tau_R}^{(n+1)\tau_R} \nonumber \\
& =\frac{1}{1+\frac{\gamma_2^2}{\omega_R^2}}\frac{\Exp^{-\gamma_2n\tau_R}-\Exp^{-\gamma_2(n+1)\tau_R}}{\omega_R}=\frac{\omega_R}{\omega_R^2+\gamma_2^2}\left(\Exp^{-\gamma_2n\tau_R}-\Exp^{-\gamma_2(n+1)\tau_R}\right).
\end{align}
Assuming that $\gamma_2\tau_R\ll1$, approximation of the exponential term can be made: $\Exp^{-\gamma_2(n+1)\tau_R}\approx\Exp^{-\gamma_2n\tau_R}(1-\gamma_2\tau_R)$, leading to
\begin{equation}
\int^{(n+1)\tau_R}_{n\tau_R}\left[\Exp^{-\gamma_2t}\sin(\omega_Rt)\right]\diff t=\frac{\omega_R\gamma_2\tau_R}{\omega_R^2+\gamma_2^2}\Exp^{-\gamma_2n\tau_R}\approx\frac{\gamma_2}{\omega_R}\tau_R\Exp^{-\gamma_2n\tau_R}.
\label{eq:appendix_tmp1}
\end{equation}

Similarly we can also solve Eq.~(\ref{eq:expsin2}) with integration by parts:
\begin{align}
\int^{(n+1)\tau_R}_{n\tau_R}\left[\Exp^{-\gamma_2t}\sin(\omega_Rt)\right]t\diff t & =\frac{1}{1+\frac{\gamma_2^2}{\omega_R^2}}\mybig[\left(-\frac{t\Exp^{-\gamma_2t}}{\omega_R}\cos(\omega_Rt)+\frac{1-\gamma_2t}{\omega_R^2}\Exp^{-\gamma_2t}\sin(\omega_Rt)\right)\Bigg\rvert_{n\tau_R}^{(n+1)\tau_R} \nonumber \\
&\hspace{6em} +\frac{2\gamma_2}{\omega_R^2}\int^{(n+1)\tau_R}_{n\tau_R}\left[\Exp^{-\gamma_2t}\sin(\omega_Rt)\right]\diff t\mybig] \nonumber \\
& =\frac{\omega_R^2}{\omega_R^2+\gamma_2^2}\left(\frac{n\tau_R\Exp^{-\gamma_2n\tau_R}-(n+1)\tau_R\Exp^{-\gamma_2(n+1)\tau_R}}{\omega_R}+\frac{2\gamma_2}{\omega_R^2}\frac{\omega_R\gamma_2\tau_R}{\omega_R^2+\gamma_2^2}\Exp^{-\gamma_2n\tau_R}\right) \nonumber \\
& \approx\frac{\omega_R^2}{\omega_R^2+\gamma_2^2}\left(\frac{\left[(n+1)\gamma_2\tau_R-1\right]\tau_R\Exp^{-\gamma_2n\tau_R}}{\omega_R}+\frac{2\gamma_2^2\tau_R}{\omega_R\left(\omega_R^2+\gamma_2^2\right)}\Exp^{-\gamma_2n\tau_R}\right) \nonumber \\
& =\frac{\omega_R\tau_R}{\omega_R^2+\gamma_2^2}\Exp^{-\gamma_2n\tau_R}\left((n+1)\gamma_2\tau_R+\frac{\gamma_2^2-\omega_R^2}{\gamma_2^2+\omega_R^2}\right) \nonumber \\
& \approx\frac{\tau_R}{\omega_R}\Exp^{-\gamma_2n\tau_R}\left[(n+1)\gamma_2\tau_R-1\right]\quad,\omega_R\gg\gamma_2.
\label{eq:appendix_tmp2}
\end{align}

With these two relations [Eqs.~(\ref{eq:appendix_tmp1}) and (\ref{eq:appendix_tmp2})], we can solve each of the integrals in Eq.~(\ref{eq:sinf1tmp}).
\begin{align}
\sum_{n=0}^{\infty}f(n\tau_R)\int^{(n+1)\tau_R}_{n\tau_R}\left[\Exp^{-\gamma_2t}\sin(\omega_Rt)\right]\diff t & =\sum_{n=0}^{\infty}f(n\tau_R)\frac{\gamma_2}{\omega_R}\tau_R\Exp^{-\gamma_2n\tau_R} \nonumber \\
& =\frac{\gamma_2}{\omega_R}\int_0^{\infty}f(t)\Exp^{-\gamma_2t}\diff t.
\end{align}
With weak dephasing $\gamma_2\ll1$, $\int_0^{\infty}f(t)\Exp^{-\gamma_2t}\diff t\approx\int_0^{\infty}f(t)\diff t$ is the area of the $f(t)$ function.
\begin{align}
\sum_{n=0}^{\infty}f'(n\tau_R)n\tau_R\int^{(n+1)\tau_R}_{n\tau_R}\left[\Exp^{-\gamma_2t}\sin(\omega_Rt)\right]\diff t & =\sum_{n=0}^{\infty}f'(n\tau_R)n\tau_R\frac{\gamma_2}{\omega_R}\tau_R\Exp^{-\gamma_2n\tau_R} \nonumber \\
& =\frac{\gamma_2}{\omega_R}\int_0^{\infty}tf'(t)\Exp^{-\gamma_2t}\diff t.
\end{align}
\begin{align}
\sum_{n=0}^{\infty}f'(n\tau_R)\int^{(n+1)\tau_R}_{n\tau_R}\left[\Exp^{-\gamma_2t}\sin(\omega_Rt)\right]t\diff t & =\sum_{n=0}^{\infty}f'(n\tau_R)\frac{\tau_R}{\omega_R}\Exp^{-\gamma_2n\tau_R}\left[(n+1)\gamma_2\tau_R-1\right]\quad,\omega_R\gg\gamma_2 \nonumber \\
& =\frac{1}{\omega_R}\left(\gamma_2\int_0^{\infty}tf'(t)\Exp^{-\gamma_2t}\diff t+\left(\gamma_2\tau_R-1\right)\int_0^{\infty}f'(t)\Exp^{-\gamma_2t}\diff t\right).
\end{align}

By realizing that
\begin{align}
\int_0^{\infty}f'(t)\Exp^{-\gamma_2t}\diff t & =f(t)\Exp^{-\gamma_2t}\Big\rvert_0^{\infty}+\gamma_2\int_0^{\infty}f(t)\Exp^{-\gamma_2t}\diff t \nonumber \\
& =-f(0)+\gamma_2\int_0^{\infty}f(t)\Exp^{-\gamma_2t}\diff t,
\end{align}
we eventually derive that
\begin{align}
\int_0^{\infty}\left[\Exp^{-\gamma_2t}\sin(\omega_Rt)\right]f(t)\diff t & =\frac{\gamma_2}{\omega_R}\int_0^{\infty}f(t)\Exp^{-\gamma_2t}\diff t+\frac{\gamma_2\tau_R-1}{\omega_R}\int_0^{\infty}f'(t)\Exp^{-\gamma_2t}\diff t \nonumber \\
& =\frac{\gamma_2^2\tau_R}{\omega_R}\int_0^{\infty}f(t)\Exp^{-\gamma_2t}\diff t+\frac{1-\gamma_2\tau_R}{\omega_R}f(0).
\end{align}

\clearpage
\section{Value of the \texorpdfstring{$B_d(x)$}{Bd} integral}
\label{sec:appendix_B_d}
Here, we solve the $B_d$ integral in Eqs.~(\ref{eq:SSFS_transient}) and (\ref{eq:SSFS_transient_summary}) under $\gamma_2\tau_0\ll1$:
\begin{equation}
\int_0^{\infty}B_d(x)\diff x.
\end{equation}

We follow the procedure in Appendix~\ref{sec:appendix_Derivation_Eq_I_Omega_R}:
\begin{align}
\int_0^{\infty}B_d(x)\diff x & =\int_0^{\infty}5\sum_{m=2}^{\infty}(m+1)m(m-1)\left[4x\Exp^{-2mx}+\left(x-\frac{3}{2}\right)\Exp^{-2(m-1)x}+\left(x+\frac{3}{2}\right)\Exp^{-2(m+1)x}\right]\diff x \nonumber \\
& =5\sum_{m=2}^{\infty}(m+1)m(m-1)\left[4\frac{1}{(2m)^2}+\frac{1}{\left[2(m-1)\right]^2}-\frac{3}{2}\frac{1}{2(m-1)}+\frac{1}{\left[2(m+1)\right]^2}+\frac{3}{2}\frac{1}{2(m+1)}\right] \nonumber \\
& =5\sum_{m=2}^{\infty}(m+1)m(m-1)\left[\frac{1}{m^2}+\frac{4-3m}{4(m-1)^2}+\frac{4+3m}{4(m+1)^2}\right] \nonumber \\
& =5\sum_{m=2}^{\infty}\Bigg[m-\frac{1}{m}+\left(\frac{-3m_{-1}^2}{4}-2m_{-1}-\frac{3}{4}+\frac{1}{2m_{-1}}\right)\big\rvert_{m_{-1}=m-1} \nonumber \\
&\hspace{25em} +\left(\frac{3m_{+1}^2}{4}-2m_{+1}+\frac{3}{4}+\frac{1}{2m_{-1}}\right)\big\rvert_{m_{+1}=m+1}\Bigg] \nonumber \\
& =5\sum_{m=2}^{\infty}\left[m-\frac{1}{m}+\left(\frac{-3}{4}m^2-\frac{1}{2}m+\frac{1}{2}+\frac{1}{2m_{-1}}\right)+\left(\frac{3}{4}m^2-\frac{1}{2}m-\frac{1}{2}+\frac{1}{2m_{+1}}\right)\right] \nonumber \\
& =5\sum_{m=2}^{\infty}\left[-\frac{1}{m}+\frac{1}{2m_{-1}}+\frac{1}{2m_{+1}}\right] \nonumber \\
& =5\left[-\left(\zeta(1)-1\right)+\frac{1}{2}\zeta(1)+\frac{1}{2}\left(\zeta(1)-1-\frac{1}{2}\right)\right] \nonumber \\
& =\frac{5}{4}.
\end{align}

\clearpage
\section{Derivation of \texorpdfstring{$I(\Omega_R)$}{I(ΩR)} in the small \texorpdfstring{$\Omega_R$}{ΩR} limit [Eq.~(\ref{eq:I_OmegaR_smallx})]}
\label{sec:appendix_Derivation_Eq_I_Omega_R}
First, we approximate the integral at the small $\Omega_R$ limit with
\begin{equation}
\sin(\Omega_Rx)\approx\Omega_Rx-\frac{\Omega_R^3x^3}{6},
\end{equation}
and obtain
\begin{align}
I(\Omega_R) & =\int_0^{\infty}B_d(x)\sin(\Omega_Rx)\diff x \nonumber \\
& \approx\Omega_R\int_0^{\infty}xB_d(x)\diff x-\frac{\Omega_R^3}{6}\int_0^{\infty}x^3B_d(x)\diff x.
\label{eq:I_Omega_R_tmp}
\end{align}
Therefore, we have two integrals to solve:
\begin{subequations}
\begin{align}
& \int_0^{\infty}xB_d(x)\diff x \\
& \int_0^{\infty}x^3B_d(x)\diff x.
\end{align} \label{eq:B_d_to_solve}
\end{subequations}

To solve them, we need to rewrite $B_d(x)$ as follows:
\begin{align}
B_d(x) & =\frac{15}{8}\csch^4(x)\left[4x+2x\cosh(2x)-3\sinh(2x)\right] \nonumber \\
& =\frac{15}{8}\csch^4(x)\left[4x+x\left(\Exp^{2x}+\Exp^{-2x}\right)-\frac{3}{2}\left(\Exp^{2x}-\Exp^{-2x}\right)\right] \nonumber \\
& =\frac{15}{8}\csch^4(x)\left[4x+\left(x-\frac{3}{2}\right)\Exp^{2x}+\left(x+\frac{3}{2}\right)\Exp^{-2x}\right].
\end{align}
Because
\begin{equation}
\csch^4(x)=\frac{8}{3}\sum_{m=2}^{\infty}(m+1)m(m-1)\Exp^{-2mx},
\end{equation}
we can rewrite $B_d$ in summations:
\begin{equation}
B_d(x)=5\sum_{m=2}^{\infty}(m+1)m(m-1)\left[4x\Exp^{-2mx}+\left(x-\frac{3}{2}\right)\Exp^{-2(m-1)x}+\left(x+\frac{3}{2}\right)\Exp^{-2(m+1)x}\right].
\end{equation}
Each integral in Eq.~(\ref{eq:B_d_to_solve}) can be solved by using
\begin{equation}
\int_0^{\infty}x^n\Exp^{-ax}\diff x=\frac{\Gamma(n+1)}{a^{n+1}}\xlongequal{n\in\mathbb{N}}\frac{n!}{a^{n+1}},
\end{equation}
where $\Gamma$ is the famous Gamma function. Next, we will solve them one by one.

We first tackle $\int_0^{\infty}xB_d(x)\diff x$.
\begin{align}
\int_0^{\infty}xB_d(x)\diff x & =\int_0^{\infty}5\sum_{m=2}^{\infty}(m+1)m(m-1)\left[4x^2\Exp^{-2mx}+\left(x^2-\frac{3}{2}x\right)\Exp^{-2(m-1)x}+\left(x^2+\frac{3}{2}x\right)\Exp^{-2(m+1)x}\right]\diff x \nonumber \\
& =5\sum_{m=2}^{\infty}(m+1)m(m-1)\left[4\frac{2}{(2m)^3}+\frac{2}{\left[2(m-1)\right]^3}-\frac{3}{2}\frac{1}{\left[2(m-1)\right]^2}+\frac{2}{\left[2(m+1)\right]^3}+\frac{3}{2}\frac{1}{\left[2(m+1)\right]^2}\right] \nonumber \\
& =5\sum_{m=2}^{\infty}(m+1)m(m-1)\left[\frac{1}{m^3}+\frac{1}{4(m-1)^3}-\frac{3}{8(m-1)^2}+\frac{1}{4(m+1)^3}+\frac{3}{8(m+1)^2}\right] \nonumber \\
& =5\sum_{m=2}^{\infty}(m+1)m(m-1)\left[\frac{1}{m^3}+\frac{2m^3+6m}{4(m-1)^3(m+1)^3}+\frac{-12m}{8(m-1)^2(m+1)^2}\right] \nonumber \\
& =5\sum_{m=2}^{\infty}(m+1)m(m-1)\left[\frac{1}{m^3}+\frac{-8m^3+24m}{8(m-1)^3(m+1)^3}\right] \nonumber \\
& =5\sum_{m=2}^{\infty}\frac{24m^2-8}{8m^2(m-1)^2(m+1)^2} \nonumber \\
& =5\sum_{m=2}^{\infty}\frac{3m^2-1}{m^2(m-1)^2(m+1)^2}\nonumber \\
& =5\sum_{m=2}^{\infty}\left[\frac{1}{2(m+1)^2}+\frac{1}{2(m-1)^2}-\frac{1}{m^2}\right].
\end{align}
By using the Riemann zeta function
\begin{equation}
\zeta(s)=\sum_{n=1}^{\infty}\frac{1}{n^s},
\end{equation}
where $\zeta(2)=\frac{\pi^2}{6}$,
it can be solved:
\begin{equation}
\int_0^{\infty}xB_d(x)\diff x=5\left[\frac{1}{2}\left(\zeta(2)-1-\frac{1}{4}\right)+\frac{1}{2}\zeta(2)-\left(\zeta(2)-1\right)\right]=\frac{15}{8}.
\label{eq:int_xB_d}
\end{equation}

Next we solve $\int_0^{\infty}x^3B_d(x)\diff x$.
\begin{align}
\int_0^{\infty}x^3B_d(x)\diff x & =\int_0^{\infty}5\sum_{m=2}^{\infty}(m+1)m(m-1)\left[4x^4\Exp^{-2mx}+\left(x^4-\frac{3}{2}x^3\right)\Exp^{-2(m-1)x}+\left(x^4+\frac{3}{2}x^3\right)\Exp^{-2(m+1)x}\right]\diff x \nonumber \\
& =5\sum_{m=2}^{\infty}(m+1)m(m-1)\left[4\frac{24}{(2m)^5}+\frac{24}{\left[2(m-1)\right]^5}-\frac{3}{2}\frac{6}{\left[2(m-1)\right]^4}+\frac{24}{\left[2(m+1)\right]^5}+\frac{3}{2}\frac{6}{\left[2(m+1)\right]^4}\right] \nonumber \\
& =5\sum_{m=2}^{\infty}(m+1)m(m-1)\left[\frac{3}{m^5}+\frac{3}{4(m-1)^5}-\frac{9}{16(m-1)^4}+\frac{3}{4(m+1)^5}+\frac{9}{16(m+1)^4}\right] \nonumber \\
& =5\sum_{m=2}^{\infty}(m+1)m(m-1)\left[\frac{3}{m^5}+\frac{3(7-3m)}{16(m-1)^5}+\frac{3(7+3m)}{16(m+1)^5}\right] \nonumber \\
& =5\sum_{m=2}^{\infty}\Bigg[\frac{3m^2-3}{m^4}+\frac{3(4-3m_{-1})(m_{-1}+2)(m_{-1}+1)}{16m_{-1}^4}\big\rvert_{m_{-1}=m-1} \nonumber \\
&\hspace{8.4em} +\frac{3(4+3m_{+1})(m_{+1}-2)(m_{+1}-1)}{16m_{+1}^4}\big\rvert_{m_{+1}=m+1}\Bigg] \nonumber \\
& =5\sum_{m=2}^{\infty}\left[3\left(\frac{1}{m^2}-\frac{1}{m^4}\right)+\frac{3}{16}\left(\frac{-3}{m_{-1}}+\frac{-5}{m_{-1}^2}+\frac{6}{m_{-1}^3}+\frac{8}{m_{-1}^4}\right)+\frac{3}{16}\left(\frac{3}{m_{+1}}+\frac{-5}{m_{+1}^2}+\frac{-6}{m_{+1}^3}+\frac{8}{m_{+1}^4}\right)\right] \nonumber \\
& =5\Bigg[3\left(\zeta(2)-\zeta(4)\right)+\frac{3}{16}\left(-3\zeta(1)-5\zeta(2)+6\zeta(3)+8\zeta(4)\right) \nonumber \\
&\hspace{3em} +\frac{3}{16}\left(3\left(\zeta(1)-1-\frac{1}{2}\right)-5\left(\zeta(2)-1-\frac{1}{4}\right)-6\left(\zeta(3)-1-\frac{1}{8}\right)+8\left(\zeta(4)-1-\frac{1}{16}\right)\right)\Bigg] \nonumber \\
& =5\frac{9}{8}\zeta(2)=\frac{45}{8}\zeta(2) \nonumber \\
& =\frac{15}{16}\pi^2.
\label{eq:int_x3B_d}
\end{align}

By plugging Eqs.~(\ref{eq:int_xB_d}) and (\ref{eq:int_x3B_d}) into Eq.~(\ref{eq:I_Omega_R_tmp}), we obtain Eq.~(\ref{eq:I_OmegaR_smallx}):
\begin{equation}
I(\Omega_R)\approx\frac{15}{8}\left(\Omega_R-\frac{\pi^2}{12}\Omega_R^3\right).
\end{equation}

\clearpage
\section{Derivation of Eq.~(\ref{eq:tmp7})}
\label{sec:appendix_Derivation_tmp7}
Here, we derive Eq.~(\ref{eq:tmp7}).

Let $s=t-\frac{x}{2}$, Eq.~(\ref{eq:tmp7}) becomes
\begin{align}
A(x)=\int_{-\infty}^{\infty}\sech^2(s+\frac{x}{2})\sech^2(s-\frac{x}{2})\diff s.
\end{align}

With the relation
\begin{equation}
\cosh(A)\cosh(B)=\frac{1}{2}\left[\cosh(A+B)+\cosh(A-B)\right],
\end{equation}
and having $A=s+\frac{x}{2}$ and $B=s-\frac{x}{2}$, we get
\begin{equation}
\cosh(s+\frac{x}{2})\cosh(s-\frac{x}{2})=\frac{1}{2}\left[\cosh(2s)+\cosh(x)\right],
\end{equation}
leading to
\begin{equation}
\sech^2(s+\frac{x}{2})\sech^2(s-\frac{x}{2})=\frac{4}{\left[\cosh(2s)+\cosh(x)\right]^2}.
\end{equation}
Therefore,
\begin{align}
A(x) & =\int_{-\infty}^{\infty}\frac{4}{\left[\cosh(2s)+\cosh(x)\right]^2}\diff s \nonumber \\
& =\int_{-\infty}^{\infty}\frac{2}{\left[\cosh(u)+\cosh(x)\right]^2}\diff u,\quad\text{let }u=2s
\end{align}

To solve this, let us first assume
\begin{equation}
F(c)=\int_{-\infty}^{\infty}\frac{\diff u}{\cosh(u)+c},
\end{equation}
where $c=\cosh(x)$. Substitute $w=\Exp^u$, where $\diff u=\frac{\diff w}{w}$ and $\cosh(u)=\frac{w+w^{-1}}{2}$, we have
\begin{align}
F(c) & =\int_0^{\infty}\frac{2\diff w}{w^2+2cw+1} \nonumber \\
& =\int_0^{\infty}\frac{2\diff w}{(w+c)^2-(c^2-1)}=\int_0^{\infty}\frac{2\diff w}{(w+c)^2-\sinh^2(x)} \nonumber \\
& =\frac{1}{\sinh(x)}\int_0^{\infty}\left(\frac{1}{w+c-\sinh(x)}-\frac{1}{w+c+\sinh(x)}\right)\diff w=\frac{1}{\sinh(x)}\ln\frac{w+c-\sinh(x)}{w+c+\sinh(x)}\Bigg|_0^{\infty} \nonumber \\
& =\frac{1}{\sinh(x)}\ln\frac{c+\sinh(x)}{c-\sinh(x)}=\frac{1}{\sinh(x)}\ln\frac{\Exp^x}{\Exp^{-x}}=\frac{2x}{\sinh(x)}
\end{align}
Because
\begin{equation}
\od{}{c}\frac{1}{\cosh(u)+c}=\frac{-1}{\left(\cosh(u)+c\right)^2},
\end{equation}
the autocorrelation term becomes
\begin{equation}
A(x)=-2\dod{F}{c}.
\end{equation}
With the chain rule,
\begin{align}
A(x) & =-2\dod{F}{x}\dod{x}{c} \nonumber \\
& =-4\frac{\sinh(x)-x\cosh(x)}{\sinh^2(x)}\frac{1}{\sqrt{c^2-1}}=4\frac{x\cosh(x)-\sinh(x)}{\sinh^3(x)}.
\end{align}